\documentclass[amssymb,aps,prd,floatfix,onecolumn]{revtex4}
\usepackage{hyperref}
\usepackage[intlimits]{amsmath}
\usepackage{graphicx}
\usepackage{tensor}
\usepackage[dvipsnames]{xcolor}
\usepackage[utf8]{inputenc}
\usepackage{natbib}
\usepackage{enumitem}
\usepackage{ulem}

\usepackage{listings}

\begin{document}

\newcommand{\cc}[1]{\overline{#1}}

\title{Monte Carlo methods for stationary solutions of general-relativistic Vlasov systems: Planar accretion onto a moving Schwarzschild black hole}

\author{Adam Cie\'{s}lik}
\email{adam.cieslik@doctoral.uj.edu.pl}
\affiliation{Szko{\l}a Doktorska Nauk Ścisłych i Przyrodniczych, Uniwersytet Jagiello\'{n}ski}
\affiliation{Instytut Fizyki Teoretycznej, Uniwersytet Jagiello\'{n}ski, {\L}ojasiewicza 11, 30-348 Krak\'{o}w, Poland}
\author{Patryk Mach}
\email{patryk.mach@uj.edu.pl}
\affiliation{Instytut Fizyki Teoretycznej, Uniwersytet Jagiello\'{n}ski, {\L}ojasiewicza 11, 30-348 Krak\'{o}w, Poland}
\author{Andrzej Odrzywo{\l}ek}
\email{andrzej.odrzywolek@uj.edu.pl}
\affiliation{Instytut Fizyki Teoretycznej, Uniwersytet Jagiello\'{n}ski, {\L}ojasiewicza 11, 30-348 Krak\'{o}w, Poland}

\begin{abstract}
We perform Monte Carlo simulations of stationary planar accretion of a collisionless gas onto a moving Schwarzschild black hole. In this work---a sequel to our previous paper on the Monte Carlo method for stationary general-relativistic Vlasov systems---we demonstrate that our approach can be extended beyond the simplifying assumptions of the spherical symmetry or axial symmetry in the planar case. Our method of computing observable quantities, such as the particle current density, can be regarded as a rigorous coarse graining scheme, adapted to a numerical grid. Main difficulties are related to the appropriate parametrization of particle trajectories and a selection of parameters consistent with assumed requirements on the distribution function.
\end{abstract}

\maketitle

\section{Introduction}

The general-relativistic kinetic model provides a description of gases in strong gravitational fields, complementary to hydrodynamics or magnetohydrodynamics. If collisions between individual gas particles are rare and the mean free path between collisions is large in comparison to the characteristic length scale in the system, the assumption of the local thermodynamical equilibrium may not hold and consequently the hydrodynamical approach may not be valid.

The general-relativistic kinetic description was applied to stellar systems \cite{ShapiroTeukolsky1993a,ShapiroTeukolsky1993b}, dark matter models \cite{DM_Vlasov,MachOdrzywolek2021a,MachOdrzywolek2021b,CieslikMachOdrzywolek2022,Shapiro2023} and matter flows around black holes \cite{MachOdrzywolek2021a,MachOdrzywolek2021b,CieslikMachOdrzywolek2022,Shapiro2023,RiosecoSarbach2017a,RiosecoSarbach2017b,RiosecoSarbach2018,CieslikMach2020,Gamboa2022,GabarreteSarbach2022,MachOdrzywolek2022,MachOdrzywolek2023,GabarreteSarbach2023,RiosecoSarbach2024}. It also provides a model used in mathematical general relativity \cite{AndreassonRein2007,AndreassonKunzeRein2011,AndreassonReview,AndreassonKunzeRein2014,AmesAndreassonLogg2016,AmesAndreassonRinne2021,Andreasson2021,HadzicLinRein2021,AmesAndreasson2024}. A numerical implementation of the general-relativistic kinetic model, the so-called particle-in-cell (PIC) method, is used to simulate the dynamics of magnetized accretion flows onto black holes \cite{Parfrey2019,Bransgrove2021,Crinquand2022,Galishnikova2023}.

In a recent paper \cite{MachCieslikOdrzywolek2023} we proposed a Monte Carlo method of solving the general-relativistic Vlasov (Liouville) equation describing the gas of non-colliding particles moving in a given spacetime. In the kinetic model free particles of the gas move along segments of geodesics, from one individual collision to the next, or simply along geodesics if collisions are absent. To be more precise one should say that the gas is described by a continuous one-particle distribution function which, as long as collisions are neglected, remains constant along geodesics, understood as curves in the phase space. The Monte Carlo method proposed in \cite{MachCieslikOdrzywolek2023} consists of three elements: \textit{i)} Choosing a finite random set of parameters characterizing geodesic trajectories of individual particles (initial condition and/or constants of motion) selected from a suitable continuous distribution; \textit{ii)} Solving geodesic equations for the selected sample of parameters; \textit{iii)} Computing suitable averages that approximate macroscopic observable quantities---particle current density, the energy-momentum tensor, etc. Out of these three points, only the second---solving geodesic equations---is relatively straightforward. The third point is implemented as follows. We consider observable quantities associated with a Dirac delta type distribution that corresponds to a finite number of point particles moving along given trajectories. Observable quantities associated with a continuous distribution function are then approximated by averaging over suitable hypersurfaces (either spacelike or timelike). This amounts to counting (with appropriate weight factors) the intersections of particle trajectories with given hypersurfaces. One could say that our method provides a particular coarse graining procedure, adapted to a given numerical grid. We emphasize that the novelty of counting over timelike hypersurfaces allows us to deal directly with stationary distributions.

The first point remains notoriously difficult. In principle any selection of geodesics would lead to a valid solution of the Vlasov equation. On the other hand, one would expect that physically sound solutions should correspond to distribution functions characterized by well defined properties (e.g., boundary or initial conditions). In practice, selecting an appropriate sample of geodesic parameters can be problematic. In \cite{MachCieslikOdrzywolek2023} we solved this task in particular examples, providing a collection of results related to the Bondi-type accretion onto a Schwarzschild black hole. We searched for stationary solutions, either spherically symmetric or axially symmetric and planar (all particles moving in single plane).

In this paper we extend our analysis presented in \cite{MachCieslikOdrzywolek2023} and investigate stationary solutions of a planar accretion problem with asymptotic boundary conditions that describe the gas moving uniformly in one direction. Such models constitute a planar equivalent of the relativistic accretion of the Vlasov gas onto a moving Schwarzschild black hole, derived in \cite{MachOdrzywolek2021a,MachOdrzywolek2021b}, and an intermediate step towards a similar analysis in the Kerr spacetime \cite{MachOdrzywolek2023}. While a planar model of this type might seem slightly artificial, it provides a perfect opportunity to discuss the details related to the choice of the sample of geodesics as well as the averaging procedure of our Monte Carlo method, outside the safe assumption of spherical symmetry.

The order of this paper is as follows. In Section \ref{sec:phasespace} we introduce our notation and collect basic definitions regarding the phase space description of gases in the general-relativistic kinetic theory. Section \ref{sec:planaraccretion} introduces the planar accretion model in the Schwarzschild spacetime, which we deal with in this paper, and which serves as an illustration and test of our method. Our Monte Carlo method is described in Sec.\ \ref{sec:montecarlo}. We start with a discussion of the general coarse graining scheme and then show its application on a polar grid in the Schwarzschild spacetime. A detailed description of our simulations is provided in Sec.\ \ref{sec:planarsccretionsimulation}. We discuss, in particular, the selection of particle trajectories corresponding to the assumed asymptotic distribution. Concluding remarks are given in Sec.\ \ref{sec:conclusions}. In Appendix \ref{appendix:distribution} we show a derivation of the distribution function of our planar accretion model. For the sake of generality, this distribution is derived also for a non-planar flow, in which the motion of the gas is not confined to a single plane. Appendix \ref{appendix:Currensts} provides a derivation of the particle current surface density associated with our planar model.

We used \textit{Wolfram Mathematica} \cite{Wolfram} to perform our simulations. Sample packages containing our numerical code will be publicly available in \cite{Notebooks}.

Throughout this paper we use the geometric system of units with $c = G = 1$, where $c$ is the speed of light, and $G$ denotes the gravitational constant. The signature of the metric is $(-,+,+,+)$.


\section{Phase-space description}
\label{sec:phasespace}

In this section we recall some basic notions of the general-relativistic kinetic theory, mostly to fix the notation. An excellent fresh introduction to the kinetic model in general relativity can be found in \cite{Acuna2022}. A lot of information can be found in \cite{AndreassonReview,SarbachZannias2014}. For the (special) relativistic kinetic theory one can consult \cite{Groot,Cercignani}.

Consider a spacetime manifold $(\mathcal M, g)$, where $\mathcal M$ is the spacetime and $g$ denotes the metric tensor. Let $x$ denote a point (an event) in $\mathcal M$. We will work in a Hamiltonian formulation in which the momenta are expressed as covectors. A particle at $x \in \mathcal M$ will be characterized by its four-momentum $p \in T_x^\ast \mathcal M$, the cotangent space at $x$. The one-particle phase space $\Gamma$ is a region in the cotangent bundle defined as
\begin{equation}
T^\ast \mathcal M = \{(x,p) \colon x \in \mathcal M, \, p \in T_x^\ast \mathcal M\}.
\end{equation}
Ensemble-averaged properties of the gas can be described by a one-particle distribution function $\mathcal F$ defined on $\Gamma \subseteq T^\ast \mathcal M$. In this work, we narrow our analysis to the gas of identical particles. Such a model is known as the ``simple gas'' \cite{Israel1963}. In principle, this assumption would allow us to restrict the phase space to the mass shell $\Gamma^+_m$ which we define as
\begin{eqnarray}
\Gamma^+_m & = & \{(x,p) \in T^\ast \mathcal M \colon g^{\mu \nu} p_\mu p_\nu = - m^2, \, p \text{ is future-directed}\},
\end{eqnarray}
where $m > 0$ denotes the rest mass of a single gas particle. Another possibility, which we choose in this paper, is to enforce the mass-shell condition by assuming that $\mathcal{F} \sim \delta(\sqrt{-p_\mu p^\mu} - m)$.
        
The Hamiltonian of a free-particle geodesic motion can be taken in the form
\begin{equation}
\label{Hamiltonian}
H(x^\mu, p_\nu) = \frac{1}{2} g^{\mu\nu}(x) p_\mu p_\nu,
\end{equation}
where \( (x^\mu, p_\nu) \) are regarded as canonical variables. Geodesic equations can be then written in the Hamiltonian form
\begin{equation}
\frac{dx^\mu}{d\tau} = \frac{\partial H}{\partial p_\mu}, \quad \frac{dp_\nu}{d\tau} = - \frac{\partial H}{\partial x^\nu},
\end{equation}
where the parameter $\tau$ is selected as $\tau = s/m$ and $s$ denotes the proper time. Thus $H = \tfrac{1}{2}g^{\mu\nu}p_\mu p_\nu = -\tfrac{1}{2} m^2$. For the ensemble of non-colliding particles, the distribution function $\mathcal{F}$ is conserved along geodesics, which can be expressed as
\begin{equation}\label{Vlasov}
\frac{d}{d\tau} \mathcal{F}(x^\mu (\tau), p_\nu (\tau)) = \{ H, \mathcal{F} \} = 0,
\end{equation}
where $\{ \cdot , \cdot \}$ denotes the Poisson bracket. Equation (\ref{Vlasov}) is known as the Vlasov or Liouville equation. Assuming the Hamiltonian (\ref{Hamiltonian}), one can rewrite Eq.\ (\ref{Vlasov}) in the textbook form
\begin{equation}
g^{\mu\nu}p_\nu \frac{\partial \mathcal{F}}{\partial x^\mu} - \frac{1}{2} p_\alpha p_\beta \frac{\partial g^{\alpha \beta}}{\partial x^\mu} \frac{\partial \mathcal{F}}{\partial p_\mu} = 0.
\end{equation}

The one-particle distribution function has the following statistical interpretation. Let $S$ be a spacelike hypersurface in $\mathcal M$, and let $s$ be a future-directed unit vector normal to $S$. The averaged number of particle trajectories whose projections on $\mathcal M$ intersect $S$ can be expressed as
\begin{equation}
\label{defN}
\mathcal{N}[S] = - \int_S \left[ \int_{P_x^+} \mathcal{F}(x,p) p_\mu s^\mu \mathrm{dvol}_x(p) \right] \eta_S,
\end{equation}
where
\begin{equation} 
P_x^+ = \{ p \in T_x^\ast M \colon g^{\mu \nu} p_\mu p_\nu < 0, \, p \text{ is future directed} \},
\end{equation}
$\eta_S$ denotes the induced volume element on $S$, and $\mathrm{dvol}_x(p)$ is the volume element on $P_x^+$. In local adapted coordinates $\mathrm{dvol}_x(p)$ is given as
\begin{equation} 
\mathrm{dvol}_x(p) = \sqrt{- \mathrm{det} \, g^{\mu \nu}(x)} dp_0 dp_1 dp_2 dp_3. 
\end{equation}
Equation \eqref{defN} gives rise to the following definition of the so-called particle current density:
\begin{equation} 
\label{Jmudef}
\mathcal{J}_\mu(x) = \int_{P_x^+} \mathcal{F}(x,p) p_\mu \mathrm{dvol}_x(p). 
\end{equation}
Thus, the number of particles in $S$ can be re-expressed as
\begin{equation}
\label{NJ}
\mathcal{N}[S] = - \int_S \mathcal{J}_\mu s^\mu \eta_S. 
\end{equation}
Using Eq.\ \eqref{Vlasov}, one can show that the particle current density satisfies the conservation law $\nabla_\mu \mathcal{J}^\mu = 0$, which again justifies formula \eqref{NJ}. The particle number density can be defined covariantly as 
\begin{equation}
n = \sqrt{- \mathcal{J}_\mu \mathcal{J}^\mu}.
\end{equation}
Such a definition of the particle number density---as a scalar quantity---allows for a connection with general-relativistic hydrodynamics, where the particle number density as well as the energy density and the pressure are usually defined as scalar functions. Alternatively, one can work with the components of $\mathcal{J}_\mu$.

One should also note that in integral formulas of type (\ref{defN}) or (\ref{Jmudef}) one can choose to transfer some information about the distribution function $\mathcal F$ from the distribution function itself to the specification of the integration domain (e.g., $S, P_x^+$), or vice versa---one can consider a fairly general phase space and impose restrictions directly on the distribution function. To some degree we will also make use of this freedom in this paper, e.g., in Eqs.\ (\ref{Jmuexact}), where we will specify a convenient form of the distribution function, but restrict the integrals over momenta to regions in the phase space available to the motion of particles of a given type (particles absorbed and scattered by the black hole).

In this work we deal with stationary systems. In particular, in our prescription of the Monte Carlo estimators for the particle current density $\mathcal J_\mu$ we will, at some stage, explicitly assume that both the distribution function $\mathcal F$ and the components $\mathcal J_\mu$ are stationary. A spacetime $(\mathcal M, g)$ is called stationary if there exist a Killing vector $k$ that is asymptotically timelike and generates one parameter groups of isometries. A definition of a stationary distribution function $\mathcal F$ defined on $T^\ast \mathcal M$ can be formulated in terms of the lift of the Killing vector $k$ to the cotangent bundle, defined as follows. For a general Killing vector field $\xi$ on $(\mathcal M,g)$ such that its value at point $x \in \mathcal M$ can be expressed as
\begin{equation}
\xi_x = \xi^\mu(x) \left.  \frac{\partial}{\partial x^\mu}  \right|_x,
\end{equation}
one defines an associated vector field on the cotangent bundle $T^\ast \mathcal M$, the so-called lift of $\xi$, by
\begin{equation}
\hat \xi_{(x,p)} =  \xi^\mu(x) \left.  \frac{\partial}{\partial x^\mu}  \right|_{(x,p)} - p_\alpha \frac{\partial \xi^\alpha}{\partial x^\mu}(x) \left. \frac{\partial}{\partial p_\mu} \right|_{(x,p)}.
\end{equation}
The lifts defined in this way satisfy Killing equations with respect to the so-called Sasaki metric---a natural metric on $T^\ast \mathcal M$ \cite{Acuna2022}. We say that a distribution function $\mathcal F$ is stationary if
\begin{equation}
\label{stationarygeneral}
\mathcal L_{\hat k} \mathcal F = k^\mu \frac{\partial \mathcal F}{\partial x^\mu} - p_\alpha \frac{\partial k^\alpha}{\partial x^\mu} \frac{\partial \mathcal F}{\partial p_\mu} = 0,
\end{equation}
where $\mathcal L_{\hat k}$ denotes the Lie derivative with respect to the lift $\hat k$.


\section{Planar stationary accretion in the Schwarzchild spacetime}
\label{sec:planaraccretion}

\subsection{Geodesic motion in the Schwarzschild spacetime}

For simplicity, we work in standard Schwarzshild coordinates $(t,r,\theta,\varphi)$, and use the metric in the form
\begin{equation}
\label{metric}
g = - N dt^2 +\frac{1}{N} dr^2 + r^2 (d\theta^2 + \sin^2{\theta} d\varphi^2),
\end{equation}
where 
\begin{equation}
N = 1 - \frac{2M}{r}.
\end{equation}
We have $\sqrt{-\mathrm{det}\, g^{\mu \nu}} = 1/(r^2 \sin \theta)$.

The Hamiltonian associated with a geodesic motion in the spacetime with the metric \eqref{metric}, is given by
\begin{equation}
H = \frac{1}{2} \left[ -\frac{1}{N} p_t^2 + N p_r^2 + \frac{1}{r^2} \left( p_\theta^2 + \frac{p_\varphi^2}{\sin^2{\theta}} \right) \right].
\end{equation}
It is well-known that the energy $E = -p_t$, the azimuthal component of the angular momentum $l_z = p_\varphi$, and the total angular momentum
\begin{equation}
l = \sqrt{p_\theta^2 + \frac{p_\varphi^2}{\sin^2{\theta}}}
\end{equation}
are conserved. The fourth constant of motion is the Hamiltonian itself, or equivalently the particle rest mass $m = \sqrt{-2H}$.

For geodesics with fixed values of $E$, $l$, $l_z$, and $m$, the radial momentum is determined by
\begin{equation}
\label{pr}
p_r = \frac{\epsilon_r}{N}\sqrt{E^2 - U(r)},
\end{equation}
where $\epsilon_r = \mp 1$ indicate an inward ($-$)/outward ($+$) motion and,
\begin{equation}
\label{U_criminal}
U(r) = N \left(m^2 + \frac{l^2}{r^2}\right)
\end{equation}
represents an effective radial potential. The component $p_\theta$ can be expressed as
\begin{eqnarray}\label{ptheta}
p_\theta = \epsilon_\theta \sqrt{l^2 - \frac{l_z^2}{\sin^2{\theta}}}.
\end{eqnarray}
For the motion confined to the equatorial plane we have $p_\theta = 0$ and consequently $l^2 = l_z^2$. In our planar models we will generally use the constant $l \ge 0$, and define $l_z = \epsilon_\varphi l$, where $\epsilon_\varphi = \pm 1$ corresponds to the motion with a growing ($+$)/decreasing ($-$) angle $\varphi$.

In this paper we will frequently use dimensionless quantities $\xi$, $\varepsilon$, $\lambda_z$, and $\lambda$ defined as (see, e.g., \cite{RiosecoSarbach2017a})
\begin{equation}
\label{dimensionless}
r = M \xi, \quad E = m \varepsilon, \quad l_z = M m \lambda_z, \quad l = M m \lambda.
\end{equation}
The radial potential \eqref{U_criminal} can be expressed in the form $U(r) = m^2 U_\lambda(\xi)$, where
\begin{equation}
U_\lambda(\xi) = \left( 1 - \frac{2}{\xi} \right) \left( 1 + \frac{\lambda^2}{\xi^2} \right).
\end{equation}
The expression for $p_r$ reads
\begin{equation}
p_r = \epsilon_r \frac{m}{N} \sqrt{\varepsilon^2 - U_\lambda(\xi)}.
\end{equation}

In this work we only take into account unbound orbits that can reach $\xi \to \infty$. The region in which the motion is permitted is defined by the condition $\varepsilon^2 - U_\lambda(\xi) \ge 0$. Since $U_\lambda(\xi)$ approaches 1 when $\xi \to \infty$, unbound orbits are characterized by the energy $\varepsilon \ge 1$. The properties of the radial potential $U_\lambda(\xi)$ are well known (see, for instance, the Appendix A in \cite{RiosecoSarbach2017a} for a detailed discussion), and they allow for the following further classification of unbound trajectories. Orbits with the angular momentum $\lambda < \lambda_c(\varepsilon)$, where
\begin{equation}\label{lambdac}
\lambda_c(\varepsilon)^2 = \frac{12}{1-\frac{4}{\left(\frac{3 \varepsilon }{\sqrt{9 \varepsilon ^2-8}}+1\right)^2}},
\end{equation}
plunge into the black hole. We will refer to such orbits as absorbed ones. Unbound orbits with $\lambda > \lambda_c(\varepsilon)$ are scattered by the centrifugal barrier. We will refer to these trajectories as scattered ones. The maximum angular momentum for a scattered unbound orbit reaching radius $\xi$ is given by
\begin{equation}\label{lambdaMax}
\lambda_\mathrm{max} (\varepsilon, \xi) = \xi \sqrt{\frac{\varepsilon^2}{1 - \frac{2}{\xi}} - 1},
\end{equation}
while the minimum permitted energy is expressed as
\begin{equation}
\label{limit2}
\varepsilon_\mathrm{min}(\xi) =
\begin{cases} \infty, & \xi \leq 3 ,\\
\sqrt{\left(1 - \frac{2}{\xi} \right) \left(1 + \frac{1}{\xi - 3} \right)}, & 3 < \xi < 4, \\
1, & \xi \ge 4.
\end{cases}
\end{equation}
The infinity symbol in Eq.\ \eqref{limit2} means that no unbound scattered trajectory can reach the region with $\xi < 3$. The radius $\xi = 3$ corresponds to the so-called circular photon orbit (or the photon sphere). A circular orbit at $\xi = 4$ is known as the marginally bound one. The Innermost Stable Circular Orbit (ISCO) occurs at $\xi = 6$. These orbits are, to some degree, relevant to the interpretation of our results presented in Sec.\ \ref{sec:Results}.


\subsection{Planar distribution function}

We will consider a planar accretion model, in which the gas particles are only moving within the equatorial plane. The restriction to the equatorial plane can be imposed on the form of the distribution function as follows. We set
\begin{equation}
\label{twodf}
\mathcal F(t,r,\theta,\varphi;p_\alpha) = \frac{1}{r} \delta\left(\theta - \frac{\pi}{2}\right)  F(t,r,\varphi;p_\alpha), \quad \mathcal J_\mu(t,r,\theta,\varphi) = \frac{1}{r} \delta\left(\theta - \frac{\pi}{2}\right) J_\mu(t,r,\varphi).
\end{equation}
Note that the factor $\delta(\theta - \pi/2)/r$ can be written as $\delta(\theta - \pi/2)/r = \delta(z)$, where $z = r \cos \theta$. The vector $\partial_z$, associated with the coordinate $z$, is normal to the equatorial plane $\theta = \pi/2$. A substitution of Eq.\ (\ref{twodf}) in Eq.\ (\ref{Jmudef}) implies that the particle current surface density $J_\mu$ can be written as
\begin{equation}
J_\mu(t,r,\varphi) = \int_{P^+_x} F(t,r,\varphi;p_\alpha) p_\mu \mathrm{dvol}_x(p).
\end{equation}

The restriction of the coordinates $(t,r,\theta,\varphi)$ to the equatorial plane $\theta = \pi/2$ implies a corresponding restriction in momenta ($p_\theta=0$) as well. We set
\begin{equation}
F(t,r,\varphi,p_t,p_r,p_\theta,p_\varphi) = r \delta(p_\theta) f(t,r,\varphi,p_t,p_r,p_\varphi),
\end{equation}
where again the factor $r \delta(p_\theta)$ can be written as $r \delta(p_\theta) = \delta(p_z)$ (at the equatorial plane). This follows immediately from the relation $p_z = - p_\theta/r$, holding at the equatorial plane. Note that distribution functions $\mathcal F$ and $f$ are conveniently related as
\begin{equation}
\label{dist_flat}
\mathcal F(t,r,\theta,\varphi,p_t,p_r,p_\theta,p_\varphi) = \delta\left(\theta - \frac{\pi}{2}\right)\delta(p_\theta)f(t,r,\varphi,p_t,p_r,p_\varphi).
\end{equation}

The particle current surface density can be now expressed in the form
\begin{equation}
\label{jmusurface}
J_\mu(t,r,\varphi) = \int_{P^+_x} r \delta(p_\theta) f(t,r,\varphi,p_t,p_r,p_\varphi) p_\mu \frac{dp_t dp_r dp_\theta dp_\varphi}{r^2} = \int_{\bar P^+_x} f(t,r,\varphi,p_t,p_r,p_\varphi) p_\mu \frac{dp_t dp_r  dp_\varphi}{r},
\end{equation}
where
\begin{equation}
    \bar P^+_x = \{ (p_t,p_r,p_\varphi) \colon \bar g^{\mu \nu} p_\mu p_\nu < 0, \, p \text{ is future directed} \},
\end{equation}
and where we have used the fact that at the equatorial plane the volume element $\mathrm{dvol}_x(p)$ reads
\begin{equation}
\mathrm{dvol}_x(p) = \frac{1}{r^2 \sin \theta} dp_t dp_r dp_\theta dp_\varphi = \frac{1}{r^2} dp_t dp_r dp_\theta dp_\varphi.
\end{equation}
Here
\begin{equation}
\label{gamma3D}
\bar g = - N dt^2 + N dr^2 + r^2 d\varphi^2
\end{equation}
denotes the metric induced at the equatorial plane.

In the remainder of this paper we will also use the following covariant particle number surface density
\begin{equation}
\label{n_s}
n_s = \sqrt{- \bar g^{\mu \nu}J_\mu J_\nu}.
\end{equation}


\subsection{Planar accretion model}

We consider a stationary gas of particles moving within the equatorial plane. Asymptotically, the gas is assumed to be homogeneous and described by the two dimensional Maxwell-J\"{u}ttner distribution \cite{Juttner1911a,Juttner1911b}, boosted with a constant velocity $v$ along the $x$ axis. In other words, we think of a planar distribution of gas that is in equilibrium, and moves with a uniform velocity $v$ in a given direction. This is mathematically equivalent to a model with a black hole moving with a constant velocity with respect to a gas remaining asymptotically at rest, and a majority of works dealing with the accretion onto a moving black hole assume this point of view. We show in Appendix \ref{appendix:distribution} that the distribution function corresponding to these asymptotic conditions can be written as
\begin{equation}
\label{dist}
f(\xi,\varphi,m,\varepsilon,\lambda) = \alpha \delta(m - m_0) \exp{\left\{ -\beta \gamma\left[ \varepsilon - \epsilon_r v  \sqrt{\varepsilon^2 - 1} \cos\left[ \varphi + \epsilon_\varphi \epsilon_r  X(\xi,\varepsilon,\lambda)  \right] \right] \right\} },
\end{equation}
where the elliptic function $X(\xi,\varepsilon,\lambda)$ is defined as
\begin{equation}
\label{EllipticX}
X(\xi, \varepsilon, \lambda) = \lambda \int_\xi^\infty \frac{d\xi'}{\xi'^2 \sqrt{\varepsilon^2 -U_\lambda(\xi')}} = \lambda \int_\xi^\infty \frac{d\xi'}{\xi'^2 \sqrt{\varepsilon^2 - \left( 1-\frac{2}{\xi'} \right)\left( 1 + \frac{\lambda^2 }{\xi'^2}\right)}}.
\end{equation}
Here $\gamma = 1/\sqrt{1 - v^2}$ is the Lorentz factor associated with the velocity $v$, $m_0$ denotes the rest mass of a single particle, $\alpha$ and $\beta$ are constants. The constant $\beta$ is related to the asymptotic temperature of the gas by $\beta = m_0/(k_\mathrm{B} T)$, where $k_\mathrm{B}$ denotes the Boltzmann constant. The distribution function specified in Eq.\ (\ref{dist}) is a planar equivalent of the distribution function derived in \cite{MachOdrzywolek2021a,MachOdrzywolek2021b}. 

In our models we will only take into account unbound orbits---trajectories that can reach (or originate at) infinity. The particle current surface density $J_\mu$ can be expressed as a sum $J_\mu = J_\mu^\mathrm{(abs)} + J_\mu^\mathrm{(scat)}$, where the part $J_\mu^\mathrm{(abs)}$ refers to absorbed orbits and the part $J_\mu^\mathrm{(scat)}$ refers to scattered trajectories. 
 
The components $J_\mu^\mathrm{(abs)}$ and $J_\mu^\mathrm{(scat)}$ can be computed using the following formulas (details of the drivation are given in Appendix \ref{appendix:Currensts})
\begin{subequations}
\label{Jmuexact}
\begin{eqnarray}
\label{JtExact}
J_t^\mathrm{(abs)}(\xi, \varphi) & = & 
- \frac{ 2\alpha m_0^3}{\xi} \int_1^\infty d \varepsilon  \; \varepsilon  \int_0^{\lambda_{c}(\varepsilon)} \frac{d \lambda}{\sqrt{\varepsilon^2 - U_\lambda(\xi)}}  \mathrm{e}^{-\beta \gamma\left[ \varepsilon +  v  \sqrt{\varepsilon^2 - 1} \cos (\varphi) \cos\left[ X(\xi,\varepsilon,\lambda) \right] \right] } \\ \nonumber
&& \times \cosh\left[ \beta \gamma v \sqrt{\varepsilon^2 -1} \sin (\varphi) \sin\left[ X(\xi,\varepsilon,\lambda) \right] \right] ,\\
J_t^\mathrm{(scat)}(\xi, \varphi) & = & 
- \frac{4 \alpha m_0^3}{\xi} \int_{\varepsilon_\mathrm{min}(\xi)}^\infty d \varepsilon e^{-\beta\gamma\varepsilon}  \; \varepsilon  \int_{\lambda_{c}(\varepsilon)}^{\lambda_\mathrm{max}(\xi,\varepsilon)} \frac{d \lambda}{\sqrt{\varepsilon^2 - U_\lambda(\xi)}}  \cosh\left[ \beta \gamma v \sqrt{\varepsilon^2 -1} \cos (\varphi) \cos\left[ X(\xi,\varepsilon,\lambda) \right] \right] \\ \nonumber
&& \times \cosh\left[ \beta \gamma v \sqrt{\varepsilon^2 -1} \sin (\varphi) \sin\left[ X(\xi,\varepsilon,\lambda) \right] \right], \\
\label{JrExact}
J_r^\mathrm{(abs)}(\xi, \varphi) & = & 
- \frac{2 \alpha m_0^3}{\xi-2} \int_1^\infty d \varepsilon    \int_0^{\lambda_{c}(\varepsilon)} d \lambda  \mathrm{e}^{-\beta \gamma\left[ \varepsilon +  v  \sqrt{\varepsilon^2 - 1} \cos (\varphi) \cos\left[ X(\xi,\varepsilon,\lambda) \right] \right] } \\ \nonumber
&&\times \cosh\left[ \beta \gamma v \sqrt{\varepsilon^2 -1} \sin (\varphi) \sin\left[ X(\xi,\varepsilon,\lambda) \right] \right] ,\\
J_r^\mathrm{(scat)}(\xi, \varphi) & = & 
\frac{4 \alpha m_0^3}{\xi-2} \int_{\varepsilon_\mathrm{min}(\xi)}^\infty d \varepsilon e^{-\beta\gamma\varepsilon}  \; \int_{\lambda_{c}(\varepsilon)}^{\lambda_\mathrm{max}(\xi,\varepsilon)} d \lambda \sinh\left[ \beta \gamma v \sqrt{\varepsilon^2 -1} \cos (\varphi) \cos\left[ X(\xi,\varepsilon,\lambda) \right] \right] \\ \nonumber
&& \times \cosh\left[ \beta \gamma v \sqrt{\varepsilon^2 -1} \sin (\varphi) \sin\left[ X(\xi,\varepsilon,\lambda) \right] \right], \\
\label{JphiExact}
J_\varphi^\mathrm{(abs)}(\xi, \varphi) & = &  -\frac{ 2\alpha m_0^3 M}{\xi} \int_1^\infty d \varepsilon    \int_0^{\lambda_{c}(\varepsilon)} d \lambda \frac{\lambda}{\sqrt{\varepsilon^2 - U_\lambda(\xi)}} \mathrm{e}^{-\beta \gamma\left[ \varepsilon +  v  \sqrt{\varepsilon^2 - 1} \cos (\varphi) \cos\left[ X(\xi,\varepsilon,\lambda) \right] \right] } \\ \nonumber
&&\times \sinh\left[ \beta \gamma v \sqrt{\varepsilon^2 -1} \sin (\varphi) \sin\left[ X(\xi,\varepsilon,\lambda) \right] \right] ,\\
J_\varphi^\mathrm{(scat)}(\xi, \varphi) & = &  -\frac{4 \alpha m_0^3M}{\xi} \int_{\varepsilon_\mathrm{min}(\xi)}^\infty d \varepsilon  e^{-\beta\gamma\varepsilon}  \int_{\lambda_{c}(\varepsilon)}^{\lambda_\mathrm{max}(\xi,\varepsilon)} d \lambda \frac{ \lambda}{\sqrt{\varepsilon^2 - U_\lambda(\xi)}}  \sinh\left[ \beta \gamma v \sqrt{\varepsilon^2 -1} \sin(\varphi) \sin\left[ X(\xi,\varepsilon,\lambda) \right] \right] \\ \nonumber
&& \times \cosh\left[ \beta \gamma v \sqrt{\varepsilon^2 -1} \cos (\varphi) \cos\left[ X(\xi,\varepsilon,\lambda) \right] \right].
\end{eqnarray}
\end{subequations}
        
The parameter $\alpha$ can be expressed in terms of the asymptotic surface number density, which we define as $n_{s,\infty} = \lim_{r \to \infty} n_s = \lim_{r \to \infty} \sqrt{-J_\mu J^\mu}$. Since, asymptotically, the gas is described by a two dimensional Maxwell-J\"{u}ttner distribution, one can show that
\begin{equation}
\label{nsinf}
n_{s,\infty}= 2\pi\alpha m_0^3 \frac{1+\beta}{\beta^2} e^{-\beta}.
\end{equation}
A calculation leading to Eq.\ (\ref{nsinf}) can be performed assuming the flat Minkowski metric, and it can be found in \cite{CieslikMachOdrzywolek2022}. Note that a boosted Maxwell-J\"{u}ttner distribution must satisfy
\begin{equation}
\lim_{r \to \infty} J_t = - n_{s,\infty} \gamma,
\end{equation}
which can also be checked by a direct calculation.


\section{Monte Carlo approach}
\label{sec:montecarlo}

\subsection{Coarse graining method}

We will now discuss the main ideas behind our Monte Carlo method, starting with a general averaging scheme. The choice of trajectories of individual particles will be discussed in the subsequent section in a particular example of the planar accretion model.

Consider a sample of $N$ particle trajectories in the phase space $T^\ast \mathcal M$ understood as mappings (curves) $\gamma_{(i)} \colon \tau \mapsto \left( x^\mu_{(i)}(\tau),p^{(i)}_\nu(\tau) \right)$, $i = 1, \dots, N$. The distribution function associated with this sample can be written in the form
\begin{equation}
\mathcal F^{(N)} (x^\mu, p_\nu) = \sum_{i = 1}^N \int \delta^{(4)} \left( x^\mu - x^\mu_{(i)}(\tau) \right) \delta^{(4)} \left( p_\nu - p_\nu^{(i)}(\tau) \right) d\tau,
\label{dfdiscrete}
\end{equation}
which can be understood as a general-relativistic version of an analogous expression known from the (special) relativistic kinetic theory (\cite{Groot}, p.\ 14, Eq.\ (A6)). In the special relativistic setting this function was also introduced in \cite{VanKampen1969}, although with a slightly different normalization, i.e., divided by the number of trajectories $N$. The distribution function $\mathcal F^{(N)}$ represents a collection of point-like test particles, and therefore it provides a fine-grained description of matter. The essence of the Monte Carlo method introduced in \cite{MachCieslikOdrzywolek2023} and developed in this paper is a suitable averaging method, allowing for an approximation of the coarse-grained smooth distribution function $\mathcal F$. Since ultimately one is interested in observable quantities, such as the particle current density $\mathcal J_\mu$, we apply this coarse graining procedure explicitly to the components of $\mathcal J_\mu$. A general discussion of coarse graining in the context of the special-relativistic kinetic theory can be found in \cite{VanKampen1969}.

The particle current density associated with the distribution function (\ref{dfdiscrete}) reads
\begin{equation}
\mathcal J^{(N)}_\mu(x) = \int_{P^+_x} \mathcal F^{(N)} (x,p) p_\mu \sqrt{- \mathrm{det} \, g^{\delta \kappa}(x)} dp_0 \dots dp_3 =  \sum_{i = 1}^N \int \delta^{(4)} \left( x^\alpha - x^\alpha_{(i)}(\tau) \right) p_\mu^{(i)} (\tau) \sqrt{- \mathrm{det} \, g^{\delta \kappa}(x)} d \tau,
\end{equation}
where we have assumed that all momenta $p^{(i)}_\mu(\tau)$ belong to $P^+_{x(\tau)}$. It was shown in \cite{MachCieslikOdrzywolek2023} that this expressions is compatible with Eq.\ (\ref{defN}), i.e., if $S$ is a spacelike hypersurface in $\mathcal M$ such that the projections of all trajectories $\gamma_{(i)}$, $i = 1, \dots, N$, on $\mathcal M$ intersect $S$, then $\mathcal N[S] = N$, where $\mathcal N[S]$ is computed according to Eq.\ (\ref{defN}), assuming the distribution function $\mathcal F^{(N)}$.

Let us view $\mathcal F^{(N)}$ as an approximation to a smooth distribution function $\mathcal F$. Select a hypersurface $\Sigma$ in $\mathcal M$ (not necessarily a spacelike one) and a small neighbourhood $\sigma \subset \Sigma$ (a numerical cell) such that $x \in \sigma$. We estimate $\mathcal J_\mu(x)$ corresponding to $\mathcal F$ by the average
\begin{equation}
\label{average1}
\langle \mathcal J_\mu(x) \rangle = \frac{\int_{\sigma} \mathcal J^{(N)}_\mu \eta_\Sigma}{\int_{\sigma} \eta_\Sigma},
\end{equation}
where $\eta_\Sigma$ denotes the volume element on $\Sigma$. For stationary systems, $\Sigma$ can be chosen as a timelike hypersurface, adapted to isometry groups generated by the Killing vector field $k$. We will see in the next section that the coarse graining (or smoothing) procedure given by Eq.\ (\ref{average1}) amounts to counting (with suitable weights) the intersections of particle trajectories with selected surfaces.

Equation (\ref{average1}) can be justified as follows. Suppose that $\mathcal F^{(N)}$ tends to a smooth distribution function $\mathcal F$ in the sense that
\begin{eqnarray}
\lefteqn{ \int dx^0 \dots dx^3 dp_0 \dots dp_3 \frac{\mathcal F^{(N)}(x,p) - \mathcal F(x,p)}{N} \phi (x,p)} \nonumber \\
&&  = \int \sqrt{-\mathrm{det} \, g_{\alpha \beta}(x)} dx^0 \dots dx^3 \sqrt{- \mathrm{det} \, g^{\delta \kappa}(x)} dp_0 \dots dp_3 \frac{\mathcal F^{(N)}(x,p) - \mathcal F(x,p)}{N} \phi (x,p)  \to  0,
\end{eqnarray}
as $N \to \infty$, where $\phi(x,p)$ is a test function on $T^\ast \mathcal M$. Choosing $\phi(x,p) = \chi\left(p \in P_x^+\right) p_\mu \psi(x)$, we get
\begin{equation}
\int \sqrt{-\mathrm{det} \, g_{\alpha \beta}(x)} dx^0 \dots dx^3 \frac{\mathcal J^{(N)}_\mu(x)}{N} \psi (x) \to \int \sqrt{-\mathrm{det} \, g_{\alpha \beta}(x)} dx^0 \dots dx^3 \frac{\mathcal J_\mu(x)}{N} \psi (x).
\end{equation}
Here $\chi \left( p \in P_x^+ \right)$ denotes the characteristic function of the set $P_x^+$. Suppose that $\Sigma$ is a hypersurface in $\mathcal M$ chosen as in Eq.\ (\ref{average1}). Let $\gamma$ be a three dimensional metric induced on $\Sigma$. Suppose further that $\tilde \alpha$ is a function such that $\sqrt{-\mathrm{det} \, g_{\alpha \beta}(x)} = \tilde \alpha(x) \sqrt{|\mathrm{det} \, \gamma_{\alpha \beta}(x)|}$. Let $(x^0,\dots,x^3)$ denote a coordinate system adapted to the hypersurface $\Sigma$, so that $\Sigma$ is characterised by $x^0 = \bar x^0$ (say) and $\eta_\Sigma = \sqrt{|\mathrm{det} \, \gamma_{\alpha \beta}(x)|} dx^1 dx^2 dx^3$. Choose $\psi(x) = \chi(x \in \sigma) \delta(x^0 - \bar x^0)/\tilde \alpha(x)$. This gives
\begin{equation}
\label{meanvalue}
\int_{\sigma} \frac{\mathcal J_\mu^{(N)}(x)}{N} \eta_\Sigma \to \int_{\sigma} \frac{\mathcal J_\mu(x)}{N} \eta_\Sigma = \frac{\mathcal J_\mu(x_0)}{N} \int_{\sigma} \eta_\Sigma,
\end{equation}
where $x_0 \in \sigma$. The last equality follows from the mean value theorem, provided that $\sigma$ is closed, bounded, and connected, and the components of $\mathcal J_\mu(x)$ are continuous on $\sigma$. Note that this equality holds for each component separately, i.e., $x_0$ might be different for different $\mu = 0, \dots, 3$ (the mean value theorem does not hold for vector-valued functions). Equation (\ref{average1}) follows directly from the limit (\ref{meanvalue}).

In the analysis of planar models investigated in this paper, we will rather use the estimate of the particle current surface density
\begin{equation}
\label{average2}
\langle J_\mu(x) \rangle = \frac{\int_{\tilde \sigma} J^{(N)}_\mu \eta_{\tilde \Sigma}}{\int_{\tilde \sigma} \eta_{\tilde \Sigma}},
\end{equation}
defined in the analogous way, but with $\tilde \sigma$ and $\tilde \Sigma$ being two dimensional surfaces.


\subsection{Intersections of trajectories with arcs of constant radius}
\label{sec:Intersections}

\begin{figure}
    \centering
    \includegraphics[width=0.45\textwidth]{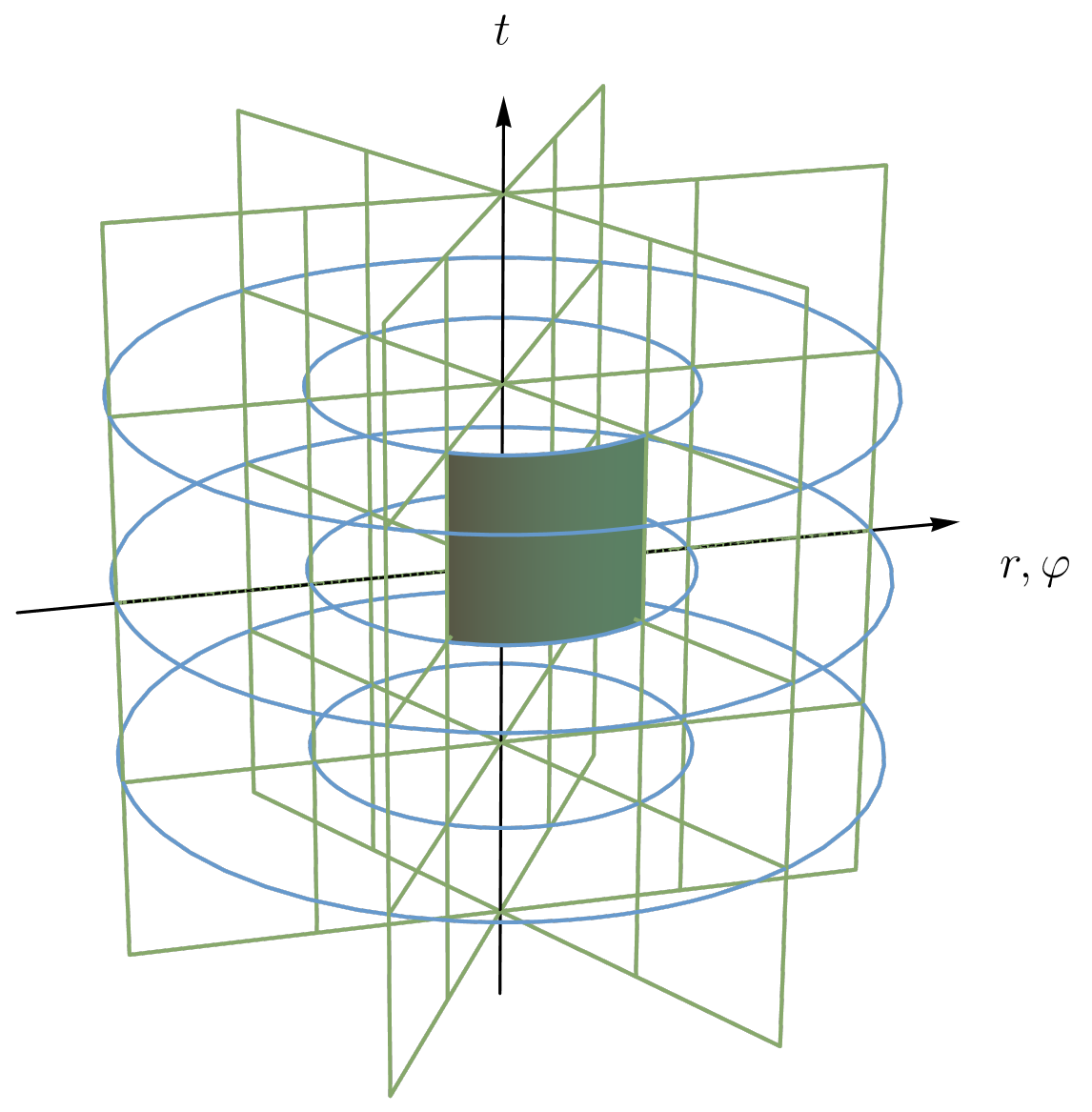}
    \caption{\label{fig:Cell_sigma}
    A visual representation of a cell $\tilde \sigma$ (green slab) defined by Eqs.\ \eqref{tildesigma} or \eqref{tildesigma2}.
    }
\end{figure}

As an example adapted to a planar stationary accretion flow in the Schwarzschild spacetime we select surfaces of constant $r = \bar r$ defined by
\begin{equation}
\tilde \Sigma = \{ (t, r,\theta,\varphi) \colon t \in \mathbb R, \, r = \bar r, \, \theta = \pi/2, \, \varphi \in [0,2\pi)  \}
\end{equation}
and cells (cf.\ Fig.~\ref{fig:Cell_sigma})
\begin{equation}
\label{tildesigma}
\tilde \sigma = \{ (t, r,\theta,\varphi) \colon  t_1 \le t \le t_2, \, r = \bar r, \, \theta = \pi/2, \, \varphi_1 \le \varphi \le \varphi_2 \}.
\end{equation}
It is also convenient to denote
\begin{equation}
\label{defS}
S = \{ (r,\theta,\varphi) \colon  r = \bar r, \, \theta = \pi/2, \, \varphi_1 \le \varphi \le \varphi_2 \}.
\end{equation}
The choice defined by Eq.\ (\ref{defS}) is different than the one made in \cite{MachCieslikOdrzywolek2023}. In our previous work, we chose $S$ as segments of constant angle $\varphi$. Both choices are adapted to stationary flows in the sense that $\tilde \sigma$ can be foliated by orbits of the timelike Killing vector field $k^\mu = (1,0,0,0)$. More precisely Let $\Phi_\tau(x_0^i)$ denote the orbit of $k$, passing through $x_0^i$ at $\tau = 0$, i.e., $\Phi_0(x_0^i) = x_0^i$. Then $\tilde \sigma$ can be expressed as the image
\begin{equation}
\label{tildesigma2}
\tilde \sigma = \Phi_{[t_1,t_2]}(S).
\end{equation}
Both choices are also compatible with a standard grid of polar coordinates (Fig.~\ref{fig:Cell_sigma}).
        
For the planar model (particles moving in the equatorial plane) the fine-grained particle current density can be expressed as
\begin{equation}
\mathcal J_\mu^{(N)} (t,r,\theta,\varphi) = \sum_{i = 1}^N \int \delta \left( \theta - \frac{\pi}{2} \right)\delta^{(3)} \left( x^\alpha - x^\alpha_{(i)}(\tau) \right)  p_\mu^{(i)}(\tau) \sqrt{- \mathrm{det} \, g^{\delta \kappa}(x)} d \tau,
\end{equation}
where
\begin{equation}
\delta^{(3)} \left( x^\alpha - x^\alpha_{(i)}(\tau) \right) = \delta \left( t - t_{(i)} (\tau) \right) \delta \left( r - r_{(i)}(\tau) \right) \delta \left( \varphi - \varphi_{(i)}(\tau) \right).
\end{equation}
A corresponding particle current surface density $J_\mu^{(N)}$ is defined as in Eq.\ (\ref{twodf}), i.e.,
\begin{equation}
\mathcal J_\mu^{(N)}(t,r,\theta,\varphi) = \frac{1}{r} \delta \left( \theta - \frac{\pi}{2} \right) J_\mu^{(N)}(t,r,\varphi).
\end{equation}
Thus,
\begin{eqnarray}
J_\mu^{(N)} (t,r,\varphi) & = & \sum_{i = 1}^N  \int r \delta^{(3)} \left( x^\alpha - x^\alpha_{(i)}(\tau) \right) p_\mu^{(i)}(\tau) \sqrt{- \mathrm{det} \, g^{\delta \kappa}(x)} d \tau,
\end{eqnarray}
where, from \eqref{metric} at the equatorial plane, $\sqrt{- \mathrm{det} \, g^{\delta \kappa}(x)} = \left( r^2 \sin \theta \right)^{-1} = 1/r^2$. The volume element on $\tilde \Sigma$ reads 
\begin{equation}
\eta_{\tilde \Sigma} =\sqrt{- \mathrm{det} \, \tilde g_{\mu \nu} (x)} \; dt d\varphi = \sqrt{1-\frac{2M}{r}} r \;dt d\varphi,
\end{equation}
where $\tilde g$ denotes the metric induced on the surface $\tilde \Sigma$:
\begin{equation}
\tilde g = -\left( 1-\frac{2M}{r}\right) dt^2 + r^2 d\varphi^2. 
\end{equation}
A direct calculation gives
\begin{eqnarray}
\label{^a57e51}
\int_{\tilde \sigma} J_\mu^{(N)} \eta_{\tilde \Sigma} & = & \int_{t_1}^{t_2} dt \int_{\varphi_1}^{\varphi_2} d\varphi \sum_{i=1}^N \int d \tau \sqrt{1-\frac{2M}{r}} \delta^{(3)} \left( x^\alpha - x^\alpha_{(i)}(\tau) \right) p_\mu^{(i)}(\tau) \nonumber \\
& = & \sum_{i \in I(\tilde \sigma)} \int d \tau \sqrt{1-\frac{2M}{r}} \delta \left( \bar r - r_{(i)}(\tau) \right) \; p_\mu^{(i)}(\tau),
\end{eqnarray}
where $I(\tilde \sigma) \subseteq \{1, \dots, N \}$ denotes the set of indices corresponding to trajectories that intersect $\tilde \sigma$. The integral with respect to $\tau$ can be computed by writing
\begin{equation}
\label{deltart}
\delta \left( \bar r - r_{(i)}(\tau) \right) = \sum_{k} \frac{\delta (\tau - \tau_k)}{ \left| \left. d r_{(i)} / d \tau \right|_{\tau = \tau_k} \right|},
\end{equation}
where $r_{(i)}(\tau_k) = \bar r$. The sum in Eq.\ (\ref{deltart}) runs over all intersections of the $i$-th trajectory with $\tilde \sigma$. Note that
\begin{equation}
\frac{d r_{(i)}}{d \tau} = g^{rr} p_r^{(i)} = \left( 1-\frac{2M}{\bar r} \right)  p_r^{(i)} = \epsilon_r \sqrt{E_{(i)}^2-\left(1-\frac{2M}{\bar r}\right)\left(m^2 +\frac{l_{(i)}^2}{\bar r^2} \right)}.
\end{equation}
Returning to Eq.\ (\ref{^a57e51}), we get
\begin{eqnarray}
\int_{\tilde \sigma} J_\mu^{(N)} \eta_{\tilde \Sigma} & = & \sum_{i \in I(\tilde \sigma)}  \sum_{k} \int d \tau \sqrt{1-\frac{2M}{\bar r}} \frac{\delta (\tau - \tau_k)}{ \left| \left. d r_{(i)} / d \tau \right|_{\tau = \tau_k} \right|} \; p_\mu^{(i)}(\tau) \nonumber \\
&=& \sqrt{1-\frac{2M}{\bar r}} \sum_{j = 1}^{N_\mathrm{int}} \frac{p_\mu^{(j)} }{\sqrt{E_{(j)}^2 - \left( 1 - 2M / \bar r \right) \left( m^2 + l_{(j)}^2 / \bar r^2 \right)}},
\end{eqnarray}
where the index $j$ in the last sum runs over all $N_\mathrm{int}$ \textit{intersections} of trajectories with the cell $\tilde \sigma$. The area (cf. Fig.~\ref{fig:Cell_sigma}) of $\tilde \sigma$ reads
\begin{equation}
\int_{\tilde \sigma} \eta_{\tilde \Sigma} =  \int_{t_1}^{t_2} dt \int_{\varphi_1}^{\varphi_2} d \varphi \; \bar r \sqrt{1-\frac{2M}{\bar r}} = \bar r \sqrt{1-\frac{2M}{\bar r}} (t_2 - t_1) (\varphi_2 - \varphi_1).
\end{equation}
This gives the Monte Carlo estimator of $J_\mu$ in the form
\begin{eqnarray}
\langle J_\mu \rangle & = & \frac{\int_{\tilde \sigma} J_\mu \eta_{\tilde \Sigma} }{\int_{\tilde \sigma} \eta_{\tilde \Sigma}} = \frac{1}{\bar r (t_2 -t_1) (\varphi_2 - \varphi_1)} \sum_{j = 1}^{N_\mathrm{int}}\frac{p_\mu^{(j)} }{\sqrt{E_{(j)}^2 - \left( 1 - 2M / \bar r \right) \left( m^2 + l_{(j)}^2 / \bar r^2 \right)}} \nonumber \\
&=& \frac{1}{M m \bar \xi (t_2 - t_1) (\varphi_2 - \varphi_1)} \sum_{j = 1}^{N_\mathrm{int}}\frac{p_\mu^{(j)} }{\sqrt{\varepsilon_{(j)}^2 - \left(1- 2 / \bar \xi \right)\left(1 +\lambda_{(j)}^2 / \bar \xi^2 \right)}},
\label{jmuMCestimator}
\end{eqnarray}
where we changed to dimensionless variables \eqref{dimensionless}.

For a stationary distribution in the Schwarzschild spacetime Eq.\ (\ref{stationarygeneral}) reduces to the condition $\partial_t \mathcal F = 0$, which implies a traslational invariance of the distribution function with respect to time $t$. Consequently, instead of computing all spacetime components of trajectories $x^\mu_{(i)}(\tau) = \left( t_{(i)}(\tau),r_{(i)}(\tau),\theta_{(i)}(\tau),\varphi_{(i)}(\tau) \right)$, $i = 1, \dots, N$, including the time coordinate $t_{(i)}(\tau)$, it suffices to consider their projections $\left( r_{(i)}(\tau),\theta_{(i)}(\tau),\varphi_{(i)}(\tau) \right)$ on hypersurfaces of constant time. The condition that a trajectory intersects $\tilde \sigma$ present in Eq.\ (\ref{jmuMCestimator}) can be then replaced with a simpler condition that its projection onto a hypersurface of constant time intersects a segment $S$. Also, instead of the normalization by $(t_2 - t_1)$ in Eq.\ (\ref{jmuMCestimator}), one can normalize by a factor proportional to the total number of trajectories $N$. The details of our implementation are given in Eqs.\ (\ref{JmuMCsim}) below.


\section{Monte Carlo method for the planar accretion problem}
\label{sec:planarsccretionsimulation}

\subsection{Selection of geodesic parameters}
\label{geodesicparameters}

In the first step of our simulations we select a set of particle trajectories, corresponding to the assumed distribution function. In our simulations of the planar accretion in the Schwarzschild spacetime we use the parameters $\{ \xi_0, \varphi^\mathrm{(init)}_i, \varepsilon_i, \lambda_i \}$, representing the radial and the azimuthal coordinates of the initial position, the energy, and the total angular momentum of $i$-th particle, respectively. The first coordinate is the same for all trajectories---all particles start at a fixed radius $r_0 = M \xi_0$. It is important to ensure that this value is sufficiently large. The coordinate values $\varphi^\mathrm{(init)}_i$ and $\varepsilon_i$ are sampled from the planar asymptotic ($\xi \to \infty$) distribution function \eqref{MinkDist} (derived in Appendix \ref{appendix:distribution}):
\begin{equation}\label{MinkDist2}
f(x, p) = \alpha \delta \left( \sqrt{-p_\mu p^\mu} - m_0 \right) \exp{ \left[ -\beta \gamma\left( \varepsilon - \epsilon_r v  \sqrt{\varepsilon^2 - 1} \cos\varphi  \right) \right] },
\end{equation}
where we assumed that all particles have the same mass $m_0$.

For a given speed $v$, distribution (\ref{MinkDist2}) depends on three parameters: $\varphi^\mathrm{(init)}_i, \varepsilon_i$ and $\epsilon_r$. The parameter $\epsilon_r$ determines the direction of the radial motion: $\epsilon_r = -1$ means that a particle moves towards the center of the coordinate system, $\epsilon_r = +1$  means that a particle moves in the opposite direction. See also definition \eqref{pr}. Absorbed particles start at $\xi = \xi_0$, and are characterized by $\epsilon_r = -1$. Our treatment of scattered particles is different. In principle we could say that all scattered particles (moving along unbound orbits) originate also at $\xi = \xi_0$ with $\epsilon_r = -1$, they reach a pericenter and then move to $\xi = \xi_0$ again, this time with $\epsilon_r = +1$. That means that a scattered trajectory of a particle originating at $(\xi_0,\varphi^\mathrm{(init)})$ with $\epsilon_r = -1$ contributes also to the distribution function at $(\xi_0,\varphi^\mathrm{(end)})$ with $\epsilon_r = +1$. While for an individual trajectory $\varphi^\mathrm{(end)}$ can be easily related to $\varphi^\mathrm{(init)}$, adjusting the overall selection procedure to the distribution function (\ref{MinkDist2}) is difficult. In practice we circumvent this difficulty by dividing each of the scattered trajectories in half. We consider segments running from $\xi_0$ to the pericenter with $\epsilon_r = -1$ and those running from the pericenter to $\xi_0$ with $\epsilon_r = +1$. Both segments are selected \textit{separateley} from the distribution function (\ref{MinkDist2}).

To randomize the parameters $\varphi_i, \varepsilon_i$ according to the distribution function \eqref{MinkDist2}, we use the Markov Chain Monte Carlo (MCMC) method, implemented in the \textit{Wolfram Mathematica} \cite{Wolfram}. A sample of the \textit{Wolfram Mathematica} code used to select the parameters of trajectories (with $\epsilon_r = -1$) looks as follows:
\begin{lstlisting}[language=Mathematica]
num = 10^6;        
normM = NIntegrate[ 
            Exp[-(*@$\beta$@*)/Sqrt[1 - v^2]*((*@$\varepsilon$@*) + v*Sqrt[(*@$\varepsilon$@*)^2 - 1]*Cos[(*@$\varphi$@*)])], 
        {(*@$\varepsilon$@*), 1, energycutoff},{(*@$\varphi$@*), 0, 2*Pi}];
distM = ProbabilityDistribution[
            Exp[-(*@$\beta$@*)/Sqrt[1 - v^2]*((*@$\varepsilon$@*) + v*Sqrt[(*@$\varepsilon$@*)^2 - 1]*Cos[(*@$\varphi$@*)])]/normM, 
        {(*@$\varepsilon$@*), 1, energycutoff}, {(*@$\varphi$@*), 0, 2*Pi}];        
dataM0 = RandomVariate[distM, num, Method -> {"MCMC", "Thinning" -> 2, "InitialVariance" -> 1}];    
\end{lstlisting}
Here $\mathtt{num}$ denotes the number of generated sets $\{\varphi^\mathrm{(init)}_i, \varepsilon_i \}$, and $\mathtt{normM}$ is a normalization factor of the truncated distribution function specified as $\mathtt{distM}$. In the last line we generate the set of parameters $\varphi^\mathrm{(init)}_i$ and $\varepsilon_i$, using the \textit{Wolfram Mathematica} function $\mathtt{RandomVariate}$. $\mathtt{RandomVariate}$ allows one to select among multiple methods of generating data, which can be specified in the $\mathtt{Method}$ parameter. The \textit{Wolfram Mathematica} implementation of the MCMC method can be controlled by a number of parameters. In our example we  set the so-called ``thinning'' and the ``initial variance.'' 

The method of selecting the remaining parameter $\lambda_i$ is described in detail in \cite{MachCieslikOdrzywolek2023}. In our scenario, the values of $\lambda_i$ are distributed uniformly. Both the energy $\varepsilon_i$ and the angular momentum $\lambda_i$ are randomized within specific limits. Their choice is arbitrary, but the numbers should be large enough to provide a wide range of different trajectories. All selected energies $\varepsilon_i$ are smaller than a cutoff value $\varepsilon_\mathrm{cutoff}$, represented in our sample code as  $\mathtt{energycutoff}$. All angular momenta are smaller than $\lambda_\mathrm{max}(\varepsilon_\mathrm{cutoff},\xi_0)$. Once the parameters of individual orbits are selected, they are subject to a test.  Specifically, they should satisfy the inequality
\begin{equation}
\label{selectionlambda}
\lambda_i \leq \lambda_\mathrm{max} (\varepsilon_i, \xi_0),
\end{equation}
where $ \lambda_\mathrm{max} $ is given by the Eq.\ \eqref{lambdaMax}. Parameters satisfying Eq.\ (\ref{selectionlambda}) are divided into those corresponding to absorbed trajectories---if the angular momentum $\lambda_i$ is less or equal to $\lambda_c(\varepsilon_i)$ given by  Eq.\ (\ref{lambdac})---and those corresponding to scattered trajectories (if $\lambda_i > \lambda_c(\varepsilon_i)$).

       
\subsection{Geodesics}

Given initial parameters $\{ \xi_0, \varphi^\mathrm{(init)}_{i}, \varepsilon_i, \lambda_i \}$, the geodesic motion of a test particle can be determined in a couple of reasonable ways. One can either express the radial distance $r = M \xi$ as a function of the azimuthal angle $\varphi$ or use the inverted relation $\varphi(\xi)$. Both possibilities are described in \cite{CieslikMach2022}. Alternatively, it is possible to use the parametrization by the proper time \cite{CieslikMach2023}. All these methods are equivalent, and the choice of the most convenient one depends on the problem at hand. In Section \ref{sec:Intersections} we described our method of estimating $\langle J_\mu \rangle$ on surfaces $\tilde \Sigma$ with a fixed radial distance $\bar r = M  \bar \xi$. Consequently, our description of geodesic trajectories should provide the values of the angle $\varphi$ corresponding to a given radius $\tilde \xi$. This can be done with the help of the function (closely related to \eqref{EllipticX})
\begin{equation}\label{varpsi-mot}
\varphi(\xi) = \epsilon_r \epsilon_\varphi \lambda \int^\xi_{\xi_0} \frac{d\xi'}{\xi'^2 \sqrt{\varepsilon^2 - U_\lambda (\xi')}} = \epsilon_r \epsilon_\varphi \left[ X(\xi_0,\varepsilon,\lambda) - X(\xi,\varepsilon,\lambda)\right],
\end{equation}
which was described in detail in \cite{CieslikMach2022} (p.\ 26, Eq.\ $(\mathrm{C}.2)$). In Equation (\ref{varpsi-mot}) we assume that $\epsilon_r$ is constant along the segment of the geodesic between $\xi_0$ and $\xi$, and that $\varphi(\xi_0) = 0$. In our previous study \cite{MachCieslikOdrzywolek2023} the surfaces $\tilde \Sigma$ were characterized by $\varphi = \mathrm{const}$. Consequenly, we used formulas for $\xi(\varphi)$, based on the Biermann-Weierstrass theorem.

It can be shown that for unbound scattered and absorbed geodesic trajectories the integral in Eq.\ (\ref{varpsi-mot}) can be expressed as
\begin{subequations}
\begin{eqnarray}
\label{phiscat}
\varphi(\xi)^\mathrm{(scat)} &=& -\frac{\epsilon_r \epsilon_\varphi}{\sqrt{y_3 - y_1}}\left[ \tilde{F}\left(\arccos{\sqrt{\frac{y_2 +\frac{1}{12}-\frac{1}{2\xi}}{y_2 - y_1}}},k \right) - \tilde{F}\left(\arccos{\sqrt{\frac{y_2 +\frac{1}{12}-\frac{1}{2\xi_0}}{y_2 - y_1}}},k \right) \right],\\
\label{phiabs}
\varphi(\xi)^\mathrm{(abs)} &=& \frac{\epsilon_\varphi}{2 \sqrt{\mu}} \left[ \tilde F\left(2 \arctan \sqrt{\frac{ \frac{1}{2 \xi}- \frac{1}{12}  - \tilde y_1}{\mu}} ,\tilde  k \right) - \tilde F \left( 2 \arctan \sqrt{\frac{\frac{1}{2 \xi_0}  -\frac{1}{12} - \tilde y_1}{\mu}} , \tilde k \right) \right], 
\end{eqnarray}
\end{subequations}
where in Eq.\ (\ref{phiscat}) $y_1$, $y_2$, and $y_3$ denote the roots of the polynomial $4y^3 - g_2 y - g_3$, satisfying $y_1 < y_2 < y_3$. In Equation (\ref{phiabs}) $\tilde y_1$ denotes a real zero of the polynomial $4y^3 - g_2 y - g_3 = 4 (y - y_1)(y^2 + py + q)$, where $p^2 - 4 q < 0$ and thus $y^2 +p y + q > 0$. The constants $g_2$ and $g_3$--- Weiestrass invariants---are given by
\begin{subequations}
\begin{eqnarray}
g_2 &=& \frac{1}{12} - \frac{1}{\lambda^2},\\
g_3 &=& \frac{1}{6^3} - \frac{1}{12 \lambda^2} - \frac{\varepsilon^2 -1}{4 \lambda^2}. 
\end{eqnarray}
\end{subequations}
The remaining parameters are defined as follows:
\begin{subequations}
\begin{eqnarray}
k^2 &=& \frac{y_2 - y_1}{y_3 - y_1},\\
\tilde k^2 &=& \frac{1}{2} \left( 1 - \frac{\tilde y_1 + p/2}{\mu} \right) \label{k2bis},
\end{eqnarray}
\end{subequations}
where $\mu = \sqrt{\tilde y_1^2 + p \tilde y_1 + q}$. Here $\tilde F(\phi,k)$ is the Legendre elliptic integral \cite{DLMF} defined by
\begin{equation}
\tilde F(\phi,k) = \int_0^\phi \frac{d \chi}{\sqrt{1 - k^2 \sin^2 \chi}}, \quad -\frac{\pi}{2} < \phi < \frac{\pi}{2}.
\end{equation}


\subsection{Monte Carlo estimators}
\label{sec:estimators}

We apply Eqs.\ (\ref{jmuMCestimator}) on a standard equidistant polar grid, with the radial coordinate in the range $2 < \xi \leq \tilde{\xi}_0$. The parameter $\tilde{\xi}_0$ is, in principle, arbitrary. The angle $\varphi$ is discretized as $\varphi_i =  i\Delta \varphi$, $i = 1, \dots N_\varphi$, where $\Delta \varphi = 2 \pi/N_\varphi$, and $N_\varphi$ denotes the number of segments in the angular direction. Similarly, the radial coordinate is discretized according to $\xi_j = 2 + j \Delta \xi$, $j = 1, \dots, N_\xi$, where $\Delta \xi = (\tilde{\xi}_0 - 2)/N_\xi$. Thus for  $j = N_\xi$, we have $\xi_j = \tilde{\xi}_0$, which corresponds to the outer boundary of the grid. Note that $\tilde \xi_0$ may be smaller than $\xi_0$, which is related to the selection of the sample of trajectories. For the results presented in Sec.\ \ref{sec:Results}, we assumed $N_\varphi=360$, $N_\xi=100$, $\tilde{\xi}_0 = 20$. 

Formula (\ref{jmuMCestimator}) is then applied to all segments $\tilde \sigma$ (Fig.~\ref{fig:Cell_sigma}) specified in Eq.\ (\ref{tildesigma}), with $\varphi_1 = \varphi_{i-1}$, $\varphi_2 = \varphi_i$, and  $\bar r = M \xi_j$, where $i = 1, \dots, N_\varphi$, $j = 1, \dots, N_\xi$, and we allow ourselves for a slight abuse of notation. For simplicity, we will keep the notation with $\varphi_1$, $\varphi_2$, and $\bar \xi$ in the remainder of this section, when referring to a single segment of the numerical grid.

Our choice of $\tilde \sigma$ (or $S$) has some advantages. An unbound absorbed trajectory crosses the circle of constant radius $\bar r = M \bar \xi$ only once. A generic unbound scattered trajectory crosses a circle of constant $\bar r = M \bar \xi$ twice, or it does not cross it at all. The precise angular coordinates of these crossing points can be computed from Eqs.\ (\ref{phiscat}) and (\ref{phiabs}). This stays in contrast to our previous choice made in \cite{MachCieslikOdrzywolek2023}, where a trajectory wrapping around the black hole could, in principle, cross a segment of a constant $\varphi$ an arbitrary number of times.

The important difference with respect to our previous work \cite{MachCieslikOdrzywolek2023} is related to the treatment of scattered trajectories. Ingoing and outgoing segments of scattered trajectories are selected separately, and in practice no scattered trajectory appears in our calculation as a whole one. As described in Sec.\ \ref{geodesicparameters}, this fact is not related to the discretization of the numerical grid, but rather to the asymptotic distribution function, which takes into account both ingoing and outgoing trajectories in the overall budget. All absorbed trajectories are characterized by $\epsilon_r = -1$. Ingoing segments of scattered trajectories should have $\epsilon_r = -1$; outgoing segments correspond to $\epsilon_r = +1$. There should be roughly the same number of ingoing and outgoing segments in a given simulation.

Let $I_\mathrm{abs}(\varphi_1,\varphi_2)$ denote the set of indices numbering absorbed trajectories crossing a segment of a circle with $\xi = \bar \xi$ and $\varphi_1 \leq \varphi < \varphi_2$. In a similar way we define the sets of indices corresponding to ingoing an outgoing segments of scattered trajectories. They are denoted by $I^-_\mathrm{scat}(\varphi_1,\varphi_2)$ and $I^+_\mathrm{scat}(\varphi_1,\varphi_2)$, respectively. The total number of absorbed trajectories is denoted by $N_\mathrm{abs}$. Symbols $N^-_\mathrm{scat}$ and $N^+_\mathrm{scat}$ denote, respectively, the total numbers of ingoing and outgoing segments of scattered trajectories in the sample. Using this notation, the Monte Carlo estimators of the particle current surface density components can be expressed as
\begin{subequations}
\label{JmuMCsim}
\begin{eqnarray}
\label{JtMCsim}
        \langle J_t^\mathrm{(abs)} \rangle &=& -\frac{2\alpha m_0^3 V_\mathrm{abs} }{N_\mathrm{abs} (\varphi_2-\varphi_1)\bar \xi} \sum_{i\in I_\mathrm{abs}(\varphi_1,\varphi_2)} \frac{\varepsilon_i}{\sqrt{\varepsilon_i^2 -\left(1 - 2 / \bar \xi \right)\left( 1 + \lambda_i^2 / \bar \xi^2 \right)}},\\
        \langle J_t^\mathrm{(scat)} \rangle &=& -\frac{2\alpha m_0^3 }{ (\varphi_2-\varphi_1) \bar \xi} \left[ \frac{V^+_\mathrm{scat}}{N^+_\mathrm{scat}}\sum_{i\in I^+_\mathrm{scat}(\varphi_1,\varphi_2)} \frac{\varepsilon_i}{\sqrt{\varepsilon_i^2 -\left(1 - 2 / \bar \xi \right)\left( 1 + \lambda_i^2 / \bar \xi^2 \right)}} \right. \nonumber \\
        && \left. + \frac{V^-_\mathrm{scat}}{N^-_\mathrm{scat}}\sum_{i\in I^-_\mathrm{scat}(\varphi_1,\varphi_2)} \frac{\varepsilon_i}{\sqrt{\varepsilon_i^2 -\left(1 - 2 / \bar \xi \right)\left( 1 + \lambda_i^2 / \bar \xi^2 \right)}} \right], \\
        \label{JrMCsim}   
        \langle J_r^\mathrm{(abs)} \rangle & = & - \frac{2\alpha m_0^3 V_\mathrm{abs}}{N_\mathrm{abs} (\varphi_2 - \varphi_1) \bar \xi} \sum_{i \in I_\mathrm{abs}(\varphi_1,\varphi_2)} \frac{1}{\left(1 - 2 / \bar \xi \right)}, \\
        \langle J_r^\mathrm{(scat)} \rangle &=& \frac{2\alpha m_0^3 }{ (\varphi_2 - \varphi_1)\bar \xi} \left[  \frac{V^+_\mathrm{scat}}{N^+_\mathrm{scat}}\sum_{i\in I^+_\mathrm{scat}(\varphi_1,\varphi_2)}  \frac{1}{\left(1 - 2 / \bar \xi \right)} - \frac{V^-_\mathrm{scat}}{N^-_\mathrm{scat}}\sum_{i\in I^-_\mathrm{scat}(\varphi_1,\varphi_2)}  \frac{1}{\left(1 - 2 / \bar \xi \right)} \right], \\
        \label{JphiMCsim}
        \langle J_\varphi^\mathrm{(abs)} \rangle & = & \frac{2\alpha m_0^3 M V_\mathrm{abs} }{N_\mathrm{abs} (\varphi_2-\varphi_1)\bar \xi} \sum_{i\in I_\mathrm{abs}(\varphi_1,\varphi_2)} \frac{\epsilon_{\varphi, i} \lambda_i}{\sqrt{\varepsilon_i^2 - \left( 1 - 2 / \bar \xi \right)\left( 1 + \lambda_i^2 / \bar \xi^2 \right)}},\\
        \langle J_\varphi^\mathrm{(scat)} \rangle & = & -\frac{2\alpha m_0^3 M}{ (\varphi_2 - \varphi_1)\bar \xi} \left[ \frac{V^+_\mathrm{scat}}{N^+_\mathrm{scat}}\sum_{i \in I^+_\mathrm{scat}(\varphi_1,\varphi_2)} \frac{\epsilon_{\varphi, i} \lambda_i}{\sqrt{\varepsilon_i^2 - \left( 1 - 2 / \bar \xi \right)\left( 1 + \lambda_i^2 / \bar \xi^2 \right)}} \right. \nonumber \\
        && \left. - \frac{V^-_\mathrm{scat}}{N^-_\mathrm{scat}}\sum_{i\in I^-_\mathrm{scat}(\varphi_1,\varphi_2)} \frac{\epsilon_{\varphi, i} \lambda_i}{\sqrt{\varepsilon_i^2 -\left(1 - 2 / \bar \xi \right)\left( 1 + \lambda_i^2 / \bar \xi^2 \right)}} \right],
\end{eqnarray}
\end{subequations}
where 
\begin{subequations}
\label{volumes}
\begin{eqnarray}
V_\mathrm{abs} & = &  \int_0^{2\pi}\int_1^{\varepsilon_\mathrm{cutoff}} \exp\left[- \beta \gamma \left( \varepsilon + v\sqrt{\varepsilon^2 -1} \cos{\varphi}\right) \right] \lambda_c(\varepsilon) d \varepsilon d\varphi, \\
V^\pm_\mathrm{scat} & = &  \int_0^{2\pi}\int_1^{\varepsilon_\mathrm{cutoff}} \exp\left[- \beta \gamma \left( \varepsilon \mp v\sqrt{\varepsilon^2 -1} \cos{\varphi}\right) \right][\lambda_\mathrm{max}(\varepsilon,\xi_0) - \lambda_c(\varepsilon)] d \varepsilon d\varphi,
\end{eqnarray}
\end{subequations}
and the sign $\epsilon_{\varphi, i}=\pm 1$ is selected randomly from a set $\{-1,+1\}$ using \textit{Mathematica} \cite{Wolfram} $\mathtt{RandomChoice}$ command for all trajectories.

In the plots shown in Sec.\ \ref{sec:Results}, the values the estimators $\langle J_\mu(x) \rangle$ corresponding to each cell are assigned to coordinates $\left( \xi_j, (\varphi_i + \varphi_{i-1})/2 \right)$. Consequently the angular coordinates change in the range $0.5^\circ \leq \varphi \leq 359.5^\circ$. To calculate the angular position of the particle, we use one of Eqs.\ \eqref{phiscat} and \eqref{phiabs}, depending on the type of trajectory. For a given radial coordinate $\xi_j$, we determine the corresponding $i$-th cell by using the relation $\varphi_i = \lfloor \varphi(\xi_j)/\Delta\varphi \rfloor + 1$.

  
\subsection{Results}
\label{sec:Results}

\begin{figure}
    \centering
    \includegraphics[width=0.45\textwidth]{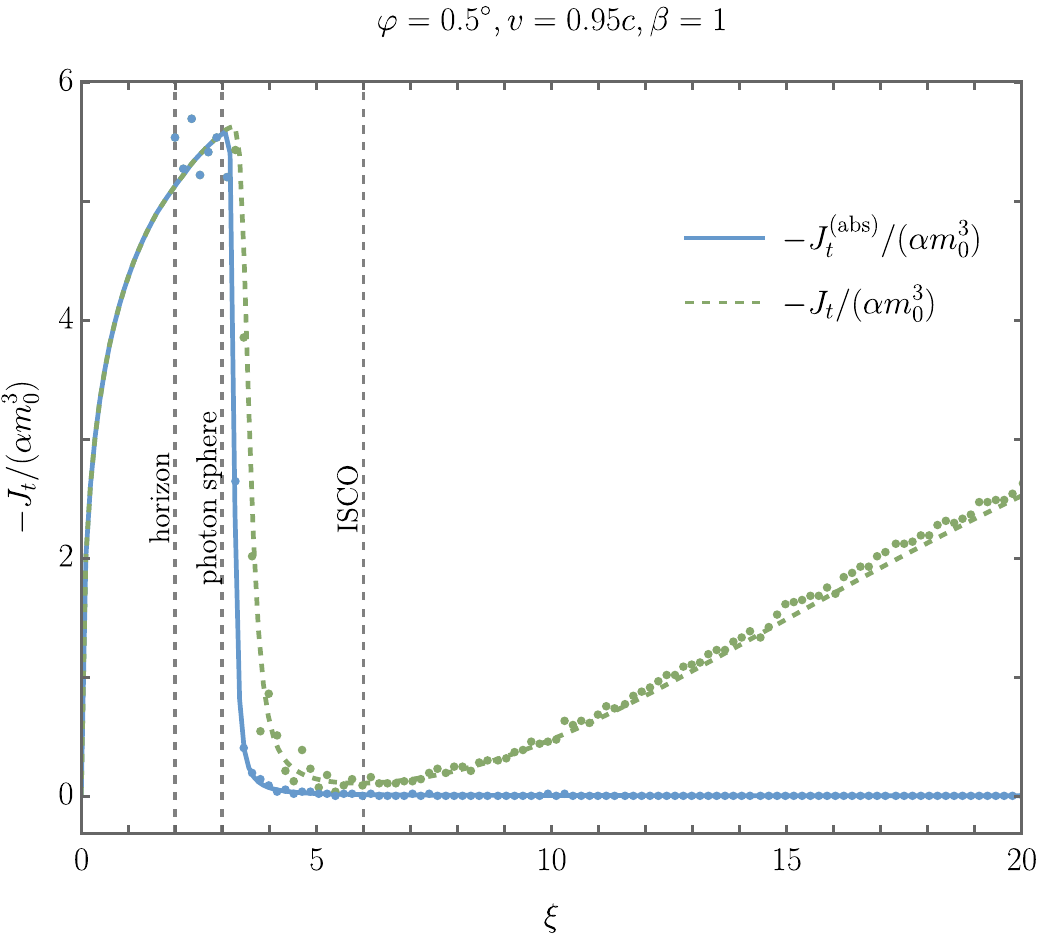}
    \includegraphics[width=0.45\textwidth]{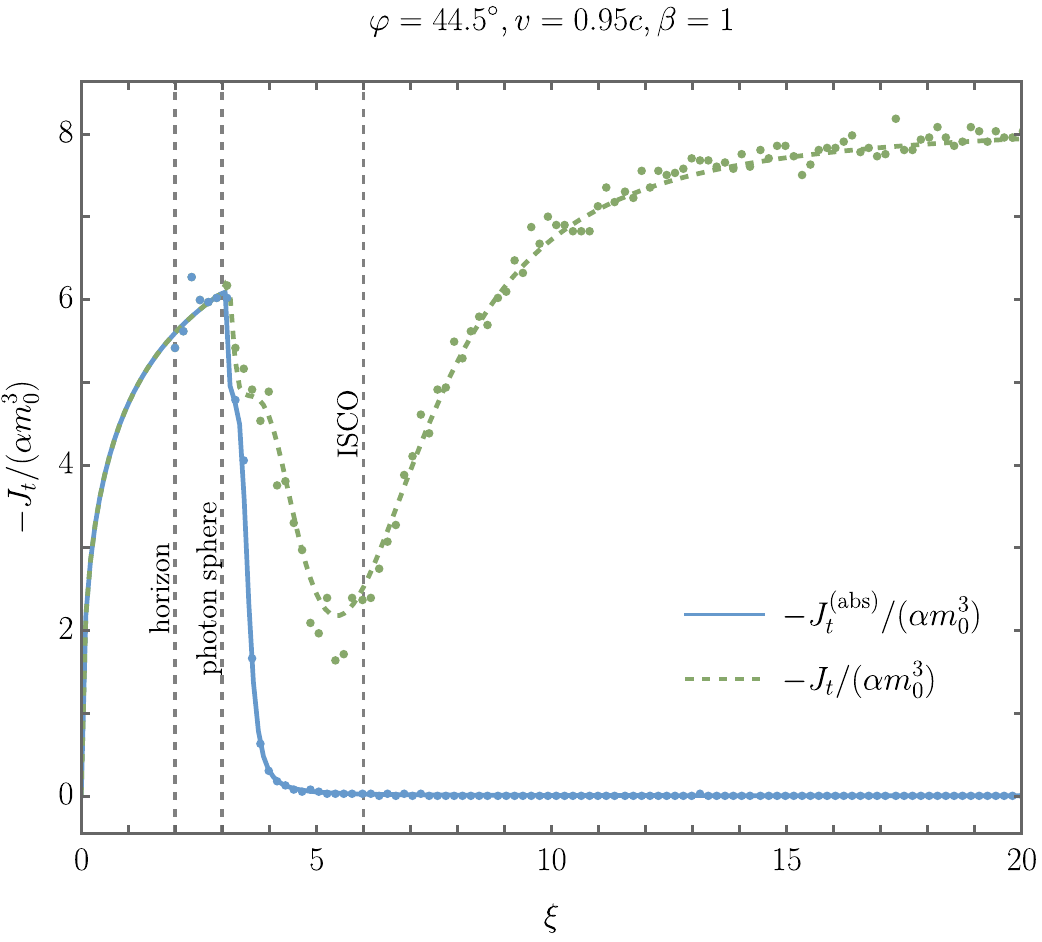}
    \includegraphics[width=0.45\textwidth]{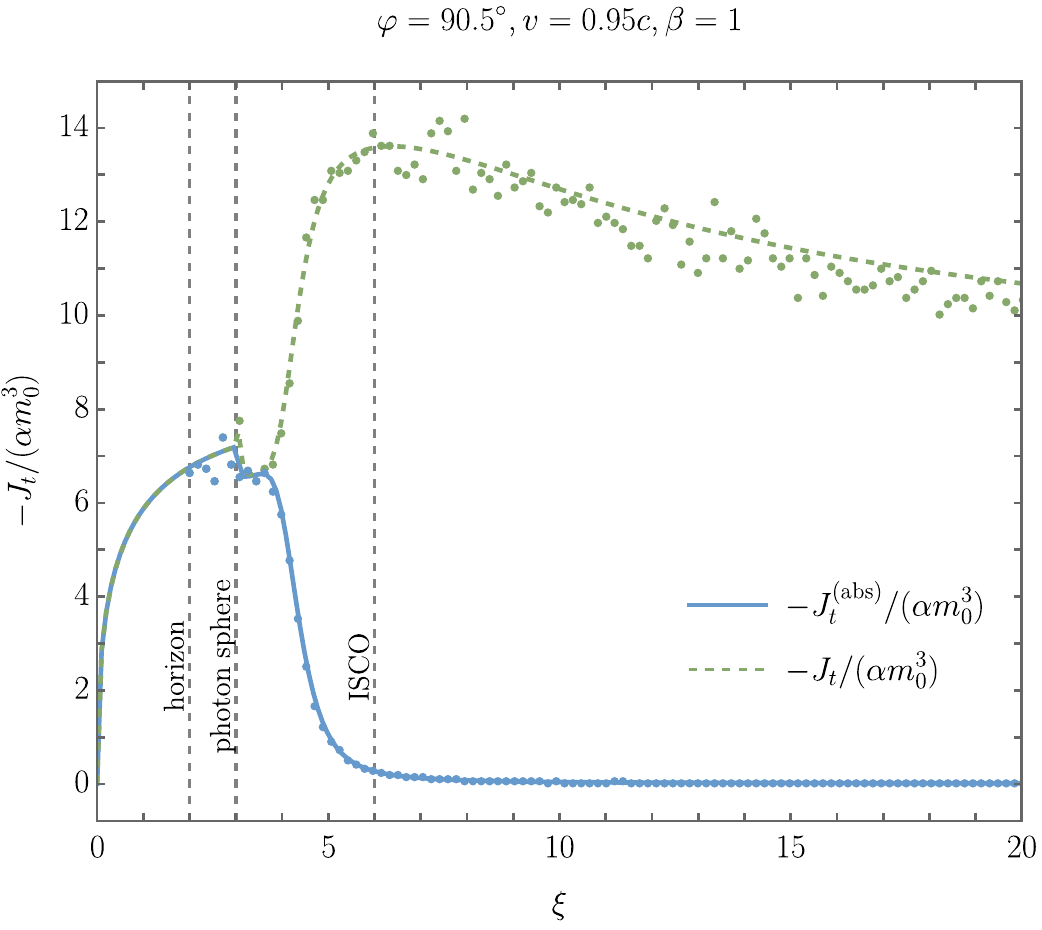}
    \includegraphics[width=0.45\textwidth]{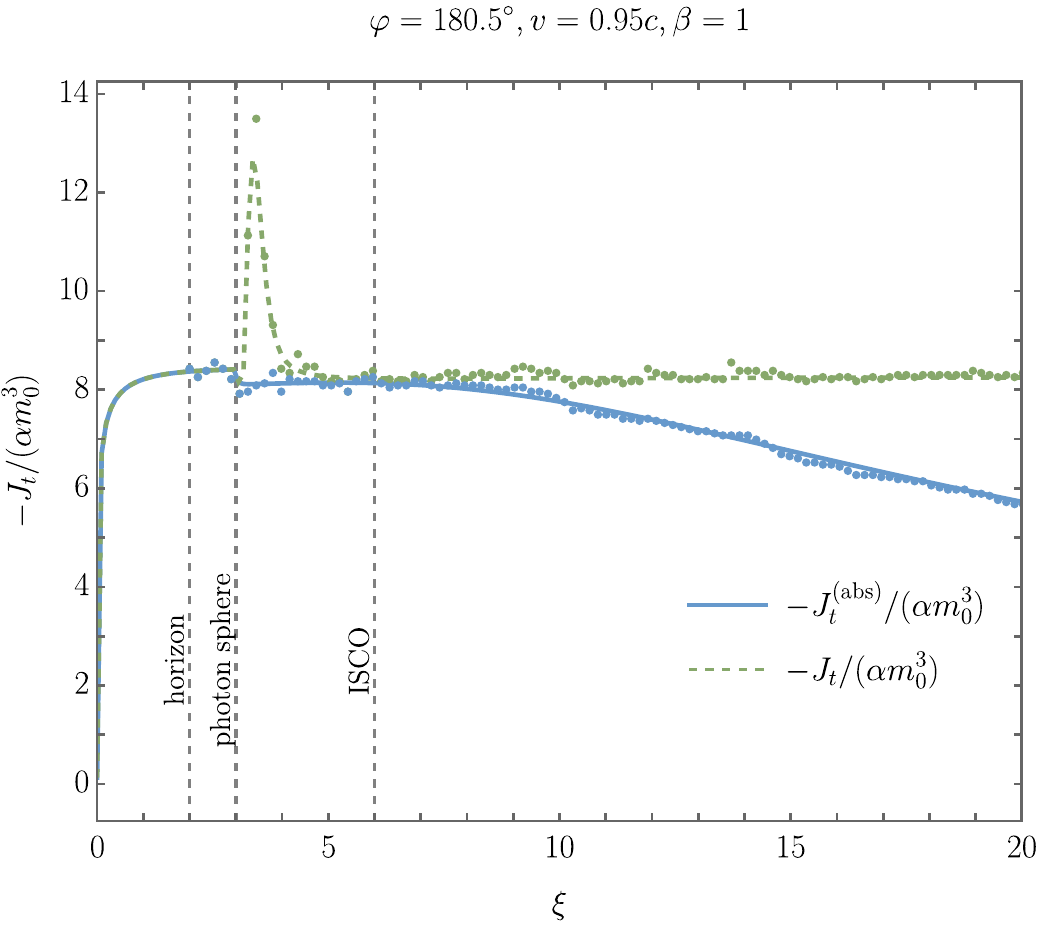}
    \caption{Time components of the particle current surface density $J_t$ in the model with $v = 0.95$, $\beta = 1$, $\varepsilon_\mathrm{cutoff} = 10$, and $\xi_0 = 1000$. Exact solutions (Eqs.\ \eqref{Jmuexact}) are plotted with solid and dashed lines. Dots (blue and green) represent sample results obtained by the Monte Carlo simulation (Eqs.\ \eqref{JmuMCsim}). There are $168\;486\;945$ trajectories: $N_\mathrm{abs} = 450\;198$, $N^-_\mathrm{scat} = 84\;010\;986$, $N^+_\mathrm{scat} = 84\;025\;761$.}
    \label{fig:J_t-v=095-b=1}
\end{figure}

\begin{figure}
    \centering
    \includegraphics[width=0.45\textwidth]{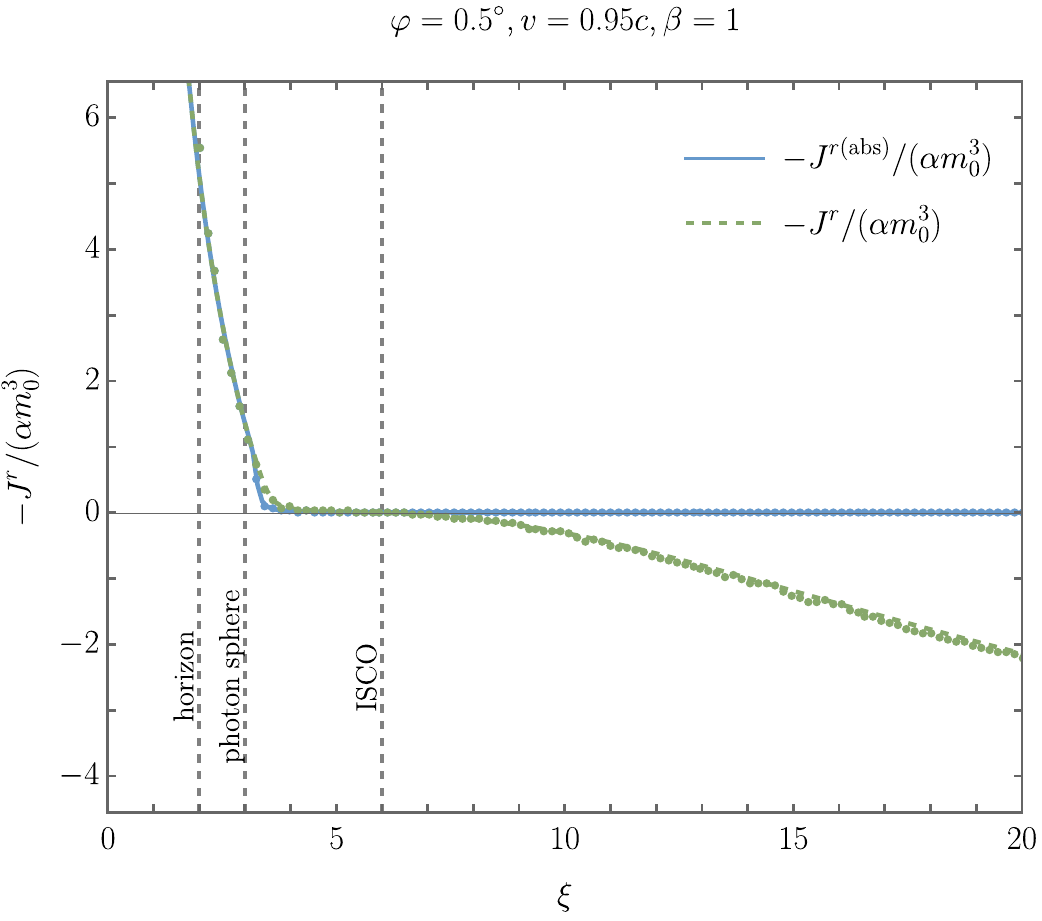}
    \includegraphics[width=0.45\textwidth]{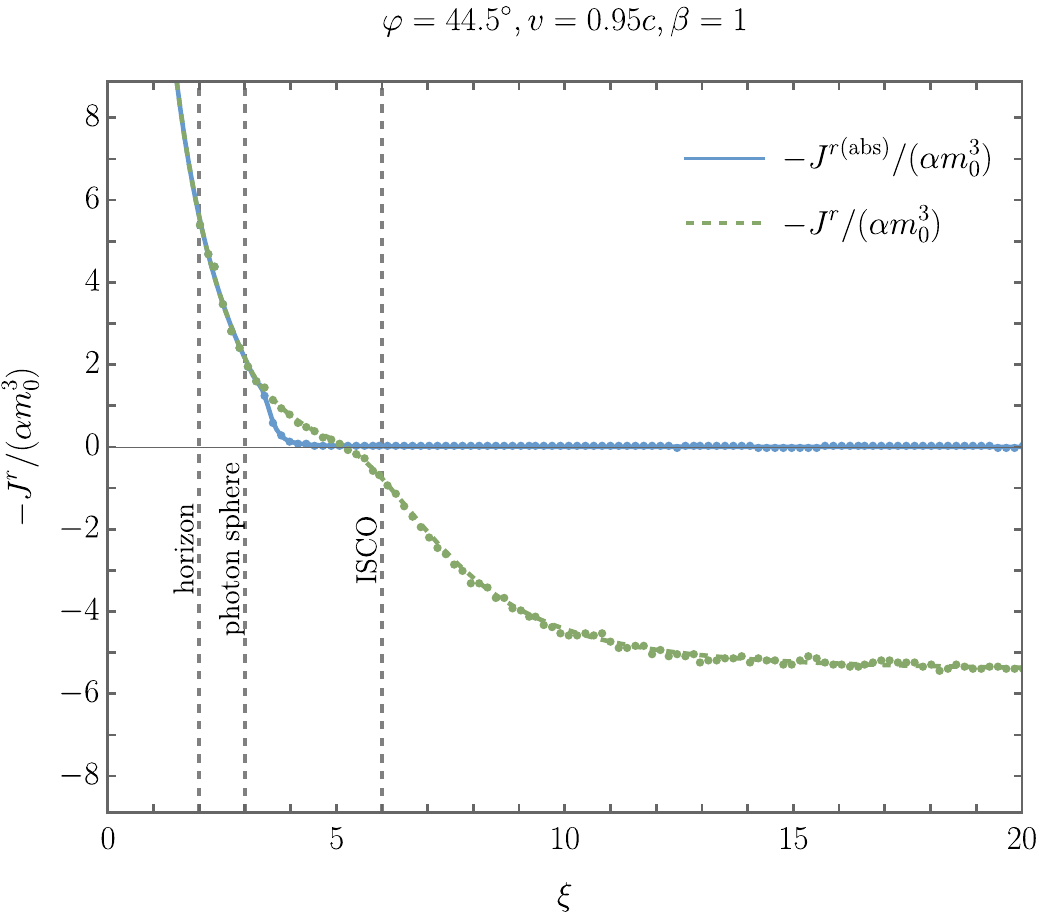}
    \includegraphics[width=0.45\textwidth]{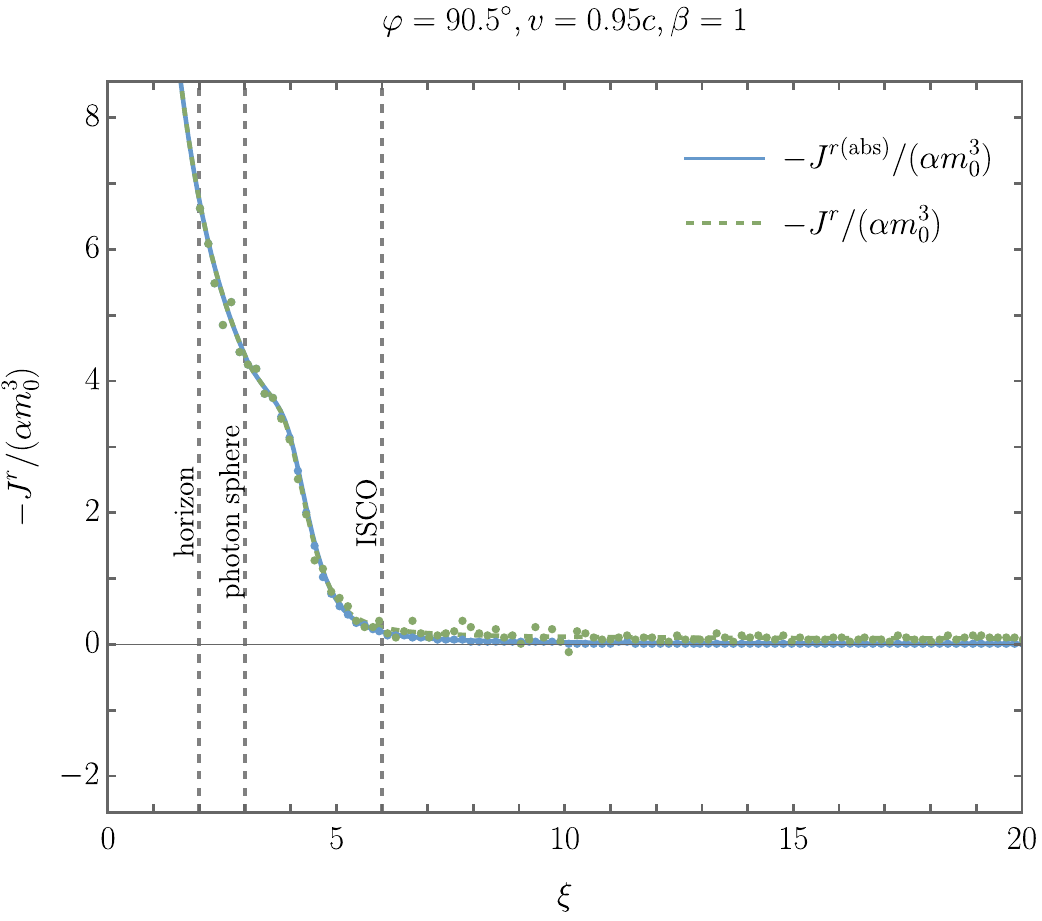}
    \includegraphics[width=0.45\textwidth]{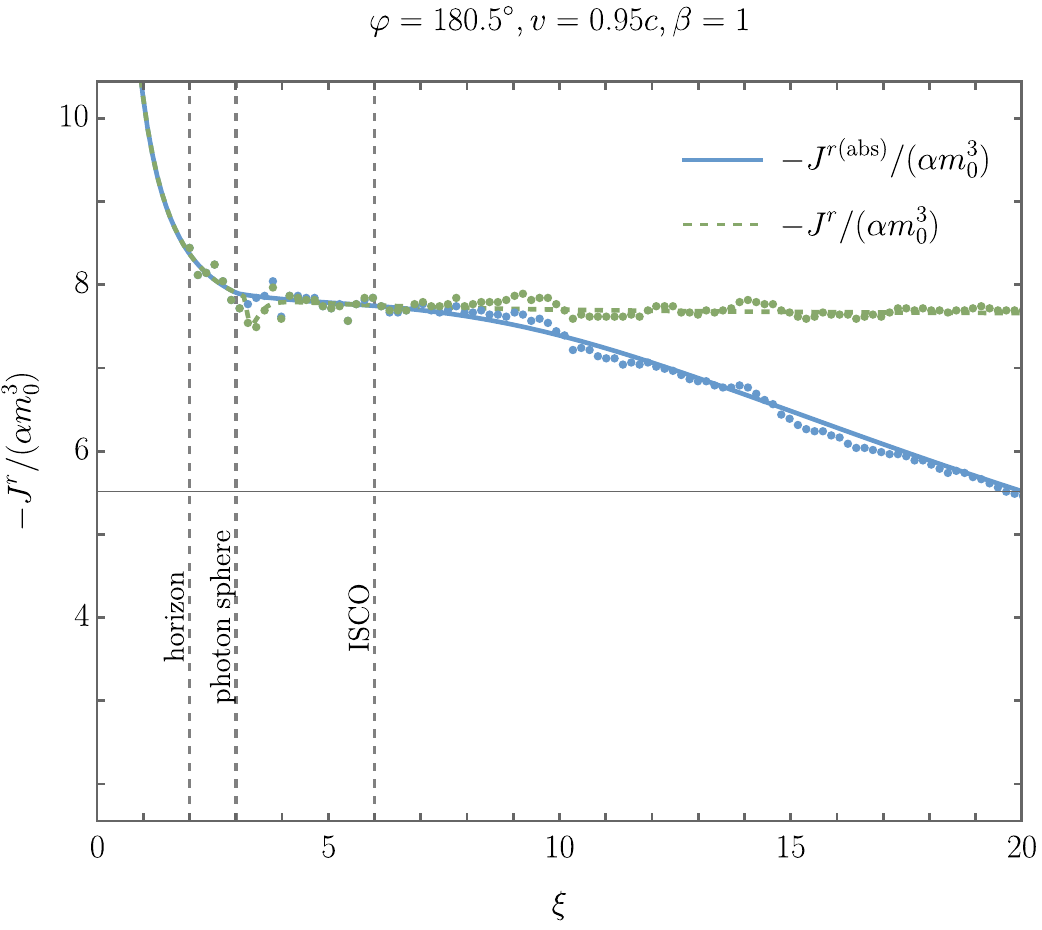}
    \caption{Radial components of the particle current surface density $J^r$ in the model with $v = 0.95$, $\beta = 1$, $\varepsilon_\mathrm{cutoff} = 10$, and $\xi_0 = 1000$. Exact solutions (Eqs.\ \eqref{Jmuexact}) are plotted with solid and dashed lines. Dots (blue and green) represent sample results obtained by the Monte Carlo simulation (Eqs.\ \eqref{JmuMCsim}). There are $168\;486\;945$ trajectories: $N_\mathrm{abs} = 450\;198$, $N^-_\mathrm{scat} = 84\;010\;986$, $N^+_\mathrm{scat} = 84\;025\;761$.}
    \label{fig:J^r-v=095-b=1}
\end{figure}

\begin{figure}
    \centering
    \includegraphics[width=0.45\textwidth]{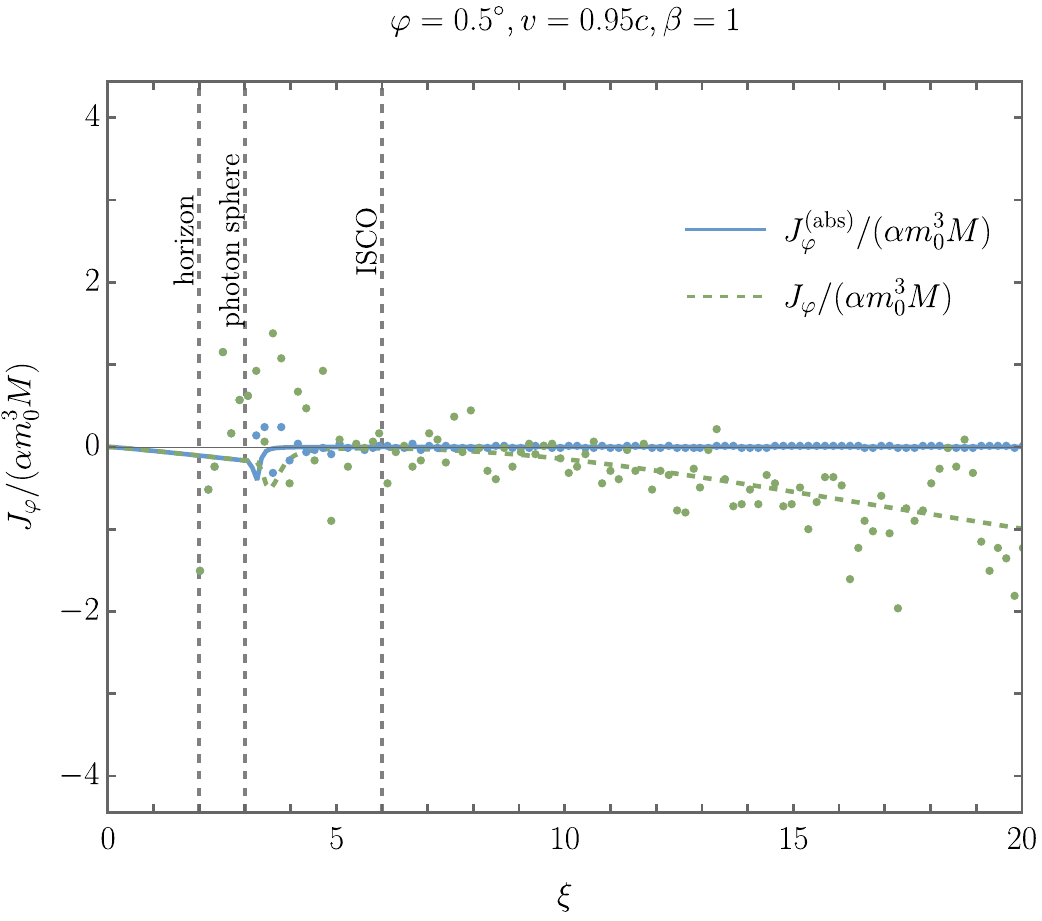}
    \includegraphics[width=0.45\textwidth]{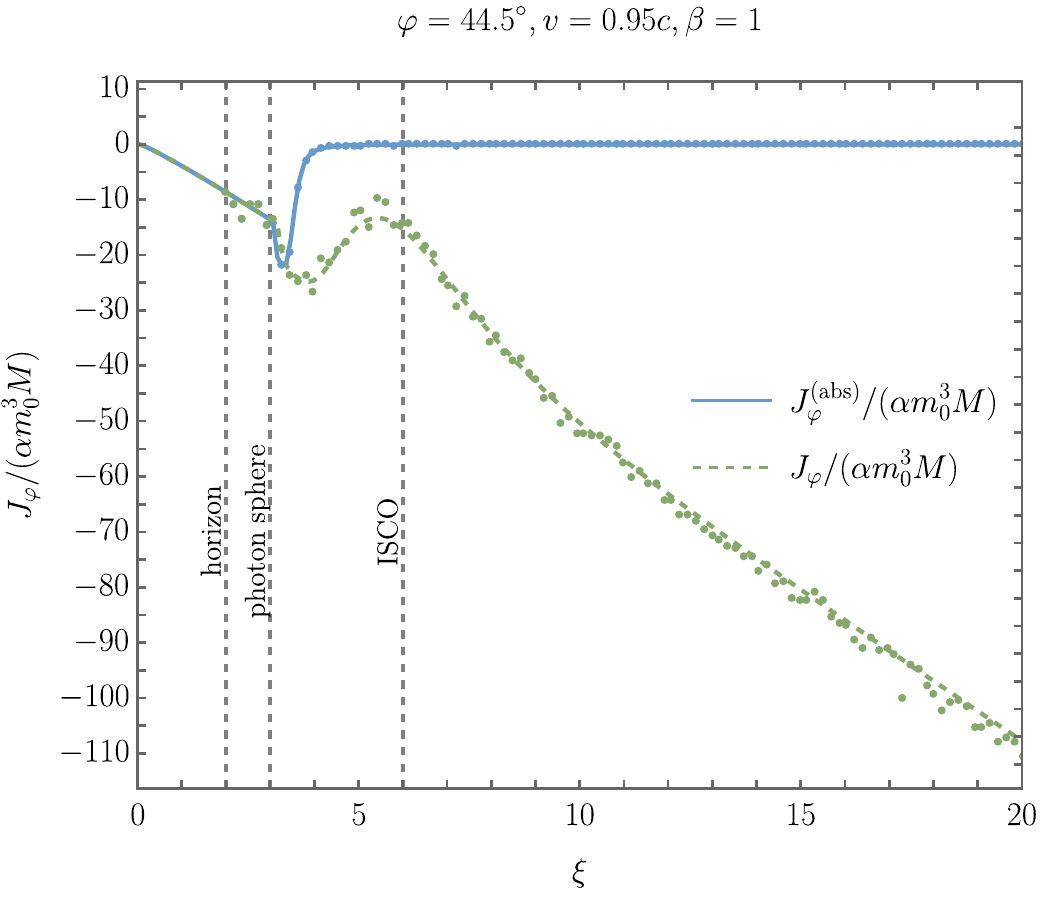}
    \includegraphics[width=0.45\textwidth]{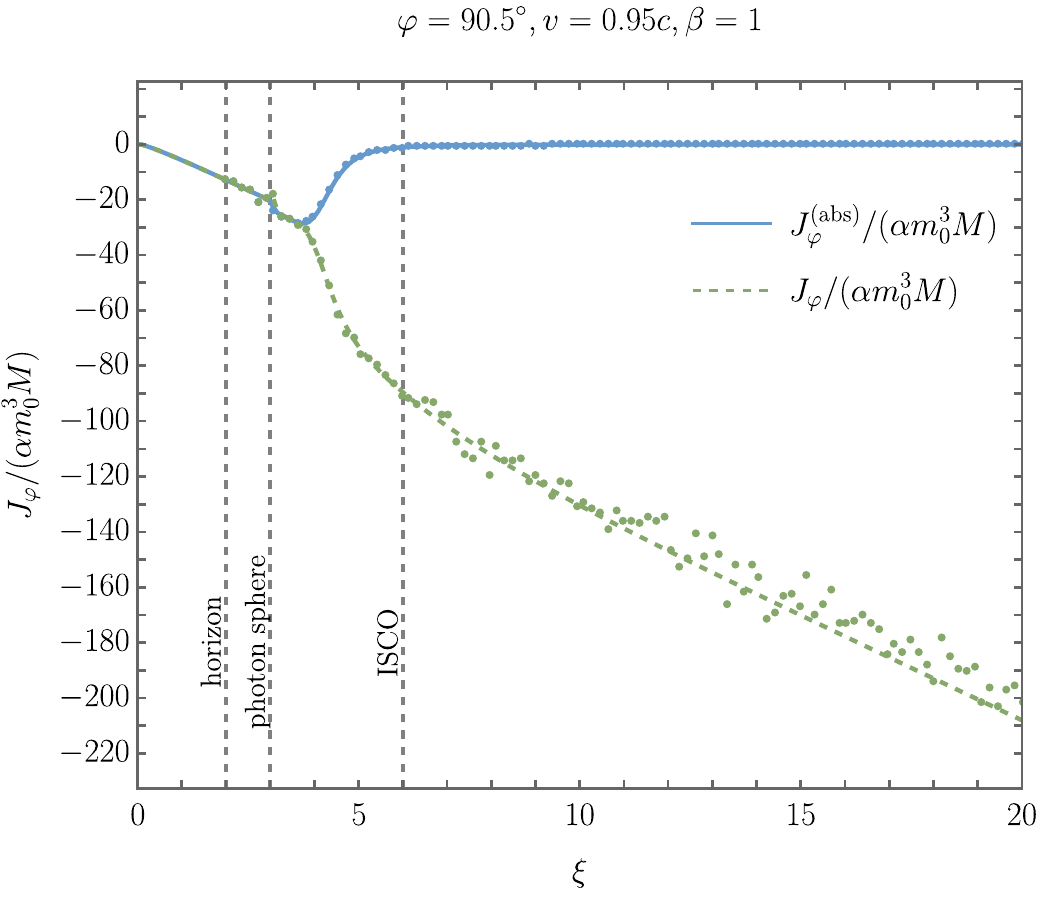}
    \includegraphics[width=0.45\textwidth]{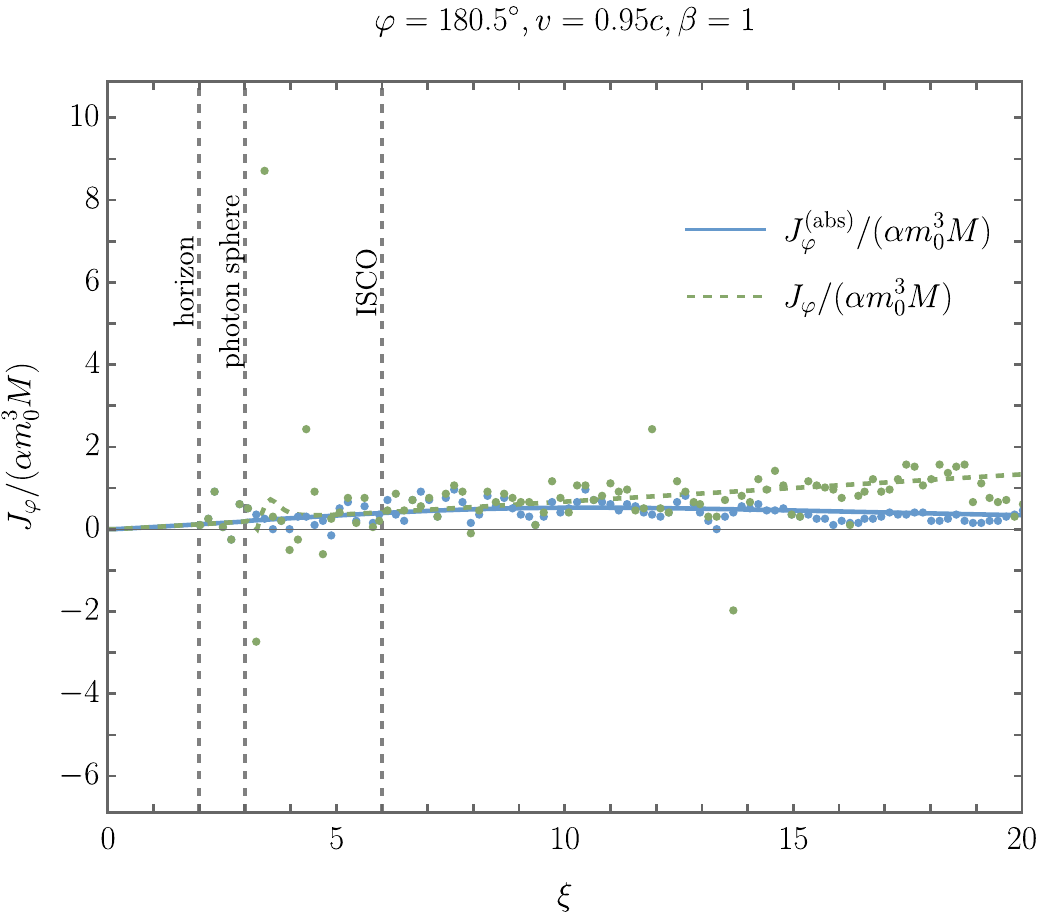}
    \caption{Angular components of the particle current surface density $J_\varphi$ in the model with $v = 0.95$, $\beta = 1$, $\varepsilon_\mathrm{cutoff} = 10$, and $\xi_0 = 1000$. Exact solutions (Eqs.\ \eqref{Jmuexact}) are plotted with solid and dashed lines. Dots (blue and green) represent sample results obtained by the Monte Carlo simulation (Eqs.\ \eqref{JmuMCsim}). There are $168\;486\;945$ trajectories: $N_\mathrm{abs} = 450\;198$, $N^-_\mathrm{scat} = 84\;010\;986$, $N^+_\mathrm{scat} = 84\;025\;761$.}
    \label{fig:J_fi-v=095-b=1}
\end{figure}

\begin{figure}
    \centering
    \includegraphics[width=0.45\textwidth]{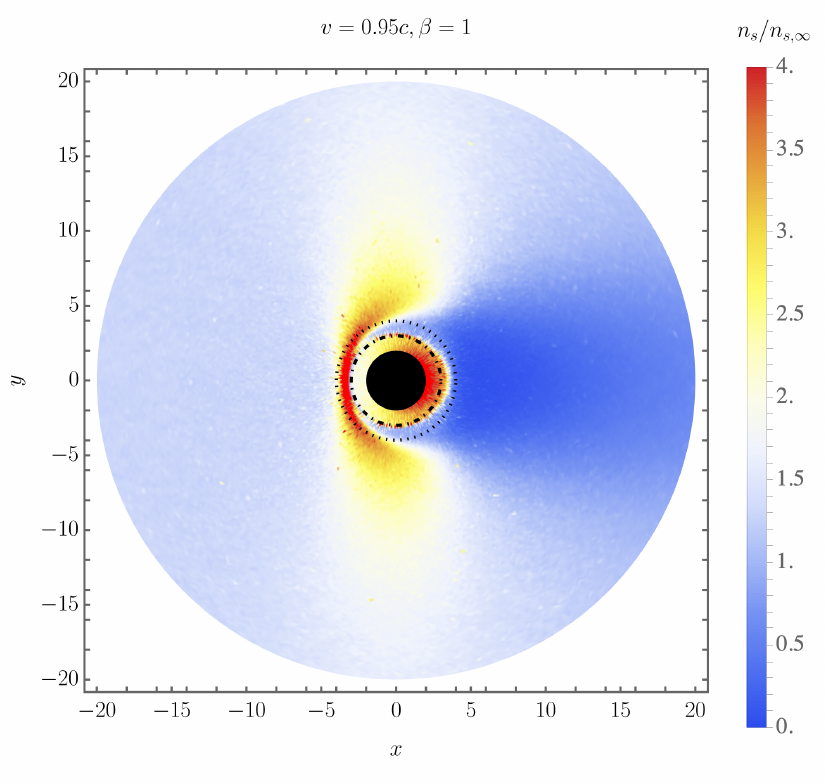}
    \includegraphics[width=0.45\textwidth]{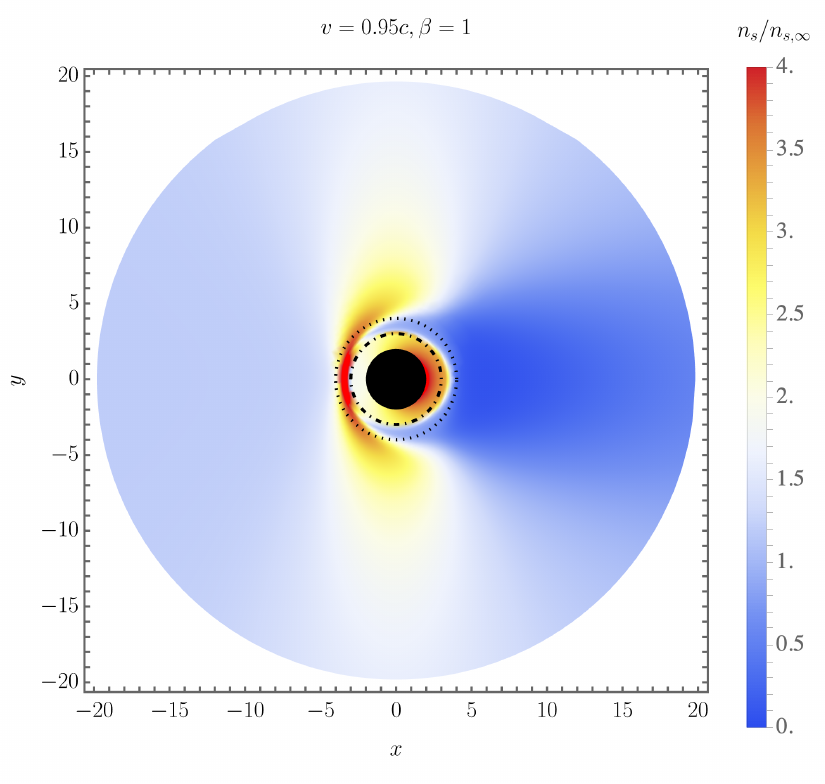}
    \includegraphics[width=0.45\textwidth]{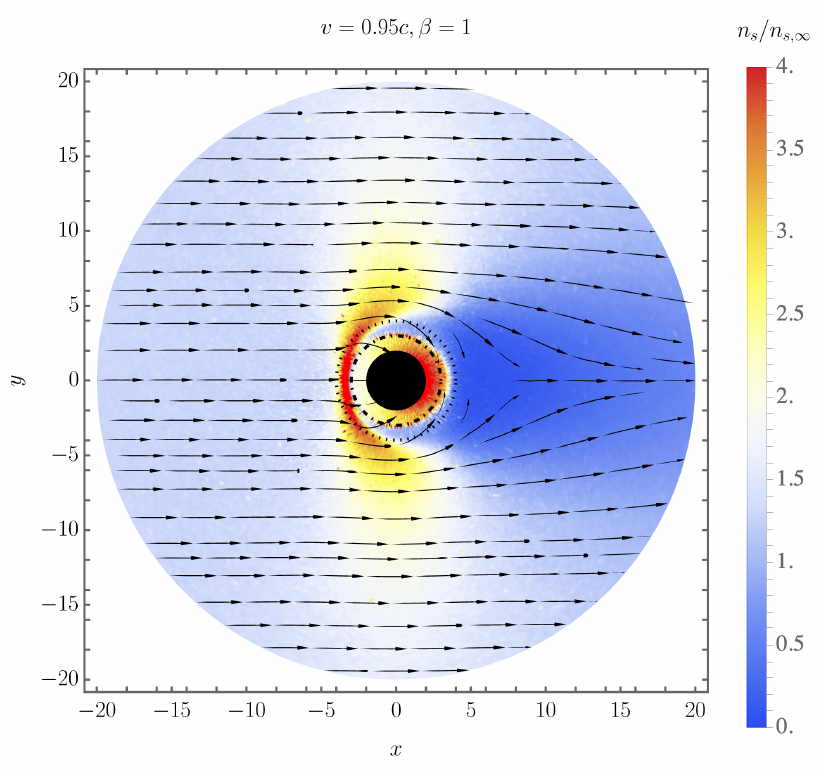}
    \includegraphics[width=0.45\textwidth]{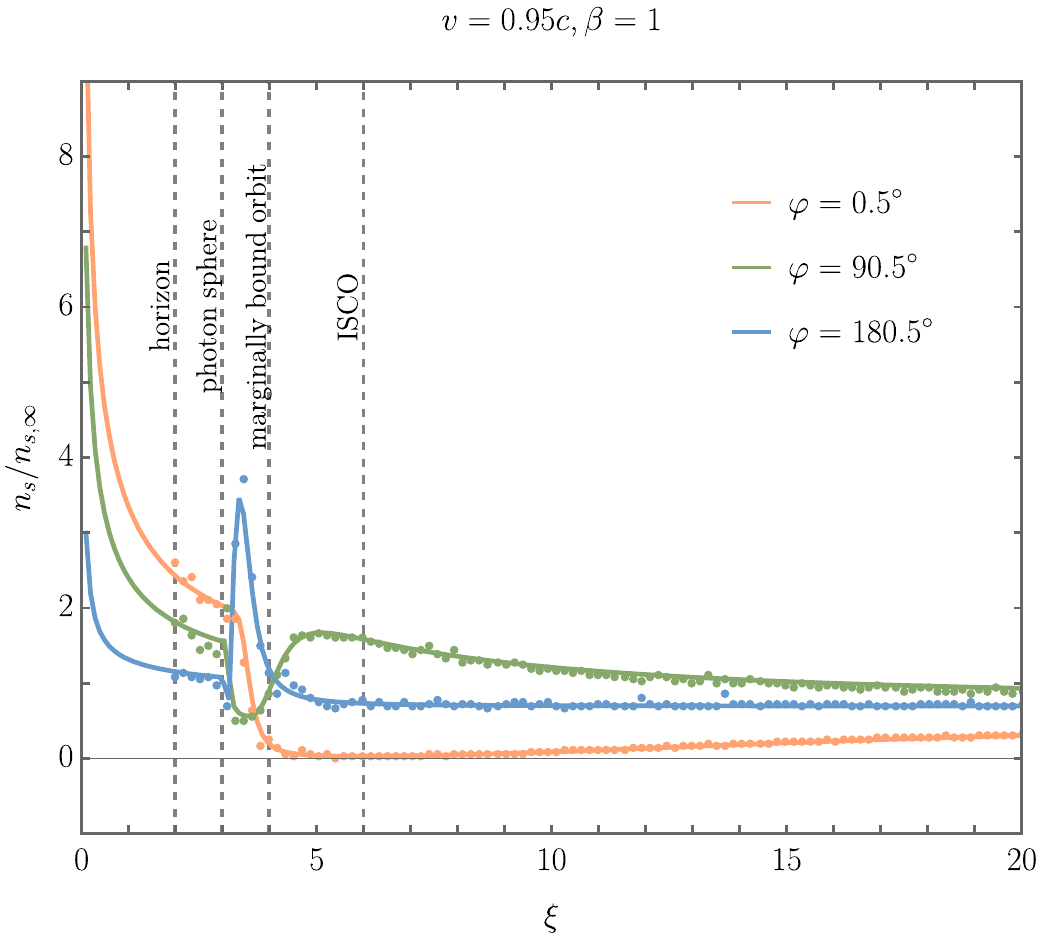}
    \caption{The particle surface density ratio $n_s/n_{s,\infty}$ in the model with $v = 0.95$, $\beta = 1$, $\varepsilon_\mathrm{cutoff} = 10$, and $\xi_0 = 1000$. The black hole moves in the negative direction of the $x$ axis, i.e, $\varphi=180^\circ$. The first three plots show the particle number surface density in the equatorial plane. Plots on the left show the results obtained in the Monte Carlo simulation. The upper right plot shows the exact result based on Eqs.\ (\ref{Jmuexact}). The remaining graph depicts the radial density profiles for three selected values of the angle $\varphi$. The simulation was prepared for $168\;486\;945$ particle trajectories: $N_\mathrm{abs} = 450\;198$, $N^-_\mathrm{scat} = 84\;010\;986$, $N^+_\mathrm{scat} = 84\;025\;761$.}
    \label{fig:ParticleDensity-v=095-b=1}
\end{figure}

\begin{figure}
    \centering
    \includegraphics[width=0.45\textwidth]{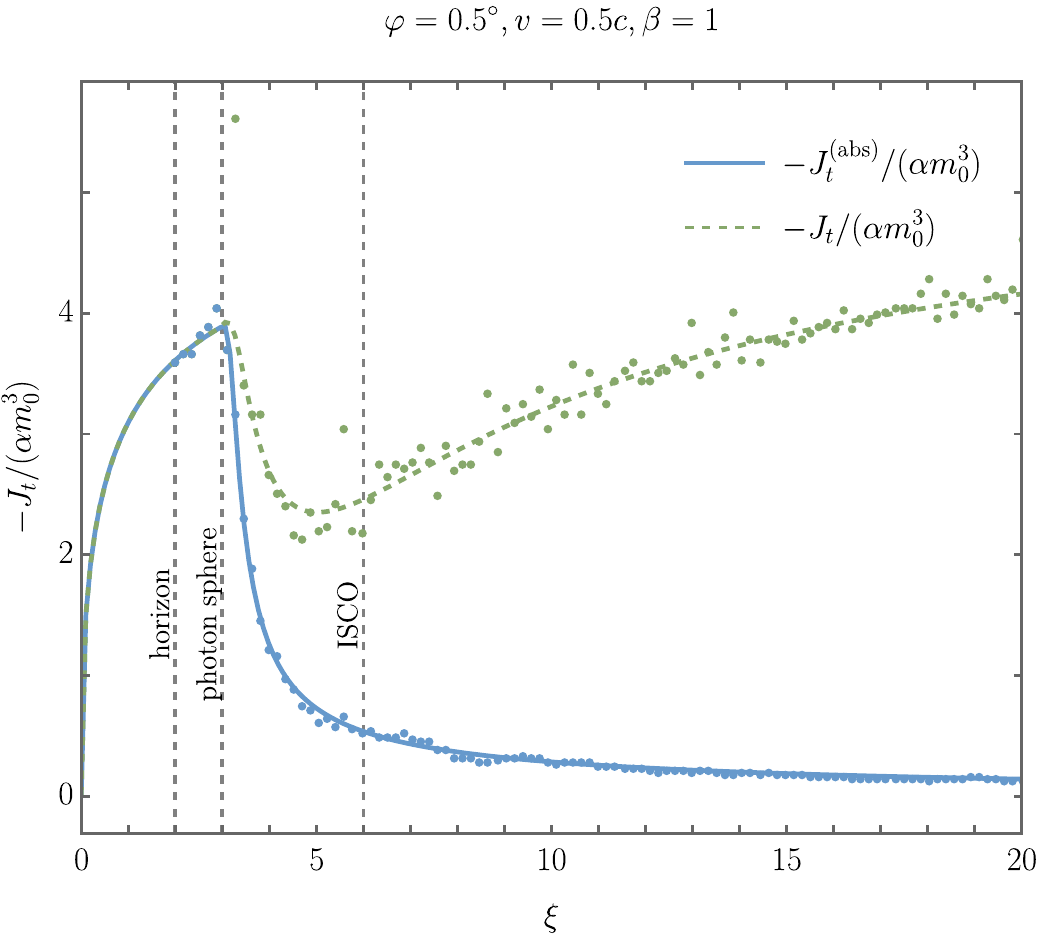}
    \includegraphics[width=0.45\textwidth]{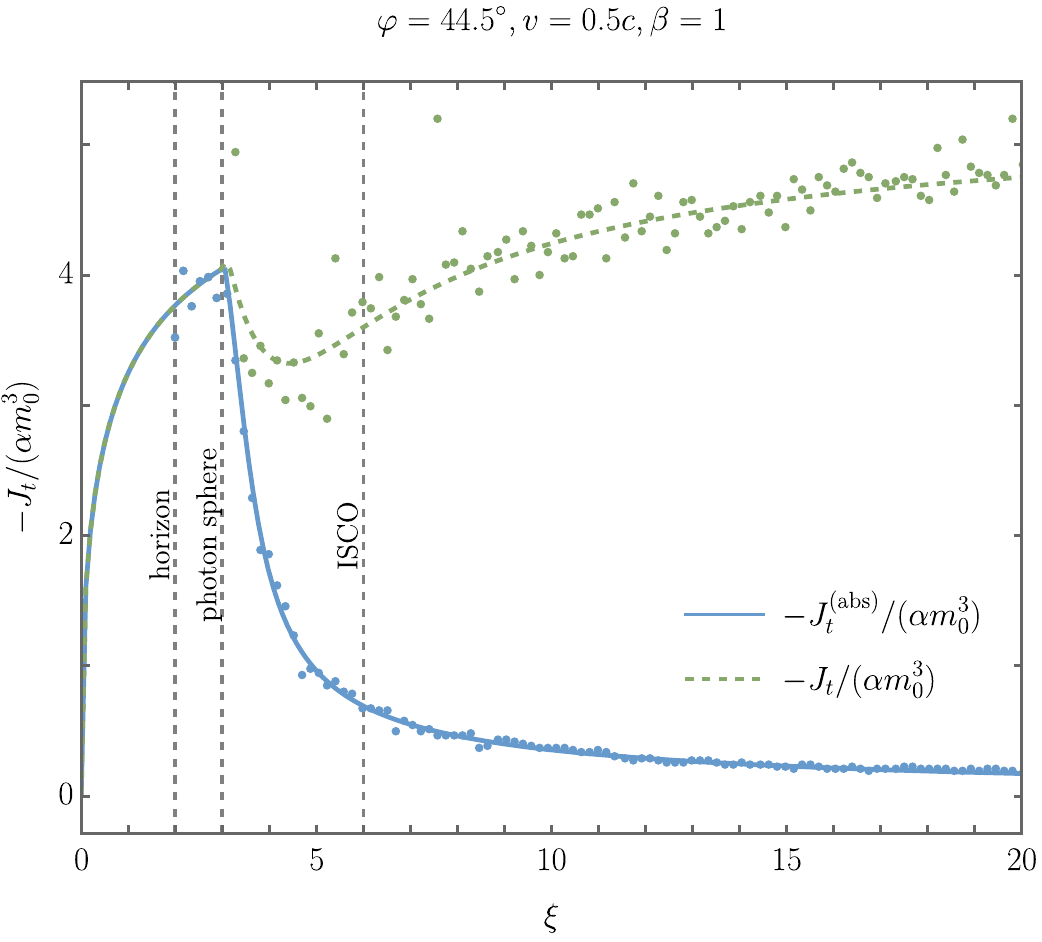}
    \includegraphics[width=0.45\textwidth]{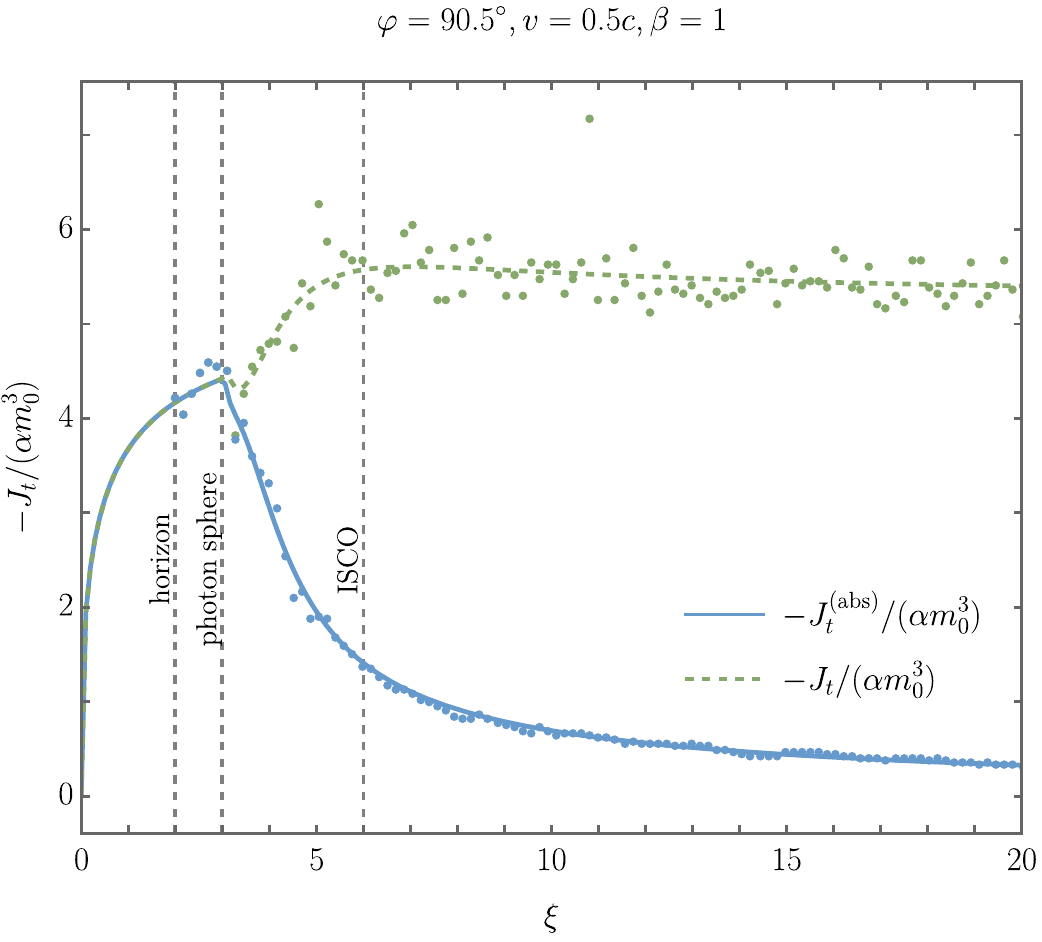}
    \includegraphics[width=0.45\textwidth]{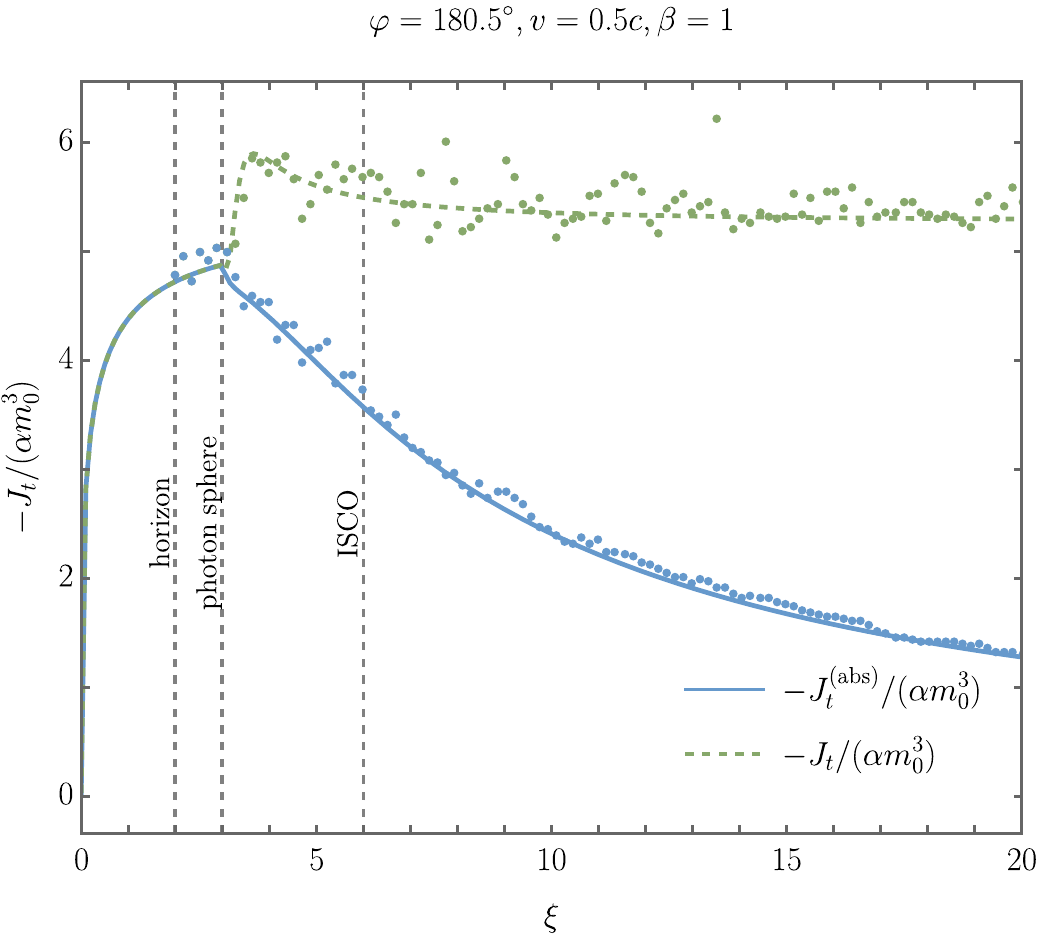}
    \caption{
    Time components of the particle current surface density $J_t$ in the model with $v = 0.5$, $\beta = 1$, $\varepsilon_\mathrm{cutoff} = 10$, and $\xi_0 = 1000$. Exact solutions (Eqs.\ \eqref{Jmuexact}) are plotted with solid and dashed lines. Dots (blue and green) represent sample results obtained by the Monte Carlo simulation (Eqs.\ \eqref{JmuMCsim}). There are $78\;361\;180$ trajectories: $N_\mathrm{abs} = 227\;889$, $N^-_\mathrm{scat} = 39\;064\;112$, $N^+_\mathrm{scat} = 39\;069\;179$.}
    \label{fig:J_t-v=05-b=1}
\end{figure}

\begin{figure}
    \centering
    \includegraphics[width=0.45\textwidth]{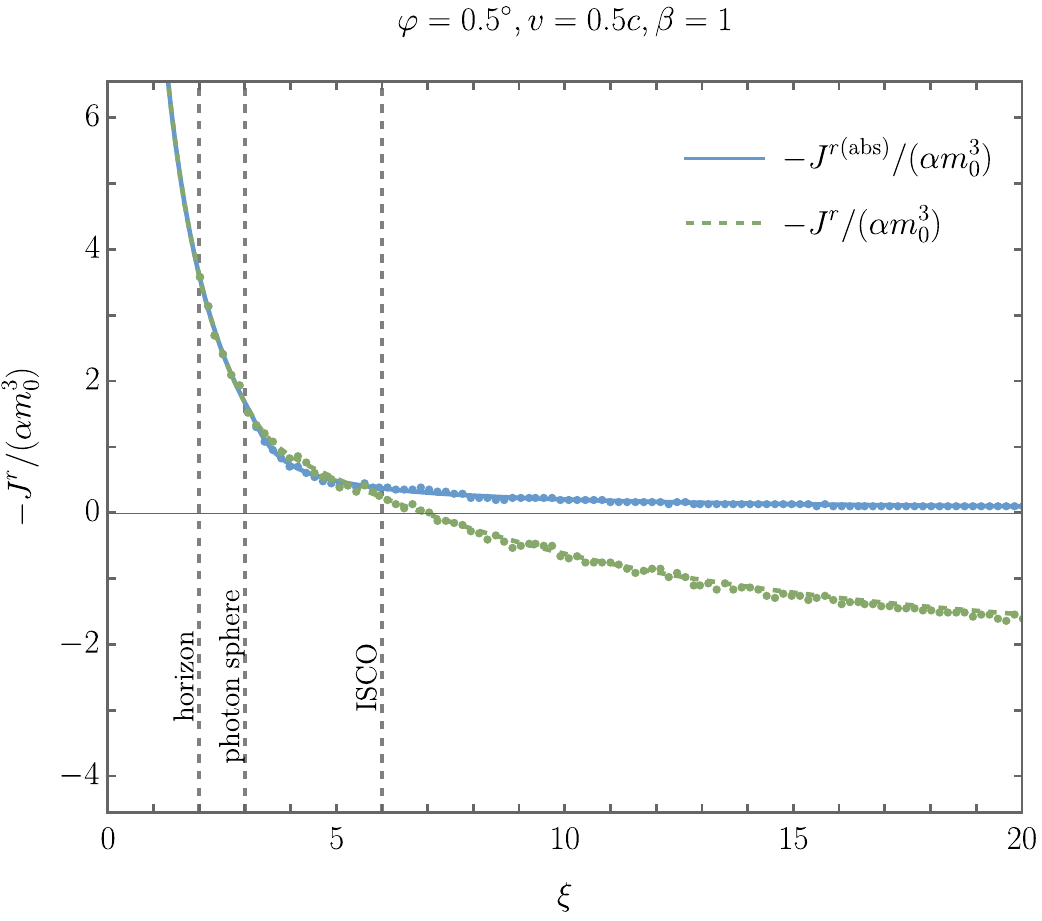}
    \includegraphics[width=0.45\textwidth]{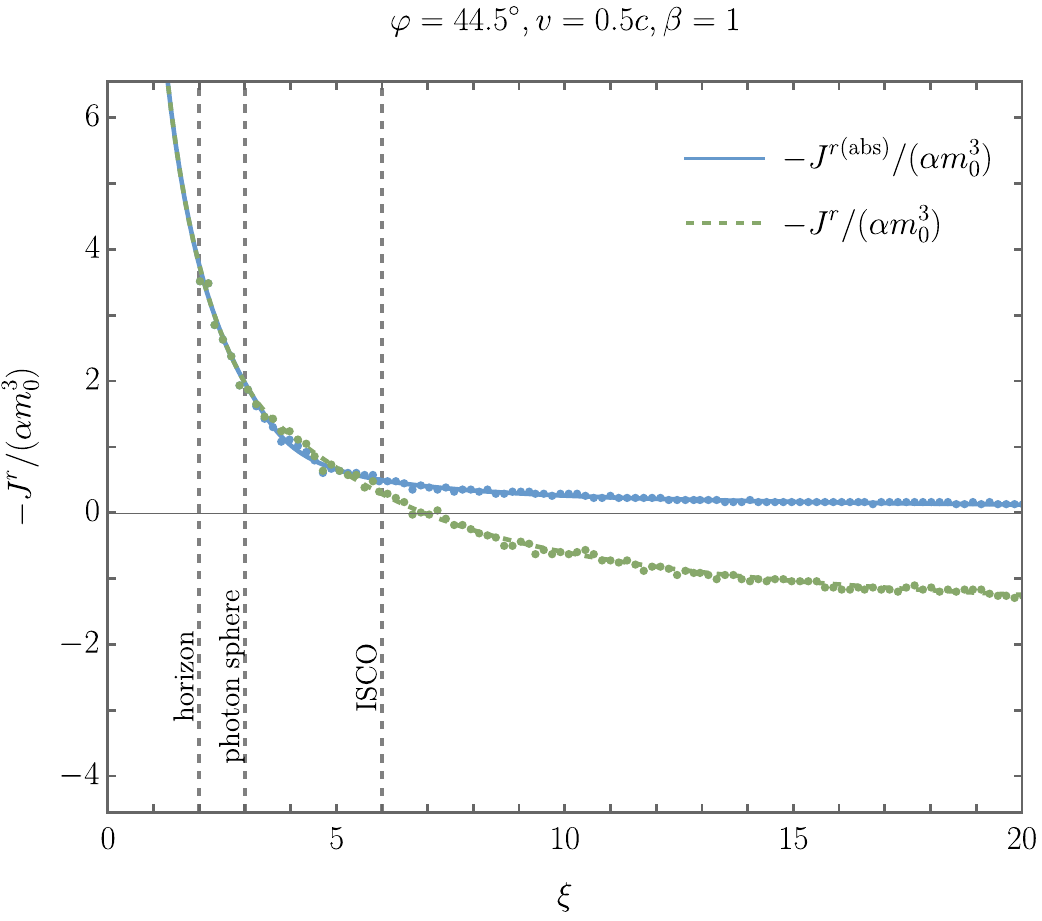}
    \includegraphics[width=0.45\textwidth]{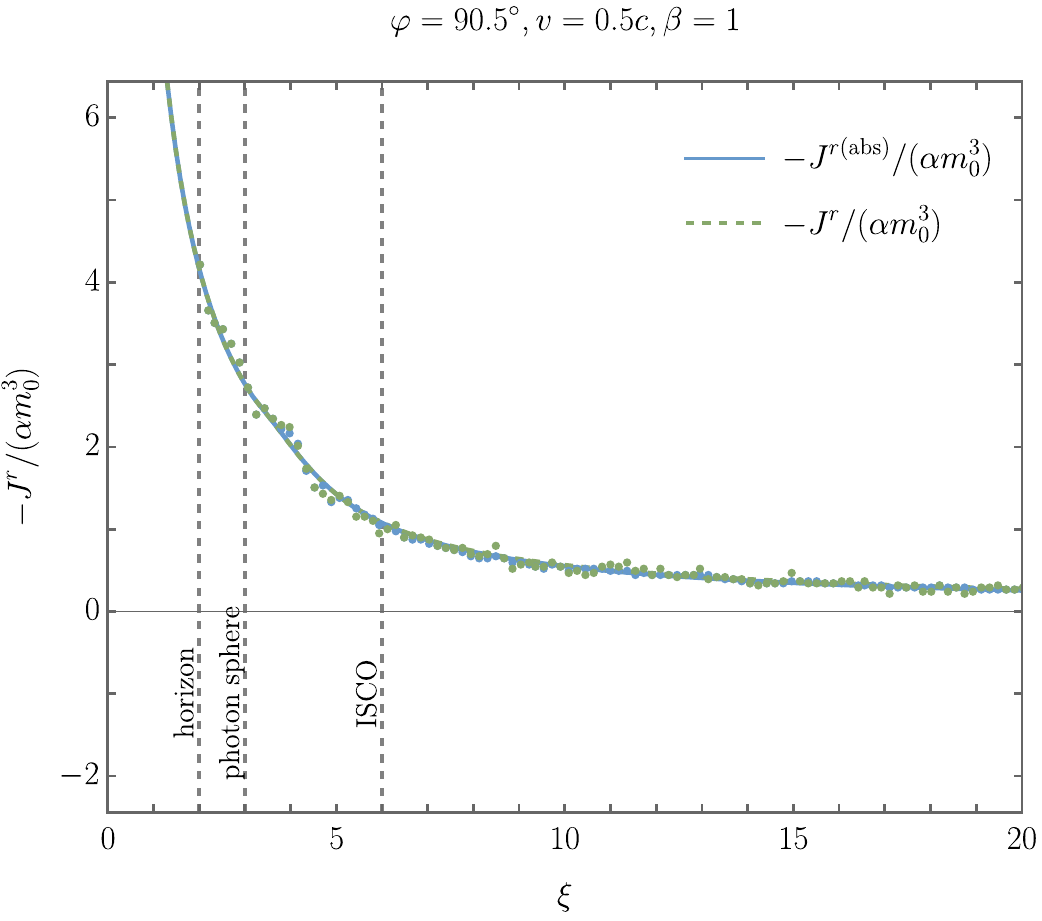}
    \includegraphics[width=0.45\textwidth]{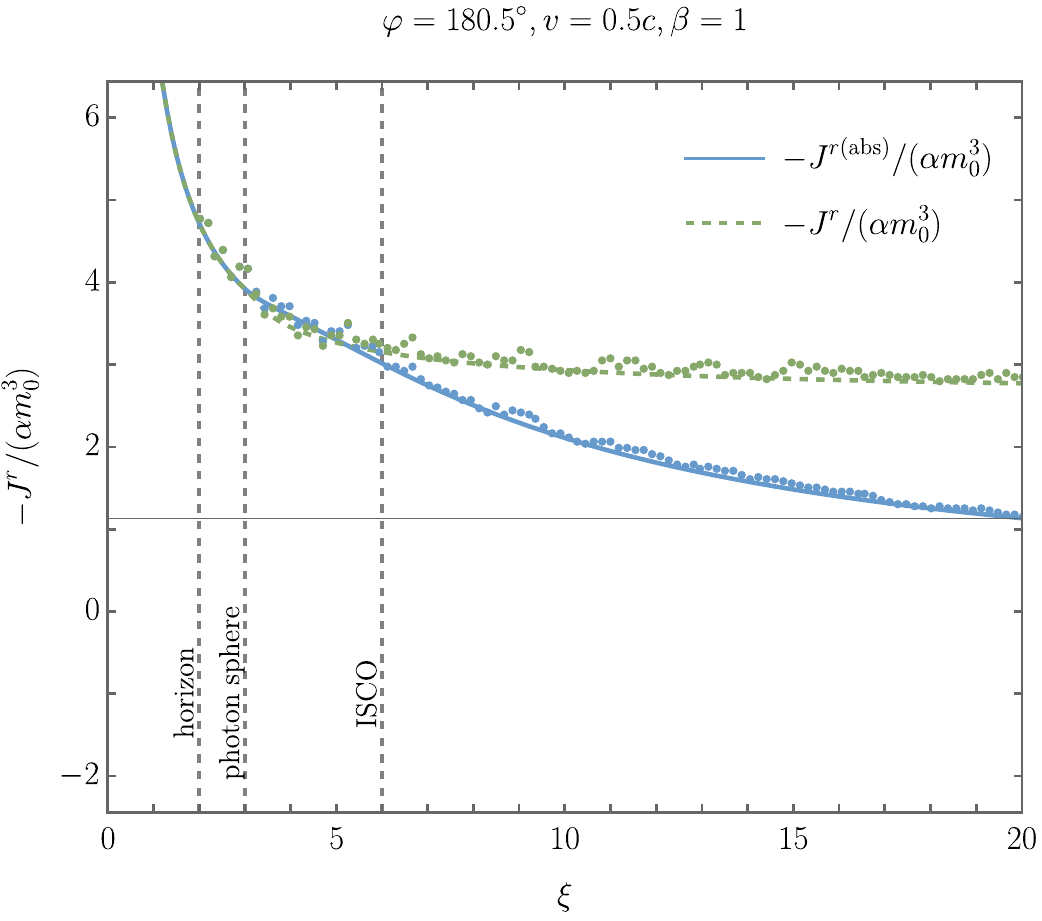}
    \caption{Radial components of the particle current surface density $J^r$ in the model with $v = 0.5$, $\beta = 1$, $\varepsilon_\mathrm{cutoff} = 10$, and $\xi_0 = 1000$. Components of the particle current surface density $J^r$ for $\xi_0 = 1000$. Exact solutions (Eqs.\ \eqref{Jmuexact}) are plotted with solid and dashed lines. Dots (blue and green) represent sample results obtained by the Monte Carlo simulation (Eqs.\ \eqref{JmuMCsim}). There are $78\;361\;180$ trajectories: $N_\mathrm{abs} = 227\;889$, $N^-_\mathrm{scat} = 39\;064\;112$, $N^+_\mathrm{scat} = 39\;069\;179$.}
    \label{fig:J^r-v=05-b=1}
\end{figure}

\begin{figure}
    \centering
    \includegraphics[width=0.45\textwidth]{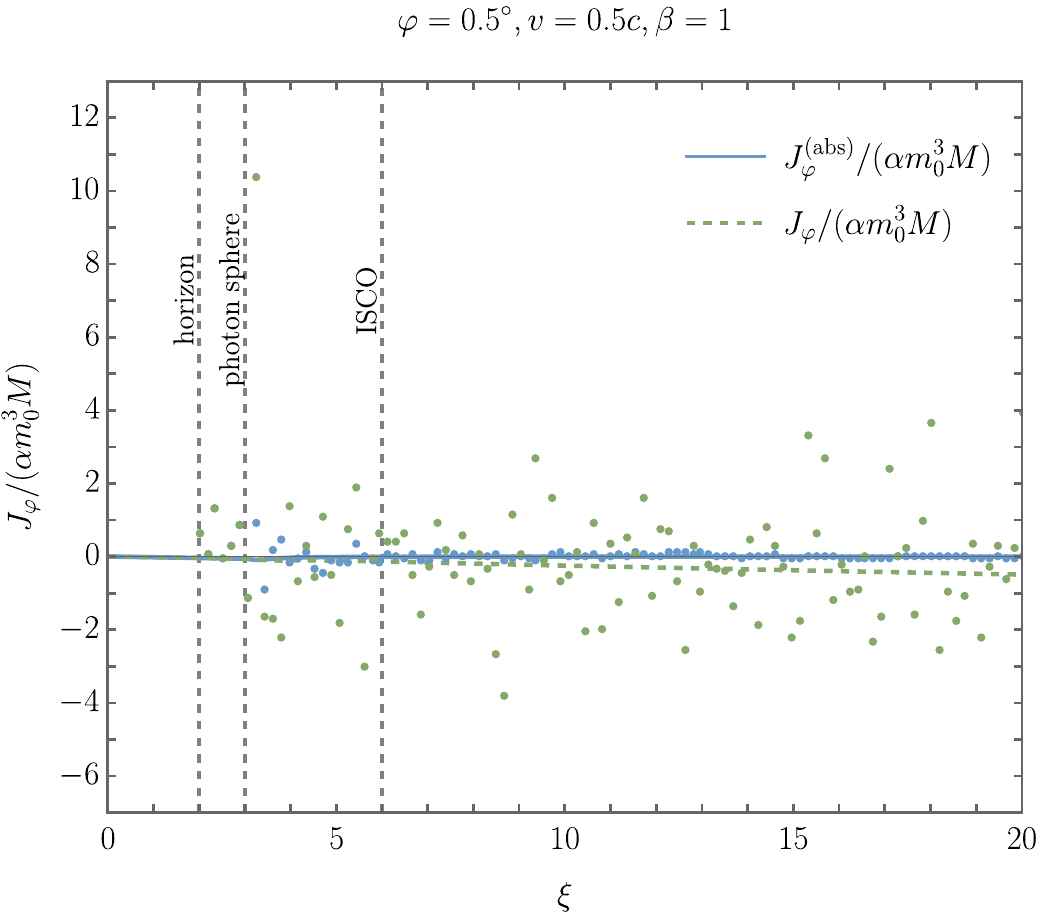}
    \includegraphics[width=0.45\textwidth]{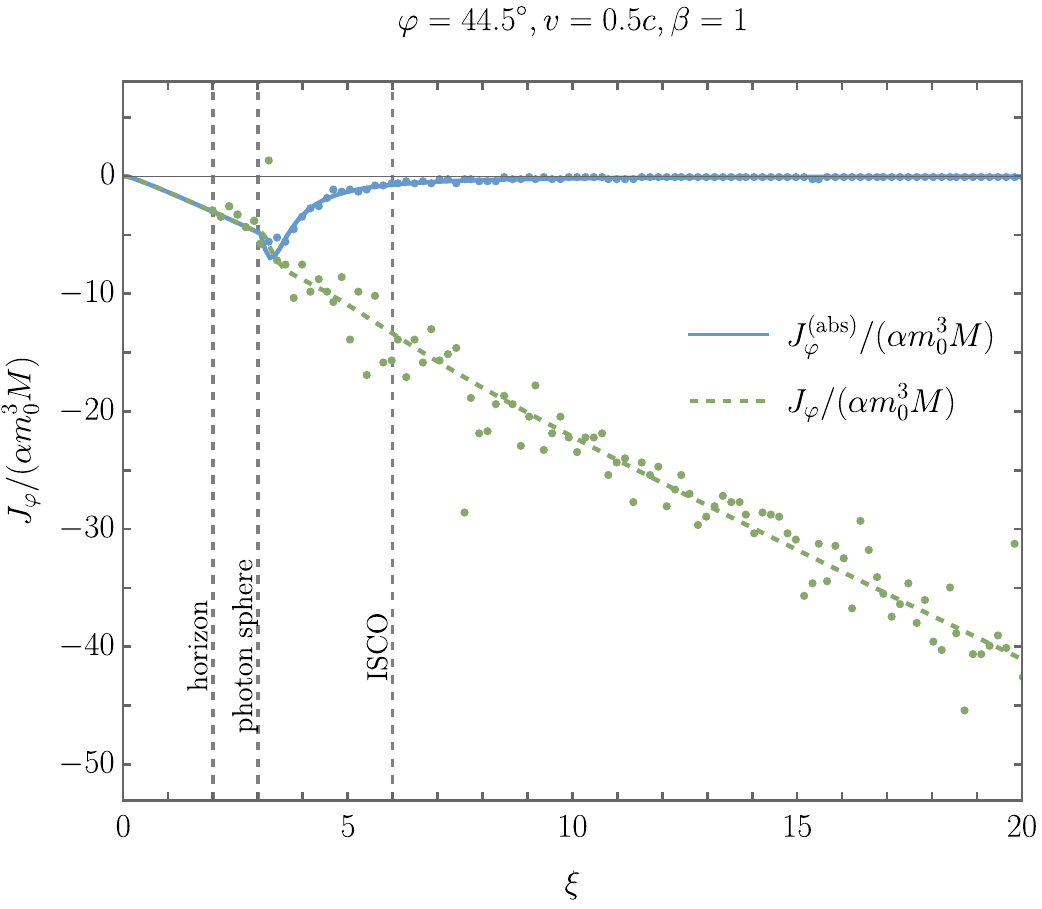}
    \includegraphics[width=0.45\textwidth]{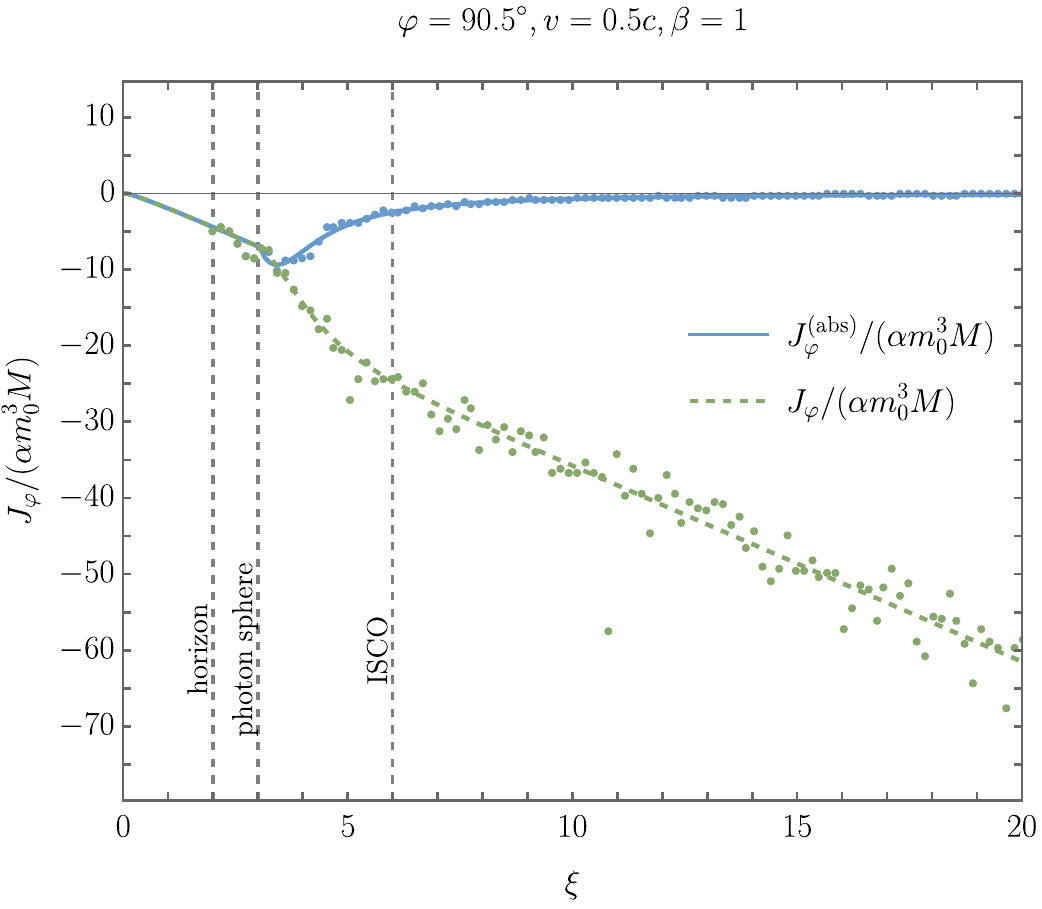}
    \includegraphics[width=0.45\textwidth]{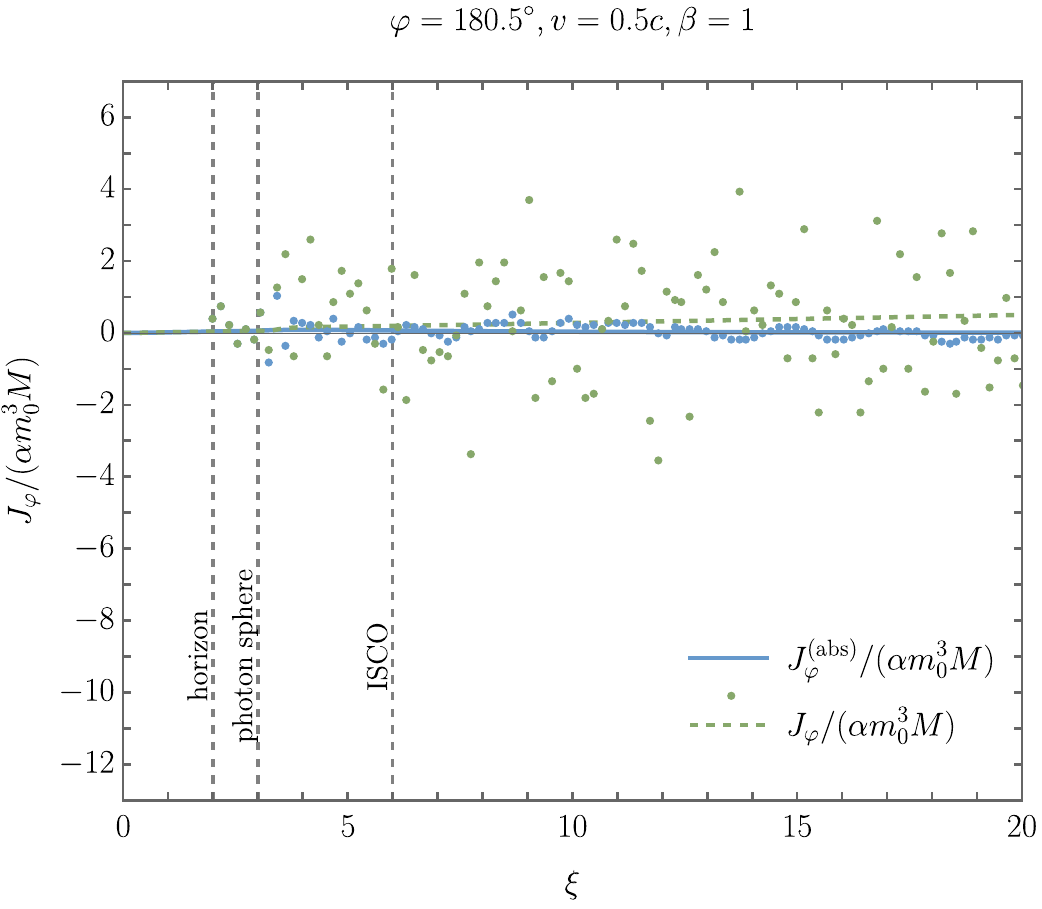}
    \caption{Angular components of the particle current surface density $J_\varphi$ in the model with $v = 0.5$, $\beta = 1$, $\varepsilon_\mathrm{cutoff} = 10$, and $\xi_0 = 1000$. Exact solutions (Eqs.\ \eqref{Jmuexact}) are plotted with solid and dashed lines. Dots (blue and green) represent sample results obtained by the Monte Carlo simulation (Eqs.\ \eqref{JmuMCsim}). There are $78\;361\;180$ trajectories: $N_\mathrm{abs} = 227\;889$, $N^-_\mathrm{scat} = 39\;064\;112$, $N^+_\mathrm{scat} = 39\;069\;179$.}
    \label{fig:J_fi-v=05-b=1}
\end{figure}

\begin{figure}
    \centering
    \includegraphics[width=0.45\textwidth]{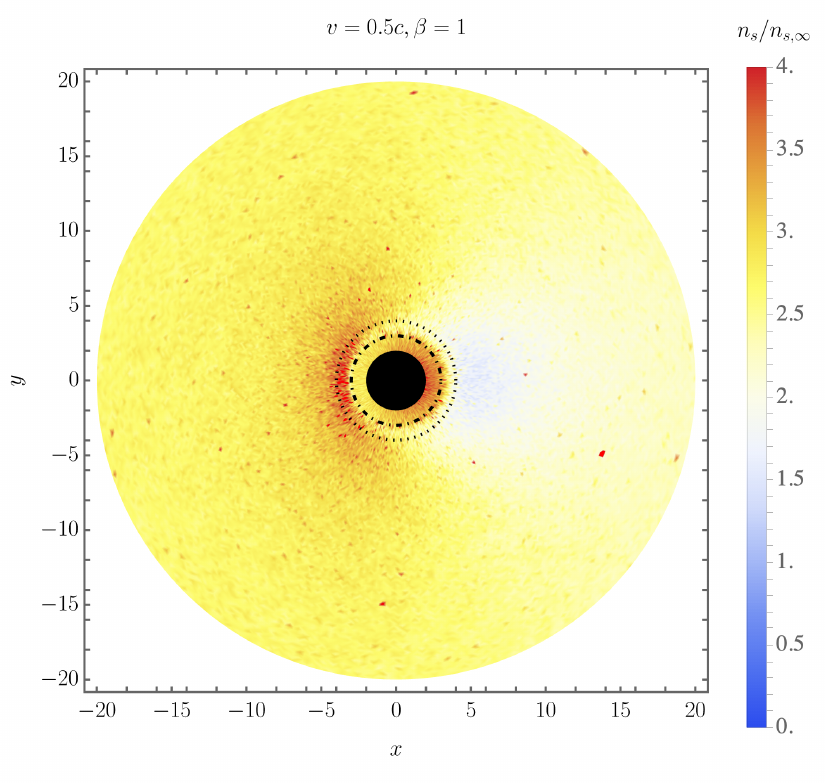}
    \includegraphics[width=0.45\textwidth]{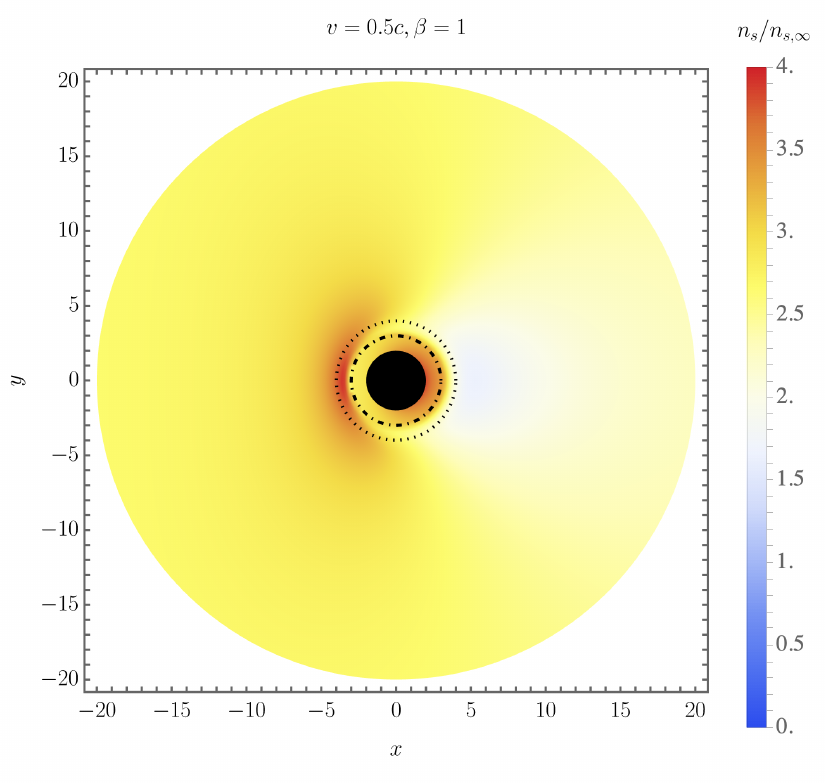}
    \includegraphics[width=0.45\textwidth]{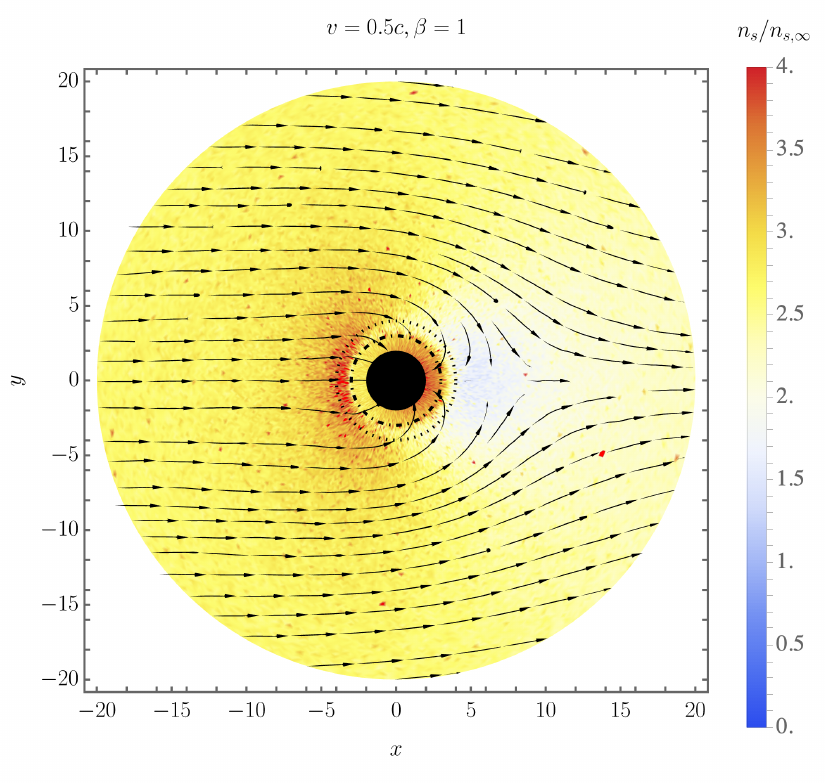}
    \includegraphics[width=0.45\textwidth]{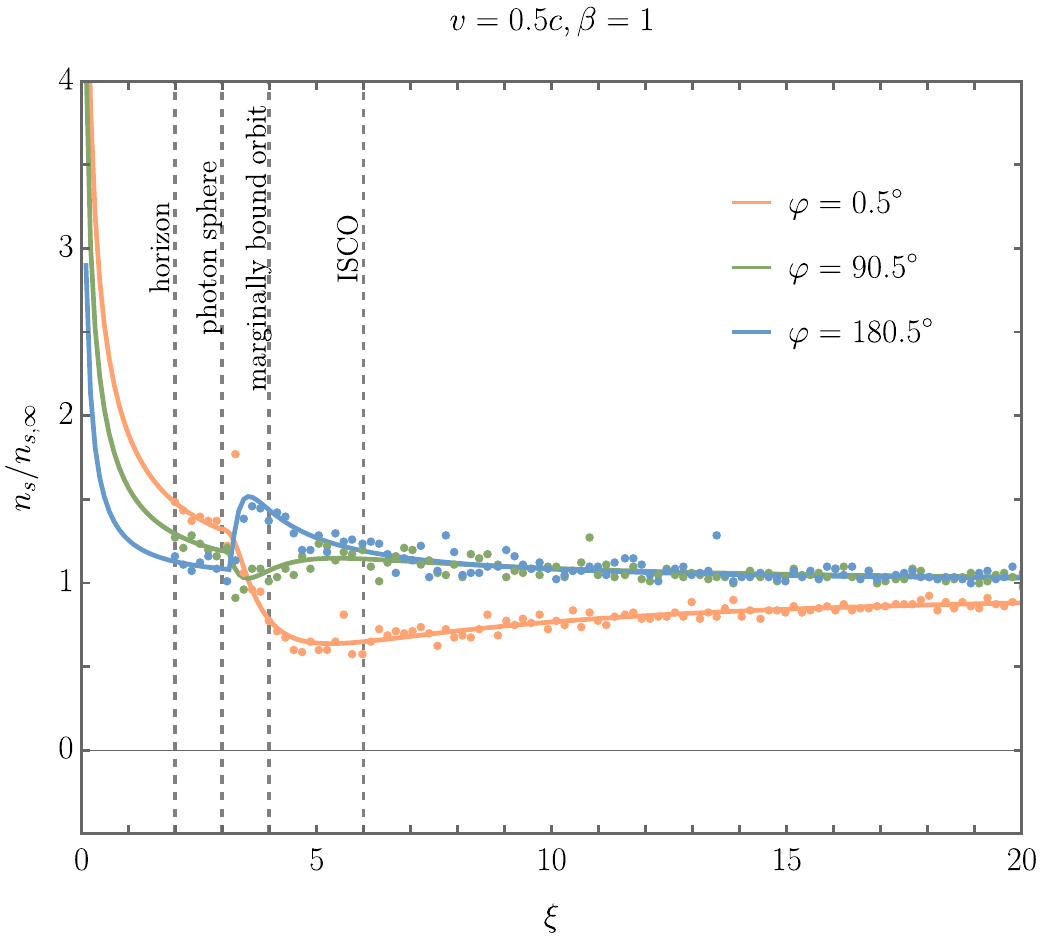}
    \caption{
    Same as in Fig.~\ref{fig:ParticleDensity-v=095-b=1} for the model with smaller velocity $v = 0.5$. Remaining parameters $\beta = 1$, $\varepsilon_\mathrm{cutoff} = 10$, and $\xi_0 = 1000$. The simulation was prepared for $78\;361\;180$ particle trajectories: $N_\mathrm{abs} = 227\;889$, $N^-_\mathrm{scat} =  39\;064\;112$, $N^+_\mathrm{scat} = 39\;069\;179$.}
    \label{fig:ParticleDensity-v=05-b=1}
\end{figure}

\begin{figure}
    \centering
    \includegraphics[width=0.45\textwidth]{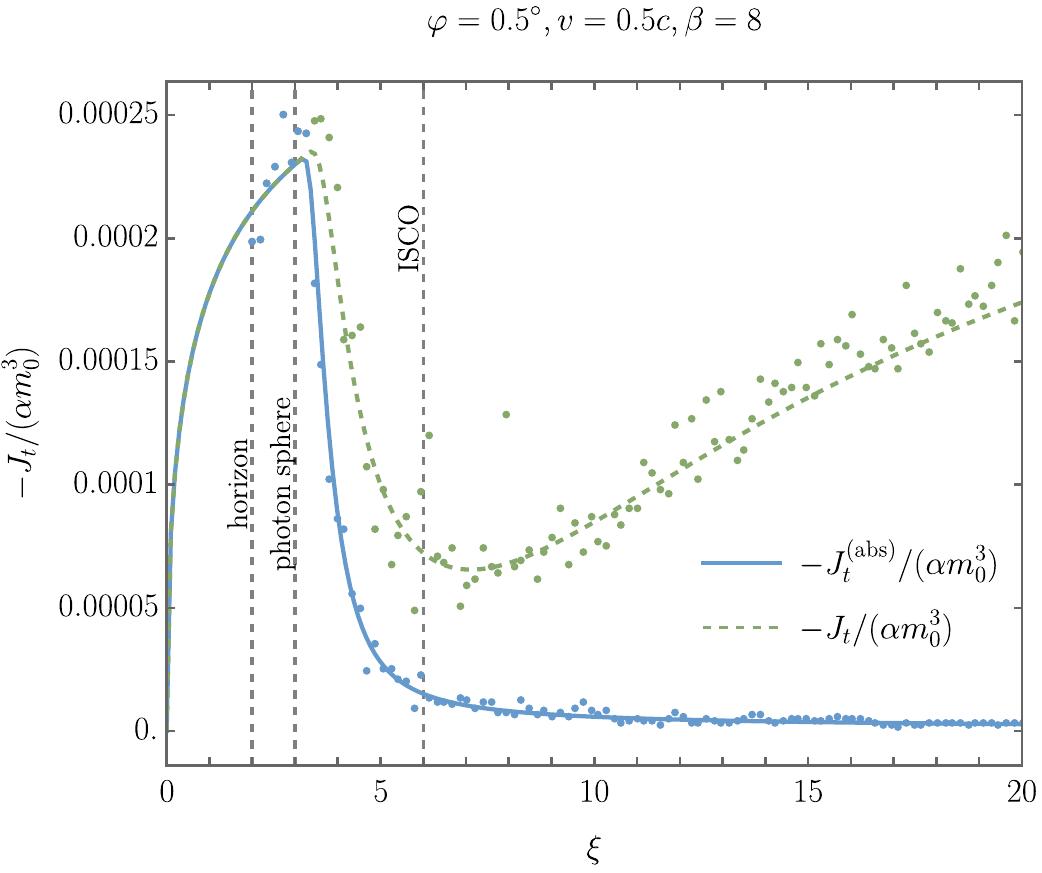}
    \includegraphics[width=0.45\textwidth]{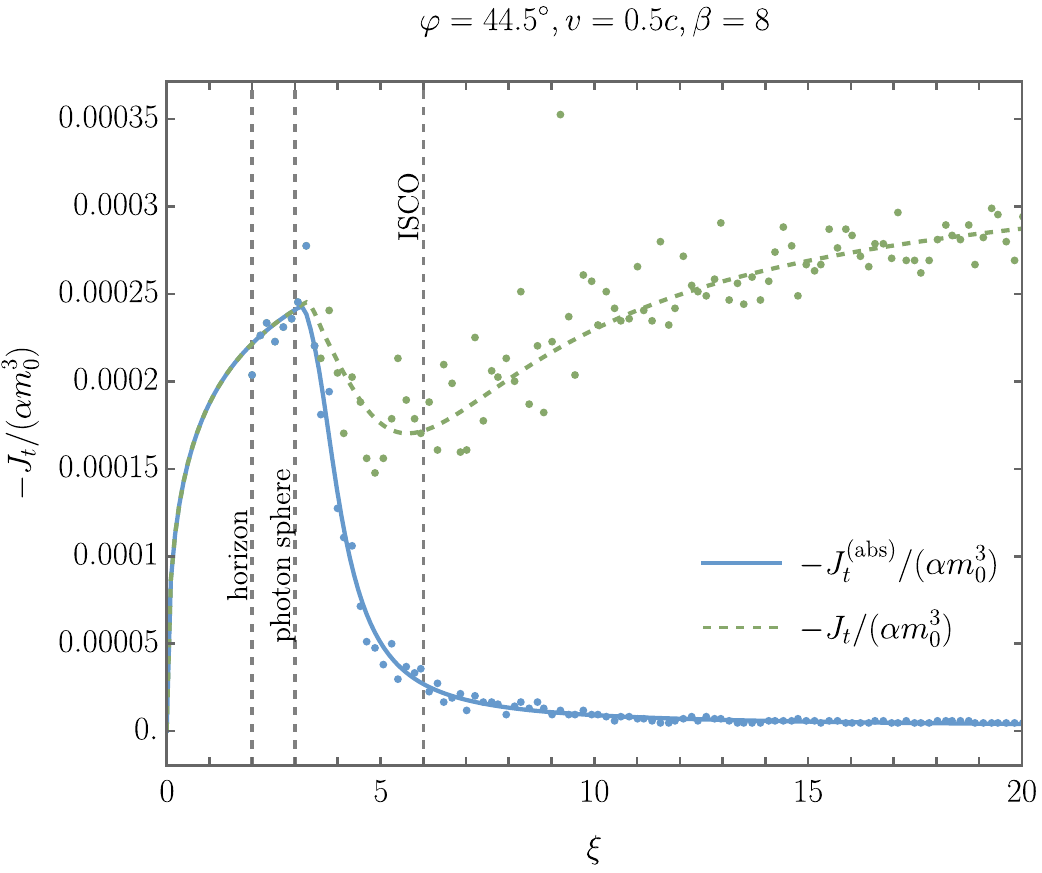}
    \includegraphics[width=0.45\textwidth]{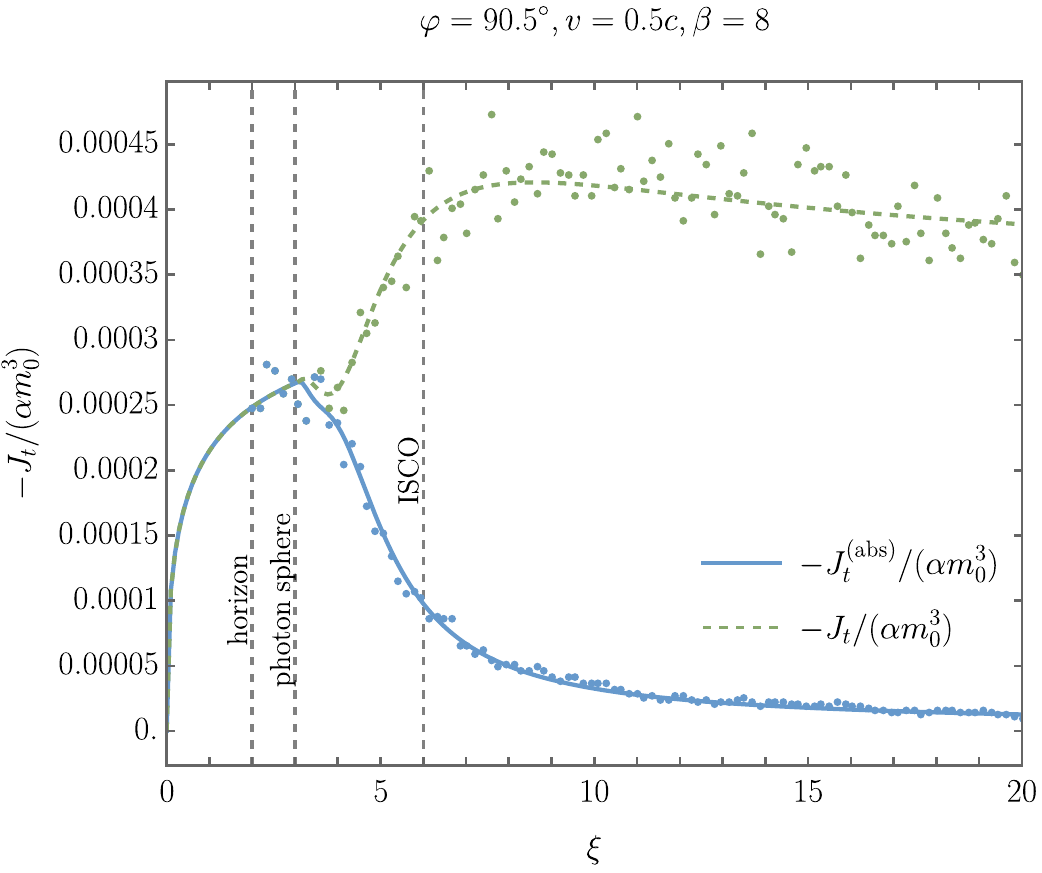}
    \includegraphics[width=0.45\textwidth]{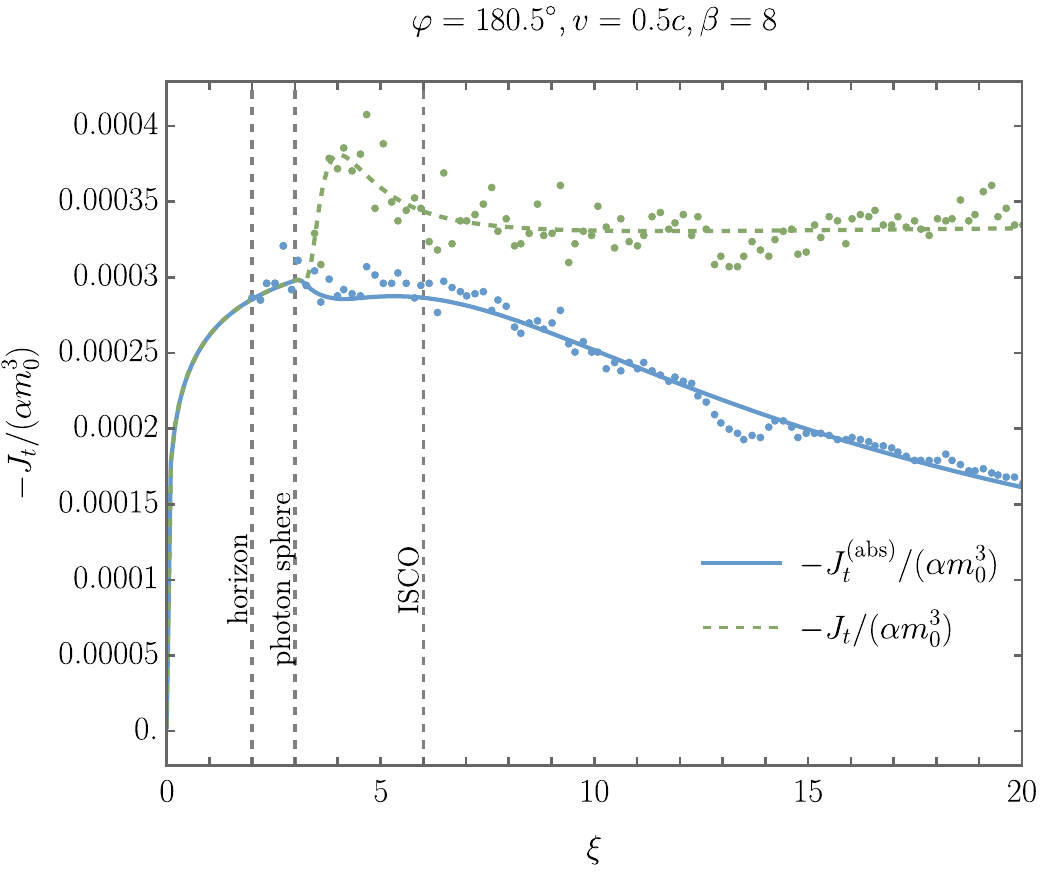}
    \caption{Time components of the particle current surface density $J_t$ in the model with $v = 0.5$, $\beta = 8$, $\varepsilon_\mathrm{cutoff} = 10$, and $\xi_0 = 1000$. Exact solutions (Eqs.\ \eqref{Jmuexact}) are plotted with solid and dashed lines. Dots (blue and green) represent sample results obtained by the Monte Carlo simulation (Eqs.\ \eqref{JmuMCsim}). There are $30\;900\;617$ trajectories: $N_\mathrm{abs} = 118\;561$, $N^-_\mathrm{scat} = 15\;391\;121$, $N^+_\mathrm{scat} = 15\;390\;935$.}
    \label{fig:J_t-v=05-b=8}
\end{figure}

\begin{figure}
    \centering
    \includegraphics[width=0.45\textwidth]{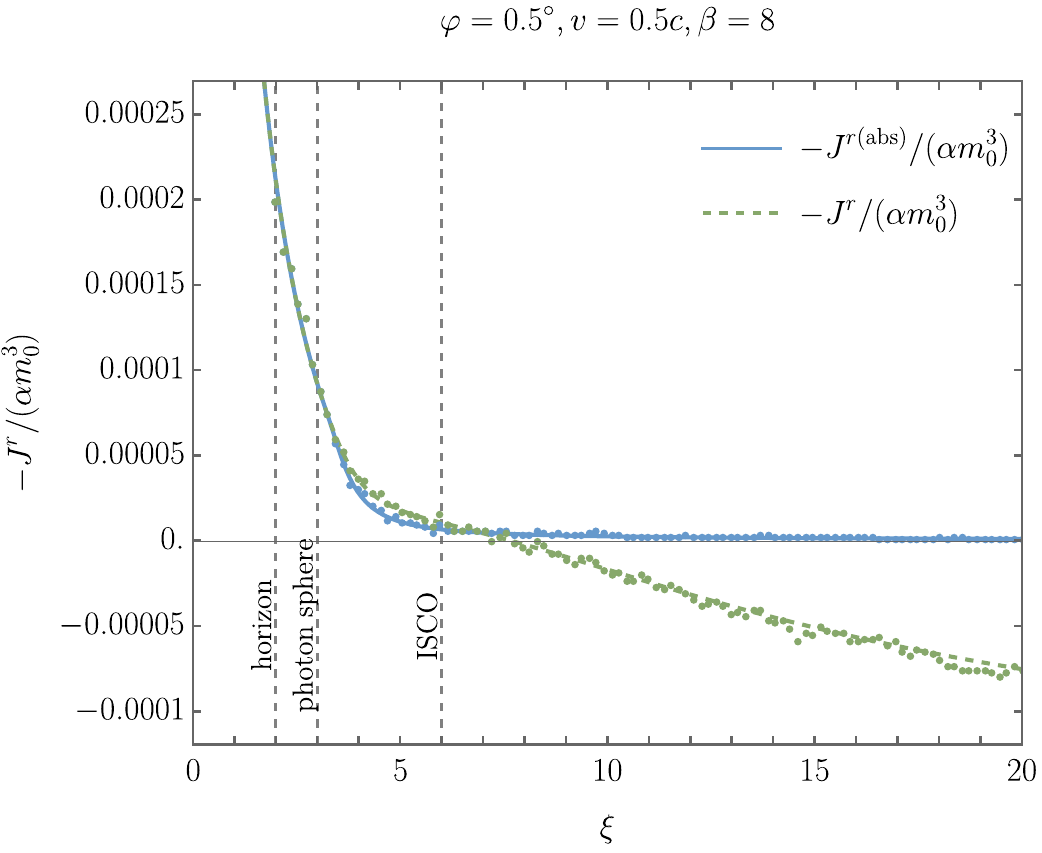}
    \includegraphics[width=0.45\textwidth]{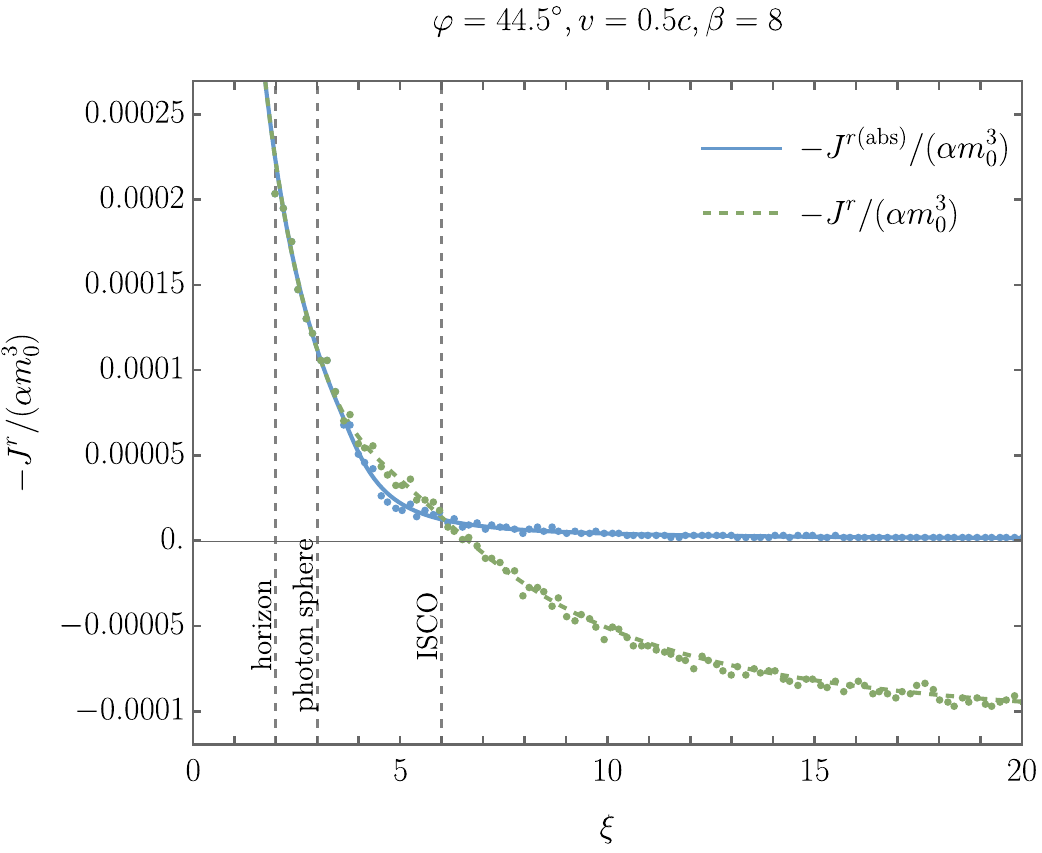}
    \includegraphics[width=0.45\textwidth]{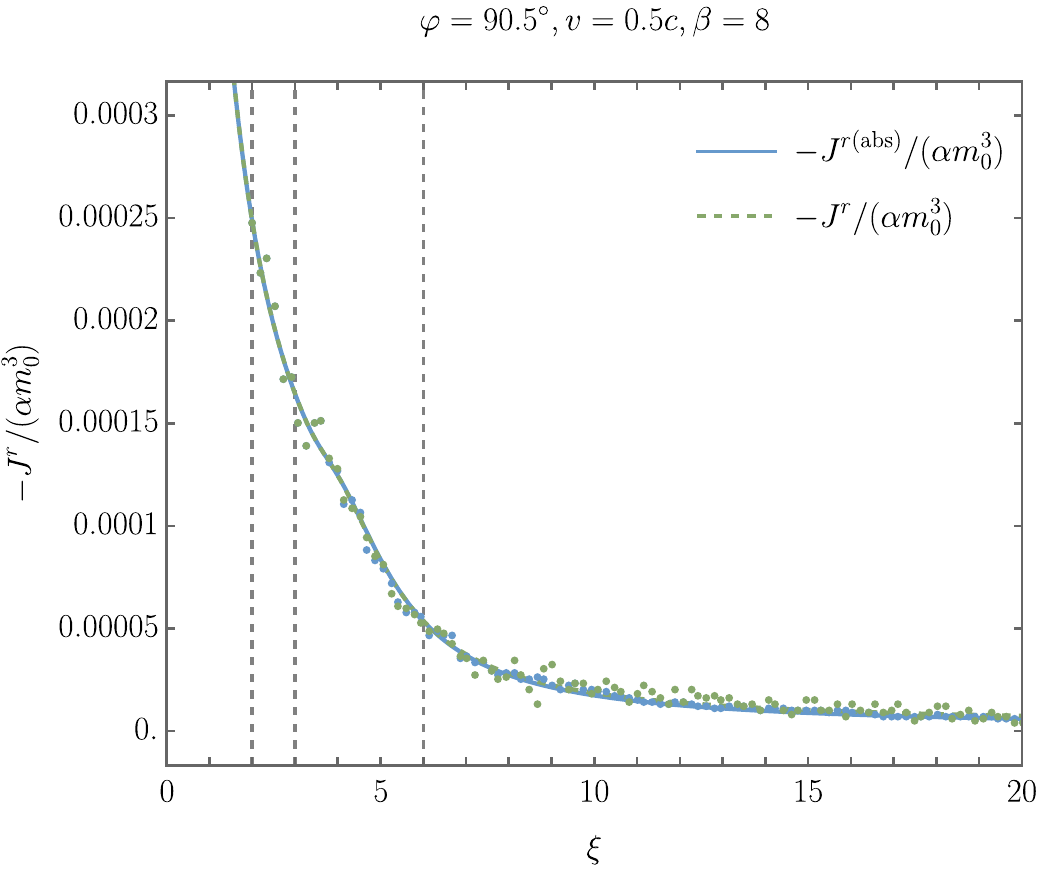}
    \includegraphics[width=0.45\textwidth]{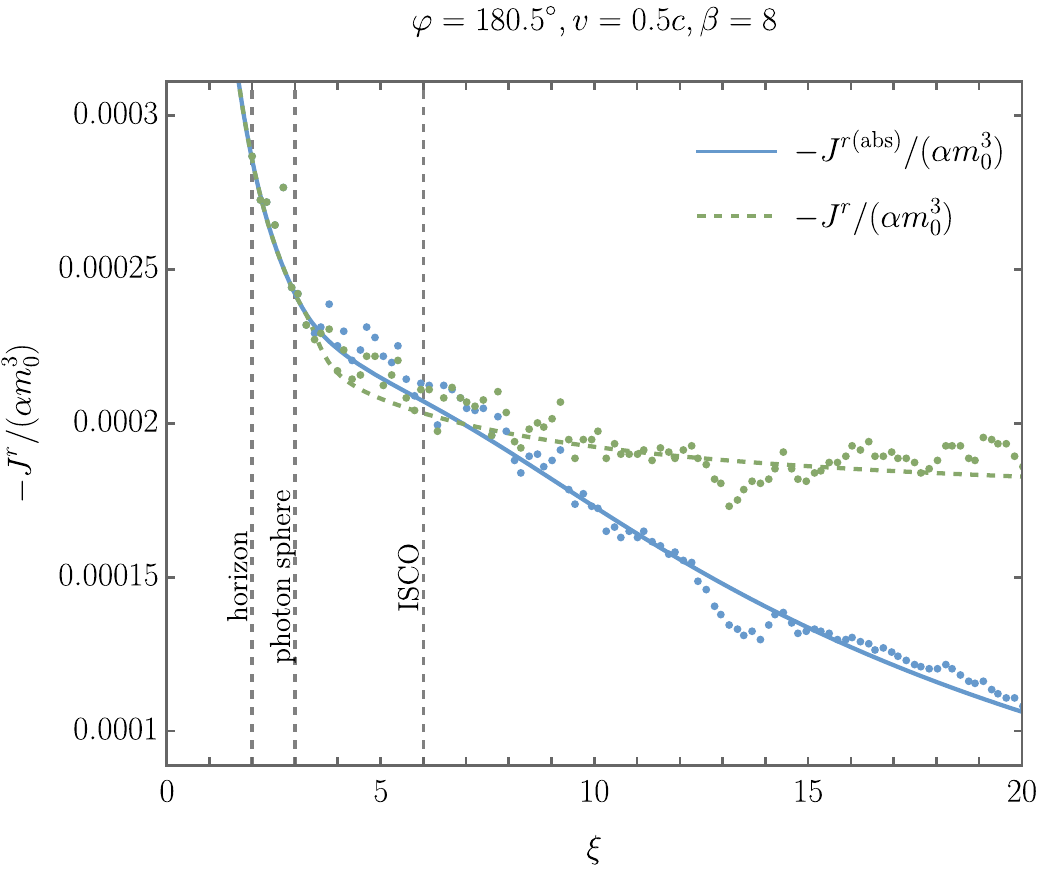}
    \caption{
    Radial components of the particle current surface density $J^r$ in the model with $v = 0.5$, $\beta = 8$, $\varepsilon_\mathrm{cutoff} = 10$, and $\xi_0 = 1000$. Exact solutions (Eqs.\ \eqref{Jmuexact}) are plotted with solid and dashed lines. Dots (blue and green) represent sample results obtained by the Monte Carlo simulation (Eqs.\ \eqref{JmuMCsim}). There are $30\;900\;617$ particles: $N_\mathrm{abs} = 118\;561$, $N^-_\mathrm{scat} = 15\;391\;121$, $N^+_\mathrm{scat} = 15\;390\;935$.}
    \label{fig:J^r-v=05-b=8}
\end{figure}

\begin{figure}
    \centering
    \includegraphics[width=0.45\textwidth]{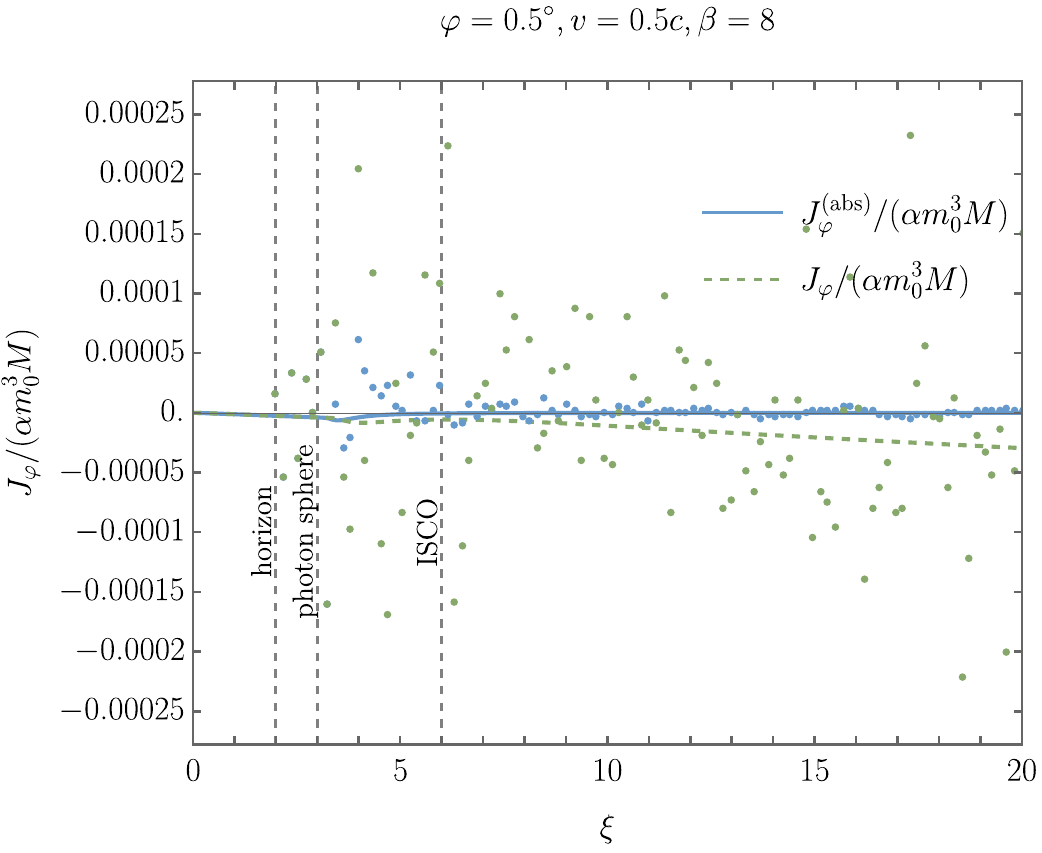}
    \includegraphics[width=0.45\textwidth]{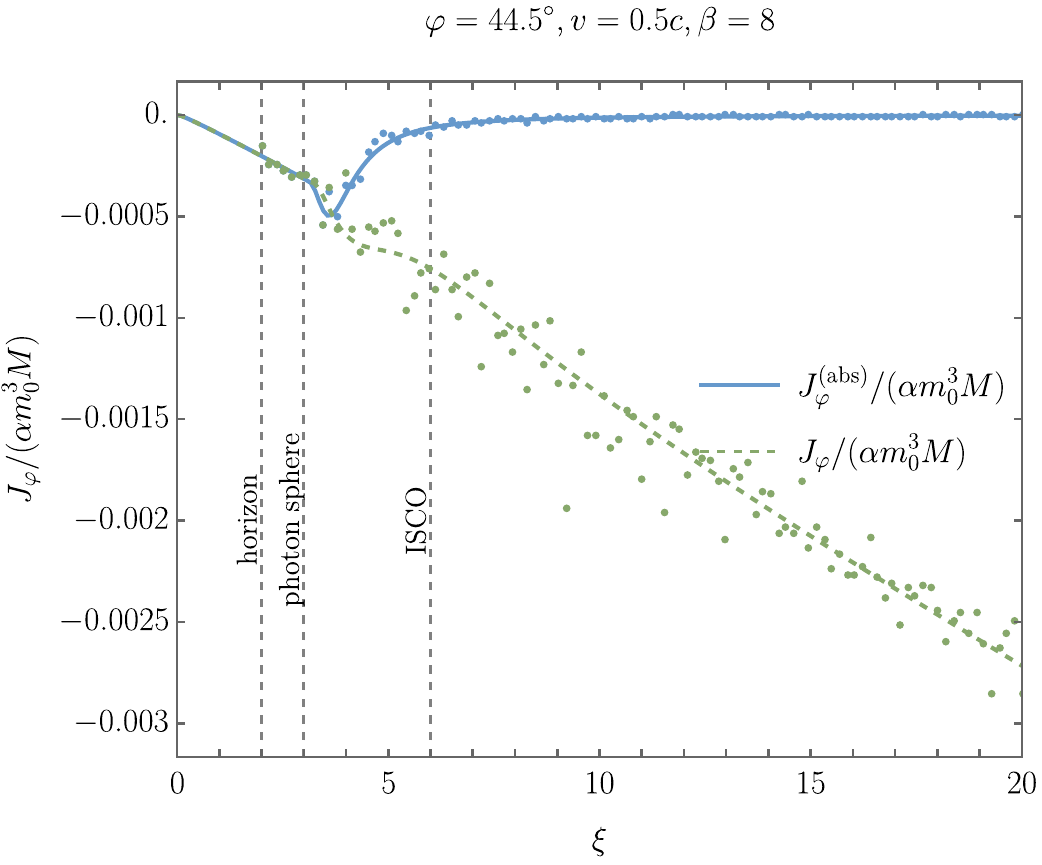}
    \includegraphics[width=0.45\textwidth]{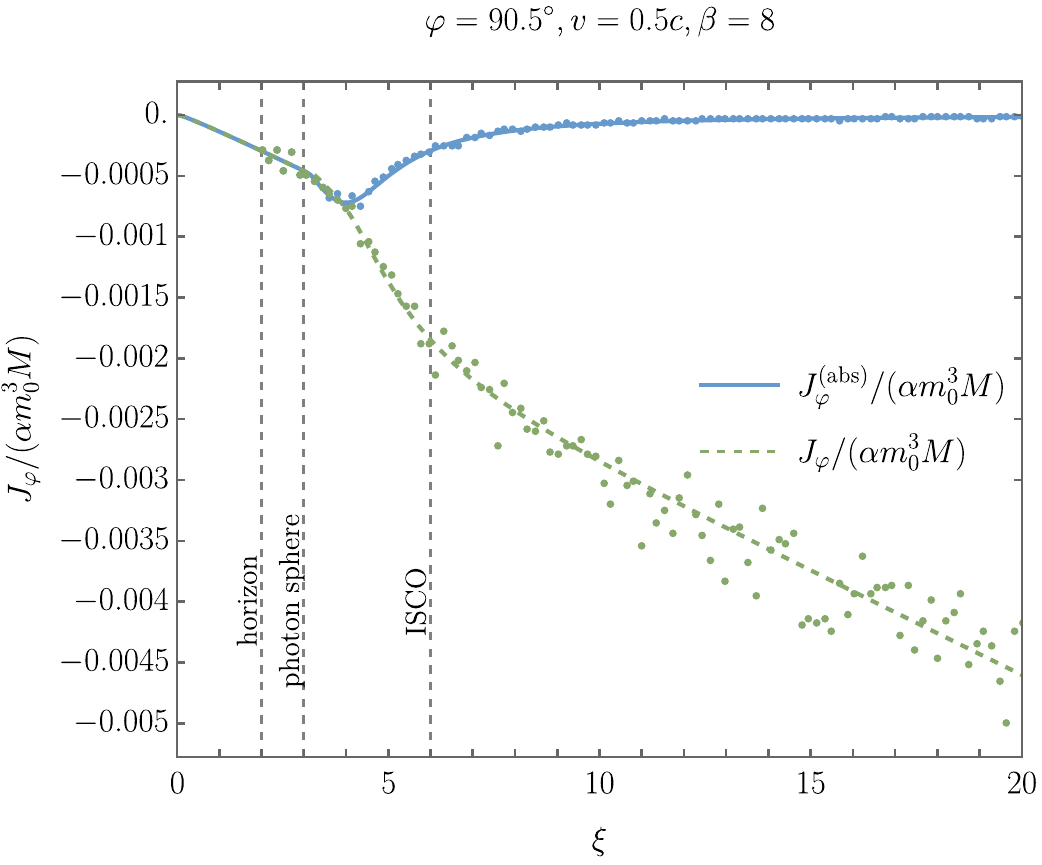}
    \includegraphics[width=0.45\textwidth]{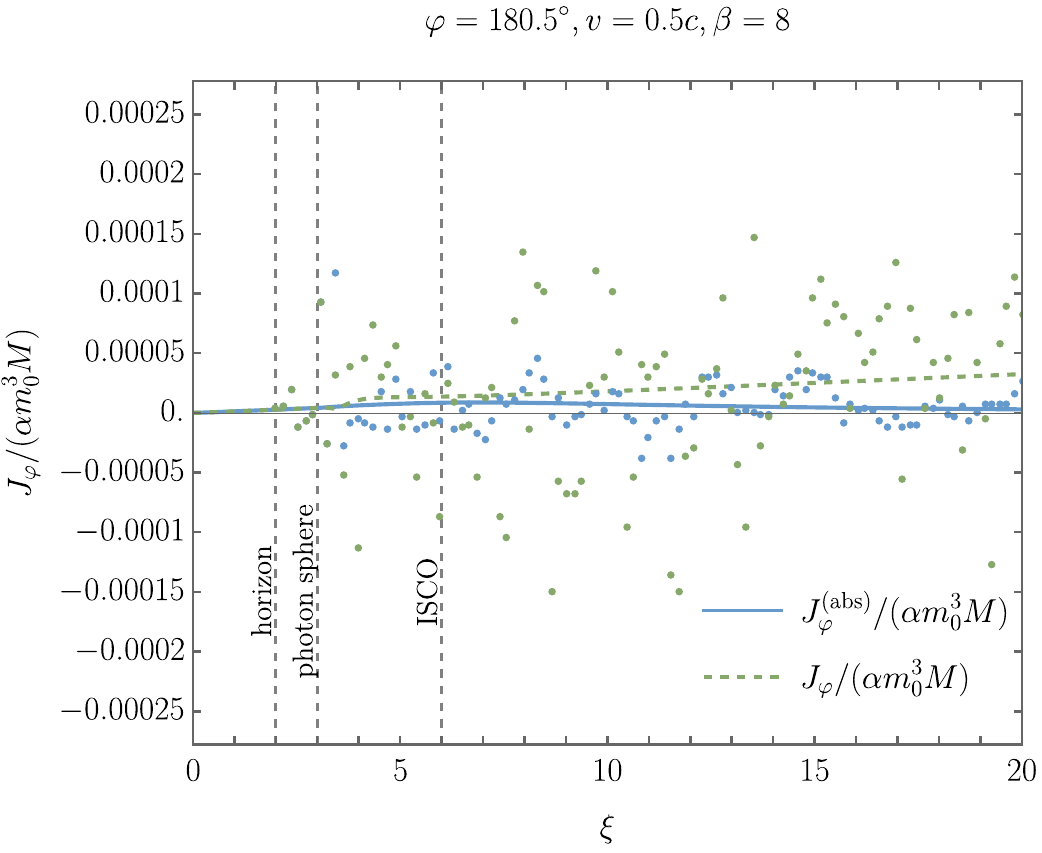}
    \caption{Angular components of the particle current surface density $J_\varphi$ in the model with $v = 0.5$, $\beta = 8$, $\varepsilon_\mathrm{cutoff} = 10$, and $\xi_0 = 1000$. Exact solutions (Eqs.\ \eqref{Jmuexact}) are plotted with solid and dashed lines. Dots (blue and green) represent sample results obtained by the Monte Carlo simulation (Eqs.\ \eqref{JmuMCsim}). There are $30\;900\;617$ trajectories: $N_\mathrm{abs} = 118\;561$, $N^-_\mathrm{scat} = 15\;391\;121$, $N^+_\mathrm{scat} = 15\;390\;935$.}
    \label{fig:J_fi-v=05-b=8}
\end{figure}

\begin{figure}
    \centering
    \includegraphics[width=0.45\textwidth]{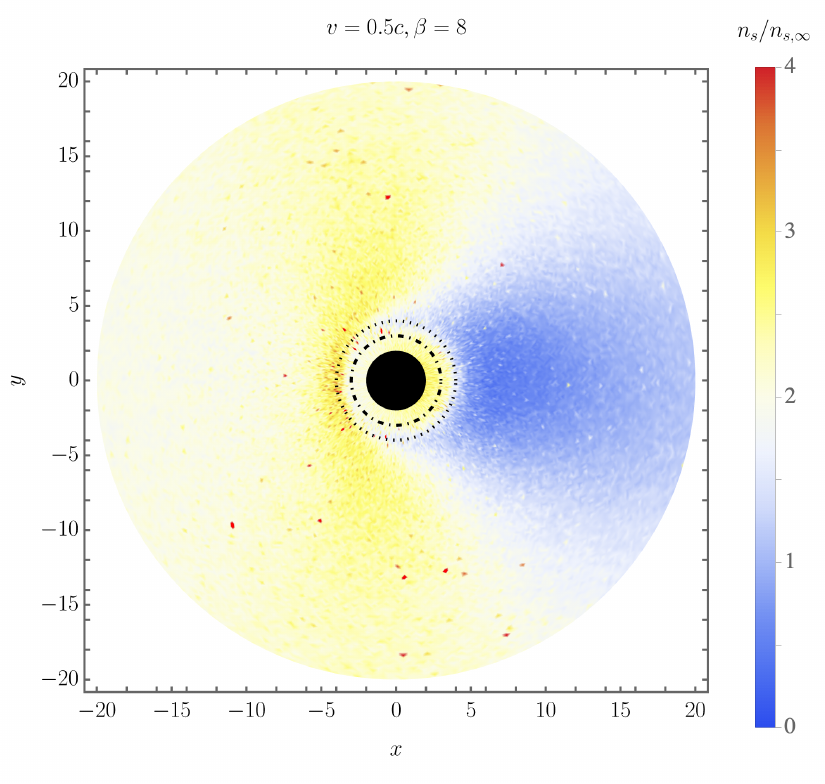}
    \includegraphics[width=0.45\textwidth]{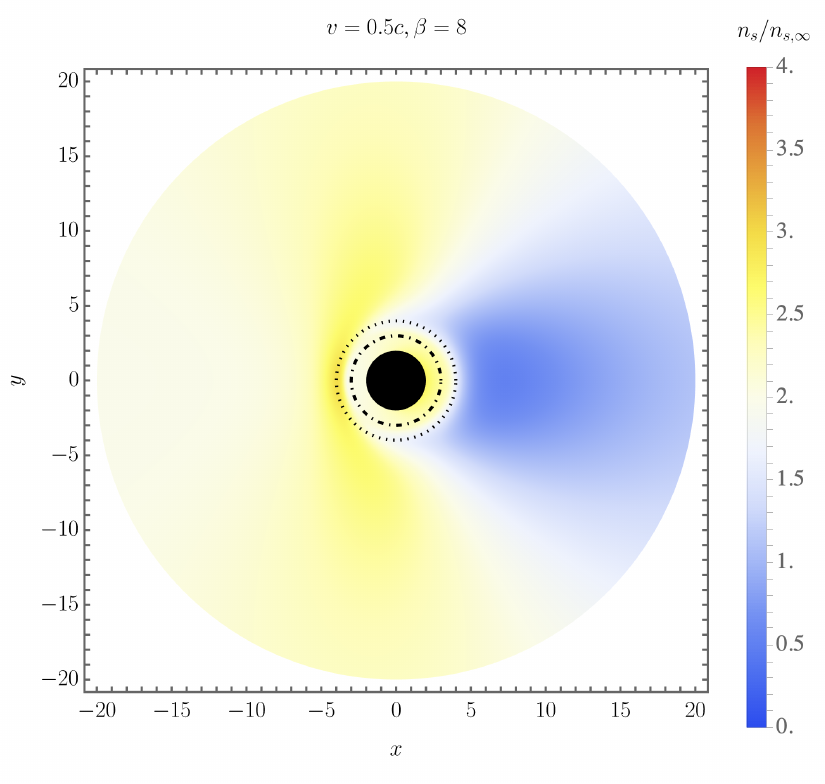}
    \includegraphics[width=0.45\textwidth]{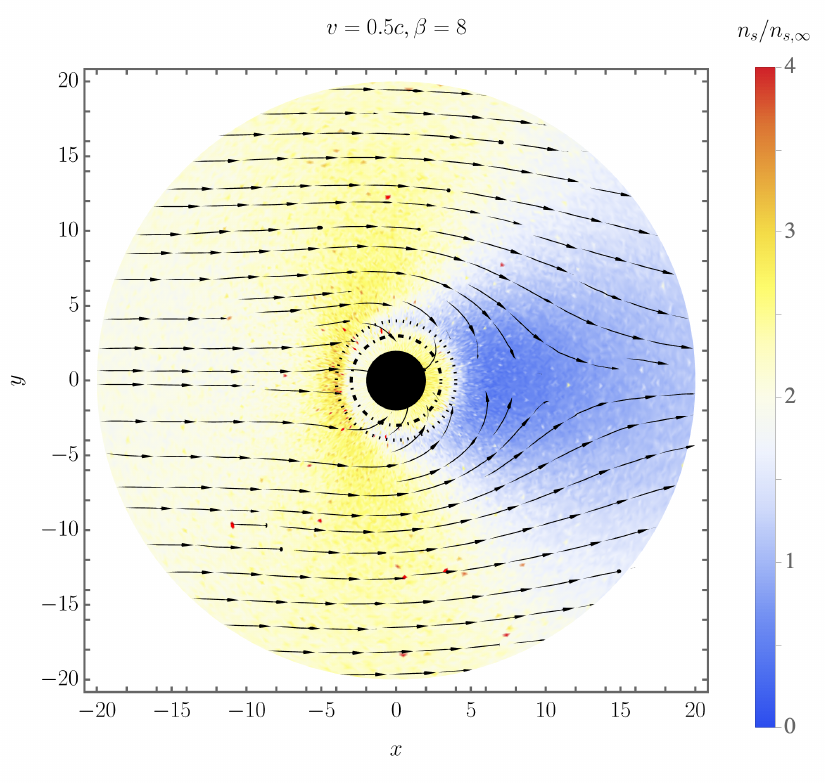}
    \includegraphics[width=0.45\textwidth]{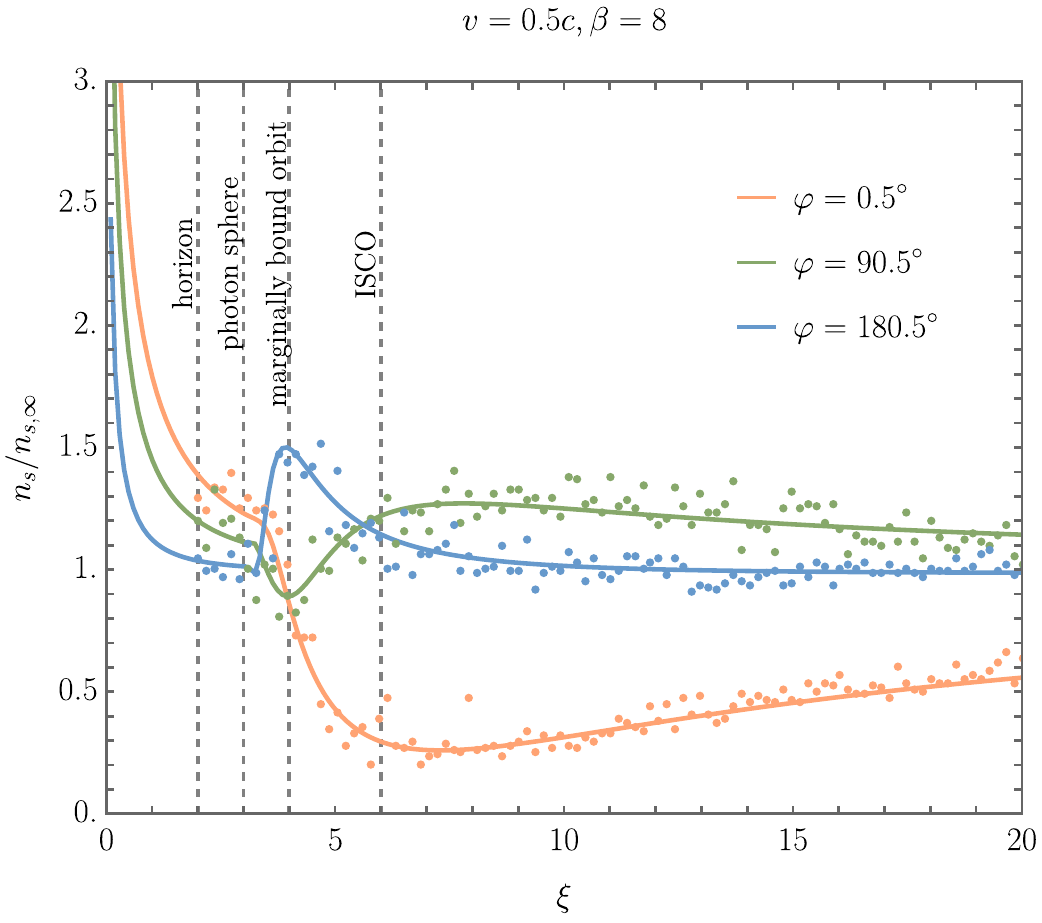}
    \caption{
    Same as in Fig.~\ref{fig:ParticleDensity-v=05-b=1} but with lower temperature.  Parameters of the model are $v = 0.5$, $\beta = 8$, $\varepsilon_\mathrm{cutoff} = 10$, and $\xi_0 = 1000$. The simulation was prepared for $30\;900\;617$ particle trajectories: $N_\mathrm{abs} = 118\;561$, $N^-_\mathrm{scat} = 15\;391\;121$, $N^+_\mathrm{scat} = 15\;390\;935$.}
    \label{fig:ParticleDensity-v=05-b=8}
\end{figure}

Results of our Monte Carlo simulations are shown in Figs.\ \ref{fig:J_t-v=095-b=1}--\ref{fig:ParticleDensity-v=05-b=8}. We compare the components of the particle current surface density $J_\mu$ and the surface density $n_s$ computed according to exact expressions (\ref{Jmuexact}) with the Monte Carlo estimators (\ref{JmuMCsim}). Figures \ref{fig:J_t-v=095-b=1}--\ref{fig:ParticleDensity-v=05-b=8} show the results obtained for 3 models with the following parameters: $v= 0.95$, $\beta = 1$ (Figs.\ \ref{fig:J_t-v=095-b=1}--\ref{fig:ParticleDensity-v=095-b=1}), $v= 0.5$, $\beta = 1$ (Figs.\ \ref{fig:J_t-v=05-b=1}--\ref{fig:ParticleDensity-v=05-b=1}), and $v= 0.5$, $\beta = 8$ (Figs.\ \ref{fig:J_t-v=05-b=8}--\ref{fig:ParticleDensity-v=05-b=8}). In all cases the maximum energy $\mathtt{energycutoff}$ was set to $\varepsilon_\mathrm{cutoff} = 10$. For a fair comparison, the same maximal energy is also used in the computation of integrals in Eqs.\ (\ref{Jmuexact})---instead of improper integrals taken over energies running up to infinity, we compute the same integrals up to $\varepsilon_\mathrm{cutoff}$. In all 3 models we assumed $\xi_0 = 1000$, as described in Sec.\ \ref{geodesicparameters}.

The graphs illustrating all 3 models share the same layout. In Figures (\ref{fig:J_t-v=095-b=1}--\ref{fig:J_fi-v=095-b=1}), (\ref{fig:J_t-v=05-b=1}--\ref{fig:J_fi-v=05-b=1}), and (\ref{fig:J_t-v=05-b=8}--\ref{fig:J_fi-v=05-b=8}) we compare the components of the particle current surface density at different angles $\varphi$. Exact solutions are shown with solid lines. Results of our Monte Carlo simulations are plotted with points. As described in Sec.\ \ref{sec:estimators}, they are only computed in a region outside the black hole horizon. Blue and green colors are used for the components corresponding to absorbed trajectories ($J_\mu^\mathrm{(abs)}$) and the total current ($J_\mu = J_\mu^\mathrm{(abs)} + J_\mu^\mathrm{(scat)}$), respectively. Figures \ref{fig:ParticleDensity-v=095-b=1}, \ref{fig:ParticleDensity-v=05-b=1}, and \ref{fig:ParticleDensity-v=05-b=8} depict the surface particle density. To illustrate the directions of the flow, we also plot directions of the vector field $(J^x, J^y)$ defined as
\begin{subequations}
\begin{eqnarray}
    J^x & = & J^r \cos{\varphi} - J^\varphi r \sin{\varphi},\\
    J^y & = & J^r \sin{\varphi} + J^\varphi r \cos{\varphi}. 
\end{eqnarray}
\end{subequations}
In each graph, we have marked the location of the horizon ($\xi=2$), the photon orbit ($\xi=3$), and the ISCO ($\xi=6$). Additionally, the location of the marginally bound orbit $(\xi=4)$ is indicated on the figures showing radial density profiles.

Numerical results shown in our examples require generating approximately $10^8$ initial geodesic parameters (this number is denoted as $\mathtt{num}$ in Sec.\ \ref{geodesicparameters}), from which we subsequently select those that satisfy Eq.\ (\ref{selectionlambda}). In the numerical examples presented in this paper we set the number of those initial parameter sets to $2\times 10^8$. This results in a different number of geodesics, depending on the assumed values of $v$ and $\beta$.

All our examples exhibit a remarkable agreement between exact expressions and results of our Monte Carlo simulations. The planar model investigated in this paper shares several common features with the three dimensional model of \cite{MachOdrzywolek2021a,MachOdrzywolek2021b}. There is a bow wave in front of the black hole, roughly between photon and marginally bound circles (cf. Figs.~\ref{fig:ParticleDensity-v=095-b=1}, \ref{fig:ParticleDensity-v=05-b=1}, \ref{fig:ParticleDensity-v=05-b=8})  and a wake behind it at ISCO distance. Around the stagnation point,  the particle number surface density drops to small values, see also Fig.~1 in \cite{MachOdrzywolek2023}. Since in this paper we mainly concentrate on the construction of our Monte Carlo scheme, we postpone the detailed discussion of this planar accretion model to some other occasion.

\section{Concluding remarks}
\label{sec:conclusions}

The Monte Carlo method introduced in \cite{MachCieslikOdrzywolek2023} and developed in this work should provide stationary solutions of the general-relativistic Vlasov equation in cases in which controlling the underlying phase-space structure is hard, and consequently exact solutions may be difficult to obtain. We have demonstrated our method by computing the particle current density, but an adaptation to other quantities, such as the components of the energy momentum tensor is straightforward.

Precise numerical simulations may, of course, be time consuming. On the other hand, two-dimensional solutions analyzed in this paper, involving the numbers of individual trajectories of the order of $10^{8}$, can be obtained using \textit{Wolfram Mathematica} \cite{Wolfram} within days on a modest computer equipped with less than 100 computing cores. We supplement this paper with a sample of \textit{Wolfram Mathematica} packages used to produce our numerical results. They will be available at \cite{Notebooks}. We believe that these packages can serve as an illustration of many details of our method, which are otherwise difficult to describe.  A clear evidence for the robustness of our Monte Carlo technique for stationary accretion problems is presented in Fig.\ \ref{fig:ParticleDensity-v=095-b=1} where, except for the statistical noise, both the surface particle density and currents are nearly identical to those obtained in a direct numerical integration. This paves a road towards problems which are unsolvable analytically at present, in particular the stationary accretion onto the Kerr black hole moving in an arbitrary direction with respect to its spin.  

We have applied our method specifically to collisionless non-magnetized systems, but its main elements are purely geometrical. We believe that it should allow for a generalization to general-relativistic Vlasov systems coupled with the electromagnetic field (see, e.g., \cite{Thaller2020}). In this case geodesic orbits should be replaced by trajectories of charged particles in a given spacetime and a given electromagnetic field (external or dynamically coupled to the gas).


\begin{acknowledgments}
A. C.\ and P.\ M.\ acknowledge a support of the Polish National Science Centre Grant No.\ 2017/26/A/ST2/00530.
\end{acknowledgments}


\appendix

\section{Boosted planar Maxwell-J\"{u}ttner distribution}
\label{appendix:distribution}

In this appendix we derive Eq.\ (\ref{dist}), which provides the distribution function corresponding to a planar stationary accretion of a collisionless gas onto a moving Schwarzschild black hole, investigated in this paper. In the asymptotic limit the distribution function given by Eq.\ (\ref{dist}) tends to the Maxwell-J\"{u}ttner distribution boosted with a constant velocity along the $x$ axis. Equation (\ref{dist}) is derived using the action-angle formalism, developed in \cite{RiosecoSarbach2017a}, and used also in \cite{CieslikMach2020, MachOdrzywolek2021a, MachOdrzywolek2021b, CieslikMachOdrzywolek2022, MachOdrzywolek2023}. In contrast to our previous works \cite{CieslikMachOdrzywolek2022, MachOdrzywolek2023} dealing with planar models, we consider a three dimensional distribution corresponding to gas particles that are not confined to the equatorial plane, and impose the restriction to a single plane at the end. This calculation follows closely the footsteps outlined in \cite{MachOdrzywolek2021a}, where the boost was applied along the $z$ axis of standard spherical coordinates \eqref{spherical_coords}. Boosting the distribution function along the $x$ axis instead allows for the subsequent adaptation to the planar problem.


\subsection{Boosted Maxwell-J\"{u}ttner distribution in the Minkowski spacetime}
        
The Maxwell-J\"{u}ttner distribution describing a gas of same rest mass particles in the Minkowski spacetime endowed with the metric
\begin{equation}
g = -dt^2 + dx^2 + dy^2 + dz^2
\end{equation}
is given by
\begin{equation}
\mathcal{F}(x, p) =  \alpha \delta \left(\sqrt{-p_\mu p^\mu} - m_0 \right) \exp{\left( \frac{\beta}{m_0} k^\mu p_\mu\right)} =  \alpha \delta \left(\sqrt{-p_\mu p^\mu} - m_0 \right)  \exp{\left( \frac{\beta}{m_0} p_t\right)},
\end{equation}
where $\alpha$ and $\beta$ are constants, $k^\mu = (1,0,0,0)$ is a timelike Killing vector, and $m_0$ denotes the rest mass of a single particle.

Applying a Lorentz boost $B^\mu{}_\nu$ with the velocity $v$ along the $x$-axis to this distribution is straightforward. Vector $k^\mu$ transforms as
\begin{equation}
k'^\mu = B^\mu{}_\nu k^\nu = (\gamma, \gamma v,0,0), \quad \gamma = \frac{1}{\sqrt{1-v^2}}
\end{equation}
in Cartesian coordinates $(t,x,y,z)$. Consequently, the boosted Maxwell-J\"{u}ttner distribution becomes
\begin{equation}
\mathcal{F}'(x, p) =  \alpha \delta \left( \sqrt{-p_\mu p^\mu} - m_0 \right)\exp{\left[ \frac{\beta}{m_0} \gamma (p_t +v p_x)\right]}.
\end{equation}
Note that the gas is boosted in the positive direction of the $x$ axis. This is equivalent to the motion of the black hole in the negative direction ($\varphi=180^\circ$) in our accretion model. By converting to spherical coordinates $(t,r,\theta,\varphi)$,
\begin{subequations}
\label{spherical_coords}
\begin{eqnarray}
x & = & r \cos{\varphi} \sin{\theta},\\
y & = & r \sin{\varphi} \sin{\theta},\\
z & = & r \cos{\theta},
\end{eqnarray}
\end{subequations}
we obtain the distribution if the form
\begin{equation}\label{F_min}
\mathcal{F}'(x, p) =  \alpha  \delta \left( \sqrt{-p_\mu p^\mu} - m_0 \right) \exp{\left\{ \frac{\beta}{m_0} \gamma\left[ p_t + v\left( \sin{\theta} \cos{\varphi} \; p_r  + \frac{\cos{\varphi} \cos{\theta}}{r} p_\theta  - \frac{\sin{\varphi}}{r \sin{\theta}} p_\varphi \right) \right] \right\} }.
\end{equation}
For simplicity, we will henceforth omit the prime in $\mathcal{F}$.


\subsection{Boosted Maxwell-J\"{u}ttner distribution and action-angle variables}

The Hamiltonian describing a geodesic motion of a free particle in Minkowski spacetime in spherical coordinates is given by
\begin{equation}
H = \frac{1}{2} \left[ - p_t^2 + p_r^2 + \frac{1}{r^2} \left( p_\theta^2 + \frac{p_\varphi^2}{\sin^2{\theta}} \right) \right].
\end{equation}
The corresponding solutions of the Hamilton equations---geodesics in the Minkowski spacetime---are simply straight lines. This formulation implies that the energy $E=-p_t$, the azimuthal component of the angular momentum $l_z = p_\varphi$, and the particle rest mass $m = \sqrt{-2H}$, are all conserved. Additionally, the total angular momentum
\begin{equation}
l = \sqrt{p_\theta^2 + \frac{p_\varphi^2}{\sin^2{\theta}}}
\end{equation}
is also conserved. For geodesics with fixed values of  $E$, $l$, $l_z$, and $m$, the radial momentum is determined by
\begin{equation}
\label{pr_min}
p_r = \epsilon_r \sqrt{E^2 -m^2 - \frac{l^2}{r^2}},
\end{equation}
where $\epsilon_r = \pm 1$ differentiates between ingoing and outgoing particles. The component $p_\theta$ is expressed as
\begin{eqnarray}
\label{ptheta_min}
p_\theta = \epsilon_\theta \sqrt{l^2 - \frac{l_z^2}{\sin^2{\theta}}},
\end{eqnarray}
with $\epsilon_\theta = \pm 1$ similarly defined.

For a motion along a segment of a geodesic $\hat \gamma$ with constant $E$, $l$, $l_z$, $m$, and constant signs $\epsilon_r$ and $\epsilon_\theta$, we define the abbreviated action as
\begin{eqnarray}
S &=& \int_{\hat \gamma} p_\mu dx^\mu = - E t + l_z \varphi + \int_{\hat \gamma} p_r dr + \int_{\hat \gamma} p_\theta d\theta \nonumber \\
&=& -Et + l_z \varphi + \epsilon_r \int \sqrt{E^2 - m^2 -\frac{l^2}{r^2}} dr + \epsilon_\theta \int  \sqrt{l^2 - \frac{l_z^2}{\sin^2{\theta}}} d\theta \nonumber \\
&=& -Et + l_z \varphi  + \epsilon_r \left[ r\sqrt{E^2 - m^2 - \frac{l^2}{r^2}} - l\arctan{\left( \frac{r}{l} \sqrt{E^2 - m^2 - \frac{l^2}{r^2}}\right)} \right] \nonumber \\
&& + \epsilon_\theta \left[ -l\arctan{\left( \frac{l\cot{\theta}}{\sqrt{l^2 - \frac{l_z^2}{\sin^2{\theta}}}} \right)} + l_z \arctan{\left( \frac{l_z \cot{\theta}}{\sqrt{l^2 - \frac{l_z^2}{\sin^2{\theta}}}} \right)} \right] + \mathrm{const}.
\label{S_min}
\end{eqnarray}
Here, $p_r$ and $p_\theta$ are defined in Eqs.\ (\ref{pr_min}) and (\ref{ptheta_min}). We introduce the following canonical transformation:  $\left( t, r, \theta, \varphi , p_t, p_r, p_\theta, p_\varphi \right) \longrightarrow \left( Q^\mu, P_\nu \right)$, where the new momenta are constants of motion: 
\begin{subequations}
\label{canon_mom_min}
\begin{align}
P_0 &= m = \sqrt{ p_t^2  - p_r^2 - \frac{1}{r^2}\left( p_\theta^2 + \frac{p_\varphi^2}{\sin^{\theta}} \right) } ,\\
P_1 &= E = -p_t,\\
P_2 &= l_z = p_\varphi,\\
P_3 &= l = \sqrt{p_\theta^2 + \frac{p_\varphi^2}{\sin^2{\theta}}},
\end{align}
\end{subequations}
and
\begin{subequations}
\label{Q_min}
\begin{align}
Q^0 &= \frac{\partial S}{\partial m} = -\frac{ \epsilon_r mr \sqrt{E^2 -m^2 - \frac{l^2}{r^2}}}{E^2 - m^2} = - \frac{rp_r}{p_r^2+ \frac{1}{r^2} \left( p_\theta^2 + \frac{p_\varphi^2}{\sin^2{\theta}} \right)}\sqrt{p_t^2 - p_r^2 - \frac{1}{r^2} \left( p_\theta^2 + \frac{p_\varphi^2}{\sin^2{\theta}} \right) },\\
Q^1 &= \frac{\partial S}{\partial E} =  -t + \frac{\epsilon_r Er\sqrt{E^2 -m^2 -\frac{l^2}{r^2}}}{E^2 - m^2} = -t - \frac{rp_t p_r}{p_r^2+ \frac{1}{r^2} \left( p_\theta^2 + \frac{p_\varphi^2}{\sin^2{\theta}} \right)},\\
Q^2 &=\frac{\partial S}{\partial l_z} = \varphi + \epsilon_\theta \arctan{\left( \frac{l_z \cot{\theta}}{\sqrt{l^2 - \frac{l_z^2}{\sin^2{\theta}}}} \right)} = \varphi + \arctan{\left( \frac{p_\varphi \cot{\theta}}{p_\theta} \right)} ,\\
Q^3 &= \frac{\partial S}{\partial l}  = -\epsilon_r \arctan{\left( \frac{r}{l} \sqrt{E^2 - m^2 - \frac{l^2}{r^2}} \right)} - \epsilon_\theta \arctan{\left( \frac{l \cot{\theta}}{\sqrt{l^2 - \frac{l_z^2}{\sin^2{\theta}}}}\right)} \nonumber \\
& = - \arctan{\left( \frac{r p_r}{\sqrt{p_\theta^2 + \frac{p_\varphi^2}{\sin^2{\theta}}}} \right)} - \arctan{\left( \frac{\sqrt{p_\theta^2 + \frac{p_\varphi^2}{\sin^2{\theta}}}\cot{\theta}}{p_\theta} \right)}.\label{Q3_min}
\end{align}
\end{subequations}       
(see \cite{MachOdrzywolek2021a}). In the above formulas, the last form is given in terms of the momenta $(p_t, p_r, p_\theta, p_\varphi)$. To derive expressions \eqref{Q_min}, we assume that the constant in Eq. \eqref{S_min} is independent of $m$, $E$, $l_z$, and $l$.
        
Basing on \cite{MachOdrzywolek2021a}, one can note several relationships between the variables $(Q^\mu,P_\nu)$ and $(x^\mu,p_\nu)$. These relationships will be useful in the following sections:
\begin{equation}
\frac{\cot{\theta}}{p_\theta} = \frac{1}{P_3} \tan\left[ -Q^3 - \arctan\left( \frac{r p_r}{\sqrt{p_\theta^2 + \frac{p_\varphi^2}{\sin^2{\theta}}}} \right) \right],
\end{equation}
\begin{equation}
\varphi = Q^2 -  \arctan{\left( \frac{p_\varphi \cot{\theta}}{p_\theta} \right)} =  Q^2 -  \arctan\left\{ \frac{P_2}{P_3} V\right\},
\end{equation}
where
\begin{equation}\label{eqV}
V =\tan\left[ -Q^3 - \arctan\left( \frac{r p_r}{\sqrt{p_\theta^2 + \frac{p_\varphi^2}{\sin^2{\theta}}}} \right) \right].
\end{equation}
In a similar fashion we get
\begin{equation}
\label{sinTh}
\sin^2{\theta} = \frac{P_3^2 + P_2^2 V^2}{P_3^2 (1+V^2)}.
\end{equation}


\subsection{Asymptotic relations}

In the asymptotic limit of $r \to \infty$, Eq.\ (\ref{F_min}) becomes
\begin{equation}
\label{dist_min_as}
\mathcal{F}(x, p) =  \alpha  \delta \left(\sqrt{-p_\mu p^\mu} - m_0 \right) \exp{\left\{ \frac{\beta}{m_0} \gamma\left[ p_t + v  \sin{\theta}\cos{\varphi} \; p_r  \right] \right\} }.
\end{equation}
Following the footsteps outlined in \cite{MachOdrzywolek2021a}, we write the following asymptotic expressions. The radial component of the four-momentum tends to
\begin{equation}
p_r = \epsilon_r \sqrt{E^2 - m^2} = \epsilon_r \sqrt{P_1^2 - P_0^2}.
\end{equation}
In the asymptotic limit, \eqref{eqV} takes the following form 
\begin{equation}
V = - \tan\left(Q^3 + \epsilon_r \frac{\pi}{2}\right)= \cot{Q^3},
\end{equation}
and Eq.\ (\ref{sinTh}) tends to
\begin{equation}
\sin{\theta} = \sqrt{\sin^2{Q^3} + \frac{P_2^2}{P_3^2} \cos^2{Q^3}},
\end{equation}
where we used the fact that  $ 0 < \theta < \pi$. Therefore $\cos{\varphi}$ can be now expressed as
\begin{equation}
\cos{\varphi} = \cos\left\{ Q^2 - \arctan\left[ \frac{P_2}{P_3} \cot{Q^3}\right] \right\} = \frac{\cos{Q^2}|\sin{Q^3}| + \frac{P_2}{P_3} \frac{|\sin{Q^3}| }{\sin{Q^3}}\sin{Q^2} \cos{Q^3}}{\sqrt{\sin^2{Q^3} + \frac{P_2^2}{P_3^2} \cos^2{Q^3}}}.
\end{equation}
Hence,
\begin{equation}
\cos{\varphi}\sin{\theta} =\cos{Q^2}|\sin{Q^3}| + \frac{P_2}{P_3} \frac{|\sin{Q^3}| }{\sin{Q^3}}\sin{Q^2} \cos{Q^3},
\end{equation}
and
\begin{equation}
\sin{Q^3} = - \sin \left[ \arctan \left( \omega \right) + \epsilon_r \frac{\pi}{2}\right] =  - \epsilon_r \cos\left[ \arctan\left(  \omega \right) \right],
\end{equation}
where \( \omega=\frac{l \cot{\theta}}{p_\theta} \). Since \( -\frac{\pi}{2}\leq\arctan \omega  \leq \frac{\pi}{2} \), it follows that \( \cos\left[ \arctan\left(  \omega \right) \right]\geq 0 \). Consequently, one obtains 
\begin{equation}
\sin Q^3 = - \epsilon_r |\sin Q^3|.
\end{equation}
Finally substituting the above expression in Eq.\ (\ref{dist_min_as}), we get
\begin{equation}\label{F_gen}
\mathcal{F}(x, p) = \alpha  \delta( P_0 - m_0 ) \exp{\left\{ -\frac{\beta}{m_0} \gamma\left[ P_1 + v  \sqrt{P_1^2 - P_0^2} \left(  \cos{Q^2}\sin{Q^3} + \frac{P_2}{P_3}  \sin{Q^2} \cos{Q^3} \right)  \right] \right\} }.
\end{equation}
It is important to note that this expression is valid for the three-dimensional motion of the gas. In the following section, we deal with asymptotic expressions corresponding to the planar motion of the gas.


\subsection{Planar motion in the asymptotic limit
\label{Appendix:Planar_asymptotic_limit}
}

To analyze the motion confined to the equatorial plane, we consider the limit $\theta \to \pi/2$  and $ p_\theta \to 0$, leading to $l_z \to \epsilon_\varphi l$, where $\epsilon_\varphi = \mathrm{sgn}(l_z)$. An application of those limits to \eqref{F_gen} is straightforward. Equations (\ref{canon_mom_min}) yield
\begin{equation}
P_2 \to \epsilon_\varphi P_3.
\end{equation}
Consequently, using the notation \eqref{dist_flat}, we have
\begin{equation}
    \mathcal F(x, p) = \delta\left(\theta - \frac{\pi}{2}\right)\delta(p_\theta)f(x, p),
\end{equation}
where
\begin{equation}
f(x, p) = \alpha  \delta( P_0 - m_0 ) \exp{\left\{ -\frac{\beta}{m_0} \gamma\left[ P_1 +  v  \sqrt{P_1^2 - P_0^2} \sin\left(  Q^3 + \epsilon_\varphi Q^2 \right)  \right] \right\} }.
\end{equation}
Additionally, from Eqs.\ \eqref{Q_min}, we obtain
\begin{equation}
Q^3 + \epsilon_\varphi Q^2 \to \epsilon_\varphi  \varphi - \epsilon_r  \frac{\pi}{2}.
\end{equation}
Therefore,
\begin{equation}
f(x, p) = \alpha  \delta( P_0 - m_0 ) \exp{\left\{ -\frac{\beta}{m_0} \gamma\left[ P_1 + v  \sqrt{P_1^2 - P_0^2} \sin\left(  \epsilon_\varphi  \varphi - \epsilon_r  \frac{\pi}{2}   \right) \right] \right\} }.
\end{equation}
Substituting \eqref{canon_mom_min}, we get
\begin{eqnarray}
f(x,p) & = & \alpha  \delta(\sqrt{-p_\mu p^\mu} - m_0 ) \exp{\left\{ -\beta \gamma\left[ \varepsilon +  \epsilon_\varphi    v  \sqrt{\varepsilon^2 - 1} \sin\left( \varphi -  \epsilon_\varphi  \epsilon_r\frac{\pi}{2}  \right)  \right] \right\} } \nonumber \\ 
& = & \alpha   \delta(\sqrt{-p_\mu p^\mu} - m_0 ) \exp{\left\{ -\beta \gamma\left[ \varepsilon - \epsilon_r v  \sqrt{\varepsilon^2 - 1} \cos\varphi  \right] \right\} }.
\label{MinkDist}
\end{eqnarray}

    
\subsection{Action-angle variables in the Schwarzschild spacetime} 
        
As previously, we consider the motion along a segment of a geodesic $\hat \gamma$ with constant $E$, $l$, $l_z$, $m$, and constant signs $\epsilon_r$ and $\epsilon_\theta$, and we define the abbreviated action as
\begin{equation}
S = \int_{\hat \gamma} p_\mu dx^\mu = - E t + l_z \varphi + \int_{\hat \gamma} p_r dr + \int_{\hat \gamma} p_\theta d\theta.
\end{equation}
In this case,  $p_r$  and $p_\theta$  are defined by Eqs.\ \eqref{pr} and \eqref{ptheta}. We now introduce the following canonical transformation:
\begin{subequations}\label{canon_mom}
\begin{align}
P_0 &= m = \sqrt{\frac{1}{N} E^2 - N p_r^2 - \frac{1}{r^2}\left( p_\theta^2 + \frac{p_\varphi^2}{\sin^2{\theta}} \right) } ,\\
P_1 &= E = -p_t,\\
P_2 &= l_z = p_\varphi,\\
P_3 &= l = \sqrt{p_\theta^2 + \frac{p_\varphi^2}{\sin^2{\theta}}},
\end{align}
\end{subequations}
and
\begin{subequations}
\begin{eqnarray}
Q^0 & = & \frac{\partial S}{\partial m} = -m \epsilon_r \int_{\hat \gamma} \frac{dr}{\sqrt{E^2 - U(r)}} = -m \int_{\hat \gamma} \frac{dr}{Np_r} = -m \int_{\hat \gamma} \frac{dr}{g^{rr} p_r} ,\\
Q^1 & = & \frac{\partial S}{\partial E} =  -t + \epsilon_r E \int_{\hat \gamma} \frac{dr}{N \sqrt{E^2 - U(r)}} = -t + \int_{\hat \gamma} \frac{\frac{1}{N} E}{Np_r} dr = -t - \int_{\hat \gamma} \frac{g^{tt} E}{g^{rr}p_r} dr,\\
Q^2 & = &\frac{\partial S}{\partial l_z} = \varphi + \epsilon_\theta \arctan{\left( \frac{l_z \cot{\theta}}{\sqrt{l^2 - \frac{l_z^2}{\sin^2{\theta}}}} \right)} = \varphi + \arctan{\left( \frac{p_\varphi \cot{\theta}}{p_\theta} \right)} ,\\
Q^3 & = & \frac{\partial S}{\partial l}  = - \epsilon_\theta \arctan{\left( \frac{l \cot{\theta}}{\sqrt{l^2 - \frac{l_z^2}{\sin^2{\theta}}}}\right)}  - l \epsilon_r \int_{\hat \gamma} \frac{dr}{r^2 \sqrt{E^2 - U(r)}} \nonumber \\
& & - \arctan{\left( \frac{\sqrt{p_\theta^2 + \frac{p_\varphi^2}{\sin^2{\theta}}}\cot{\theta}}{p_\theta} \right)} - l \int_{\hat \gamma} \frac{dr}{r^2 N p_r}.
\label{Min_Q3}
\end{eqnarray}
\end{subequations}

Using \eqref{EllipticX} we could reformulate Eq. \eqref{Min_Q3} as 
\begin{equation}
Q^3 = -  \epsilon_\theta \arctan{\left( \frac{l \cot{\theta}}{\sqrt{l^2 - \frac{l_z^2}{\sin^2{\theta}}}}\right)} + \epsilon_r X(\xi,\varepsilon,\lambda).
\end{equation}
However, $\lim_{\xi\to\infty}X(\xi, \varepsilon, \lambda) = 0$, which would disagree with the asymptotic value of $Q^3$ calculated from Eq.\ (\ref{Q3_min}), i.e.,
\begin{equation}
\lim_{r\to\infty}Q^3 = -\epsilon_r\frac{\pi}{2} - \epsilon_\theta \arctan{\left( \frac{l \cot{\theta}}{\sqrt{l^2 - \frac{l_z^2}{\sin^2{\theta}}}}\right)}.
\end{equation}
Therefore, to ensure consistency with the asymptotic behavior, we choose
\begin{equation}
Q^3 = -  \epsilon_\theta \arctan{\left( \frac{l \cot{\theta}}{\sqrt{l^2 - \frac{l_z^2}{\sin^2{\theta}}}}\right)} + \epsilon_r X(\xi,\varepsilon,\lambda) - \epsilon_r \frac{\pi}{2}.
\end{equation}


\subsection{Boosted Maxwell-J\"{u}ttner distribution on the equatorial plane}        

Equation (\ref{F_gen}) yields a general form the distribution function expressed in the action-angle coordinates
\begin{equation}
\mathcal F (x, p) = \alpha  \delta(P_0 - m_0 ) \exp{\left\{ -\frac{\beta}{m_0} \gamma\left[ P_1 +  v  \sqrt{P_1^2 - P_0^2} \left(  \cos{Q^2}\sin{Q^3} + \frac{P_2}{P_3}  \sin{Q^2} \cos{Q^3} \right)  \right] \right\} }.
\end{equation}
As in Appendix \ref{Appendix:Planar_asymptotic_limit}, we take a limit $\theta \to \pi/2$ and $p_\theta \to 0$, so that $l_z \to \epsilon_\varphi l$. Similarly
\begin{subequations}
\begin{eqnarray}
&& P_2  \to \epsilon_\varphi P_3,\\
&& Q^3 +  \epsilon_\varphi  Q^2 \to \epsilon_\varphi  \varphi + \epsilon_r X(\xi,\varepsilon,\lambda) - \epsilon_r \frac{\pi}{2},
\end{eqnarray}
\end{subequations}
and therefore 
\begin{equation}
f(x, p) = \alpha  \delta( P_0 - m_0 ) \exp{\left\{ -\frac{\beta}{m_0} \gamma\left[ P_1 +  v  \sqrt{P_1^2 - P_0^2} \sin\left[ \epsilon_\varphi\varphi + \epsilon_r X(\xi,\varepsilon,\lambda) - \epsilon_r \frac{\pi}{2}  \right] \right] \right\} }.
\end{equation}
Substituting Eq.\ (\ref{canon_mom}), one obtains the final expression
\begin{eqnarray}
f(x, p) & = & \alpha  \delta \left( \sqrt{-p_\mu p^\mu} - m_0 \right) \exp{\left\{ -\beta \gamma\left[ \varepsilon + \epsilon_\varphi  v  \sqrt{\varepsilon^2 - 1} \sin\left[ \varphi + \epsilon_\varphi \epsilon_r \left( X(\xi,\varepsilon,\lambda) - \frac{\pi}{2} \right)  \right] \right] \right\} } \nonumber \\
&=& \alpha  \delta \left( \sqrt{-p_\mu p^\mu} - m_0 \right) \exp{\left\{ -\beta \gamma\left[ \varepsilon - \epsilon_r v  \sqrt{\varepsilon^2 - 1} \cos\left[ \varphi + \epsilon_\varphi \epsilon_r  X(\xi,\varepsilon,\lambda)  \right] \right] \right\} }.
\label{F_schwar}
\end{eqnarray}


\section{Particle current surface density}
\label{appendix:Currensts}

To evaluate momentum integrals in Eq.\ (\ref{jmusurface}), we change the variables according to $(p_t,p_r,p_\varphi) \to (m^2,E,l_z)$, where
\begin{subequations}
\begin{eqnarray}
m^2 & = &\frac{1}{N} p_t^2 - N p_r^2 - \frac{1}{r^2} p_\varphi, \\
E & = & -p_t,\\
l_z & = & p_\varphi.
\end{eqnarray}
\end{subequations}
The Jacobian of this transformation is simply
\begin{equation}
\frac{\partial (m^2, E, l_z)}{\partial(p_t, p_r, p_\varphi)} = -2 N p_r = - 2 \epsilon_r \sqrt{E^2 - U(r)}  = - 2 \epsilon_r m \sqrt{\varepsilon^2 - U_\lambda}.
\end{equation}
Consequently,
\begin{equation}
\frac{dp_t dp_r dp_\varphi}{r} = \frac{dm dE dl_z}{r \sqrt{\varepsilon^2 - U_\lambda}} = \frac{m^2 dm d \varepsilon d \lambda_z}{\xi \sqrt{\varepsilon^2 - U_\lambda}},
\end{equation}
where we introduced the dimensionless variables $\varepsilon$ and $\lambda_z$. Thus Eq.\ (\ref{jmusurface}) reads
\begin{eqnarray}
J_\mu(t,r,\varphi) & = & \int f(t,r,\varphi,p_t,p_r,p_\varphi) p_\mu \frac{dp_t dp_r  dp_\varphi}{r} = \frac{1}{\xi} \int f(t,\xi,\varphi,m,\varepsilon,\lambda_z) p_\mu \frac{m^2 dm d \varepsilon d \lambda_z}{\sqrt{\varepsilon^2 - U_\lambda}} \nonumber \\
& = & \sum_{\epsilon_\varphi = \pm 1} \frac{1}{\xi} \int f(t,\xi,\varphi,m,\varepsilon,\epsilon_\varphi,\lambda) p_\mu \frac{m^2 dm d \varepsilon d \lambda}{\sqrt{\varepsilon^2 - U_\lambda}},
\end{eqnarray}
where in the last step we used the relation $\lambda_z = \epsilon_\varphi \lambda$, valid at the equatorial plane, and where we also abuse the notation by allowing ourselves to change freely the arguments in $f$.

Assuming $f$ in the form of Eq.\ (\ref{F_schwar}) and
\begin{equation}
(p_t,p_r,p_\varphi) = \left( - m \varepsilon, \frac{\epsilon_r m}{N} \sqrt{\varepsilon^2 - U_\lambda}, M m \lambda_z \right) = m \left( - \varepsilon, \frac{\epsilon_r}{N} \sqrt{\varepsilon^2 - U_\lambda} , \epsilon_\varphi M \lambda \right)
\end{equation}
we get Eqs.\ (\ref{Jmuexact}).


\end{document}